\documentclass{aa}  
\usepackage{graphicx}
\usepackage{longtable}
\usepackage[varg]{txfonts}

\begin{document}
\title{Dissecting the Planetary Nebula NGC~4361 with MUSE}
 \thanks{Based on observations collected at the European Organisation for 
 Astronomical Research in the Southern Hemisphere, Chile in Science 
 Verification (SV) observing proposal 60.A-9100(B).
}
  
\titlerunning{NGC~4361 with MUSE}

\author{J. R. Walsh\inst{1} \and  A. Monreal Ibero\inst{2} \and J.
Laging\inst{2} \and M. Romeijnders\inst{3,2} }
\institute{European Southern Observatory, Karl-Schwarzschild Strasse 2,
D-85748 Garching, Germany \\
% ORCID https://orcid.org/0000-0002-8008-910X
\email{jwalsh@eso.org}
\and
Leiden Observatory, Leiden University, P.O. Box 9513, 2300 RA Leiden,
the Netherlands
% \email{monreal@strw.leidenuniv.nl,laging@strw.leidenuniv.nl}
% ORCID https://orcid.org/0000-0002-6455-2491
\and
Department of Information and Computing Sciences, Utrecht University,
the Netherlands 
%            \email{m.c.romeijnders@uu.nl} % Optional, now commented
% ORCID: https://orcid.org/0009-0000-3605-3807
}

\date{Received: 13 March 2024; accepted: 01 August 2024}

\abstract 
% context heading (optional)
{Optical integral field spectroscopy of planetary nebulae (PNe) provides a unique tool to
explore the spatial relationships between the complex mixture of the many components -- 
neutral, low and high ionization gas, dust and central star -- and their underlying physical 
conditions. 
} 
% aims heading (mandatory)
{The optical line and continuum emission in the very high ionization Galactic PN, NGC~4361, 
has been mapped in order to study the distribution of ionization, extinction, electron 
temperature and density.
} 
% methods heading (mandatory)
{Based on commissioning data, MUSE Wide Field (60$\times$60$''$) normal mode 
(4750-9300\,\AA) observations of NGC~4361 were reduced. The PN is larger than 
a single MUSE field and only the central 1 arcmin$^{2}$ of the PN 
was observed in good conditions. Emission images in recombination and collisionally 
excited lines were extracted and line ratios provide dust extinction, electron 
density and temperature and ionic abundances using standard 
techniques. A family of compact low ionization knots (dubbed 'Freckles') was discovered 
and techniques were developed to measure their spectra, as distinct from the extended 
high ionization medium.
} 
% results heading (mandatory)
{The nebula is confirmed as optically thin in the H-ionizing continuum based
on its very low \ion{He}{i} emission, even to the edges of the field. The
electron temperature $T_{\rm e}$ is shown to have large-scale spatially 
coherent structure as indicated from a previous long-slit spectrum. 
Prior to this study, no low ionization emission had been positively detected; MUSE 
revealed both weak extended [\ion{N}{ii}] and [\ion{O}{ii}] and $>$100 spatially
unresolved knots. There are several linear associations of these
knots but none point back convincingly to the central star. They have low-moderate
ionization with $T_{\rm e}$ $\sim$~11000\,K, $N_{\rm e}$ $\sim$1500 cm$^{-3}$ 
and generally show higher extinction than the extended high ionization nebula.

Within the MUSE field, a low redshift emission line galaxy was serendipitously
found hiding behind NGC~4361. The spectrum of this dwarf galaxy was carefully
extracted from the bright foreground nebular emission and the galaxy's line and 
continuum properties were determined. 
}
% conclusions heading (optional)
{
NGC~4361 is not completely optically thin, as indicated by several extended regions 
and many compact features of lower ionization emission. The identified low ionization 
'Freckles' do not clearly appear to differ in (He, N, O, S) abundance with respect to the 
extended high ionization gas. The spatial distribution and radial velocities
of these features suggest that they belong to a thick disk oriented perpendicular to the 
large-scale nebular gas, perhaps remnants of an earlier structure.

The low luminosity disk galaxy at $\sim$87\,Mpc has bright \ion{H}{ii} regions with 
metallicity 12+log(O/H) $\cong$ 8.4 and is suggested as a Magellanic irregular or 
low-mass spiral.
}


\maketitle
%
%----------------------------------------------------------------------------------
\nolinenumbers
\section{Introduction}
\label{Sect:Intro}

NGC~4361 (PN G294.1 $+$43.6) is one of the rare planetary nebulae (PNe) with \ion{He}{II} 
4686\AA\ emission stronger than H$\beta$; its nebula ionization class defined by 
\ion{He}{II} / H$\beta$ \citep{Dopita1990} 
exceeds 10, the maximum value. Even given the temperature of hot PN central
stars, such a condition is suggestive of a nebula optically thin in the
Lyman continuum (i.e., matter bounded). The high ionization of NGC~4361 has been 
confirmed by the presence of other high ionization species from optical \citep{Heap1969,Aller1978, 
TorresPeimbert1990} and ultra-violet (UV) \citep{AdamKoeppen1985} spectroscopy. 
\citet{Aller1978} reports detection of 
[\ion{Ne}{V}]3426\,\AA\ indicating ionization by $>$120\,eV photons and \citet{TorresPeimbert1990} 
from the [\ion{O}{III}]4363/5007\,\AA\ ratio determined a high value of $T_{\rm e}$ of 
$\sim$ 18000\,K, later confirmed by \citet{Liu1998}. The nebula is also detected  
in extended X-ray emission by EXOSAT \citep{Apparaoetal1989}
and ROSAT \citep{Kreysingetal1992}, indicating a plasma at $<$2$\times$10$^{5}$K. 
\citet{Aller1978} noted the extreme weakness of the low ionization lines and to date no 
studies have detected [\ion{N}{II}] emission in NGC~4361. 

The bright central star (Gaia mean G mag. 13.0876) was observed by 
\citet{Mendezetal1981, Mendez1992} and fit by a model atmosphere $\sim$80\,kK 
but the interpretation is complicated by the strong \ion{He}{II} emission filling 
in the stellar absorption lines. \citet{Ziegleretal2012} analysed Far Ultraviolet 
Spectroscopic Explorer (FUSE) spectra finding a 126\,kK central star (CS) with
log $g$ of 6.0. Several attempts at modelling the nebular spectrum have been made: 
the early model of \citet{Aller1979} inferred typical Galactic PN metallicity 
12 + log$_{10}$ (O/H) = 8.6 ([O/H]) and a CS of $\sim$80\,kK, while 
\citet{TorresPeimbert1990} suggest a low metallicity of [O/H] = 8.2 with a 90\,kK 
CS. \citet{Howardetal1997} find [O/H] of 8.15 for a 120\,kK star. The model of
\citet{TorresPeimbert1990} showed an overall C/O ratio > 1, but with indications 
of the inner zone being C richer. While the high C/O was confirmed by the spectroscopy 
of \citet{Liu1998} from analysis of the \ion{C}{III} and \ion{C}{IV} recombination 
lines, the abundance of C was not confirmed to vary spatially. The electron density 
is measured at $\sim$1200--1500 cm$^{-3}$ \citep{Aller1979, Liu1998}.
 
% Liu gets Te=17900\,K mean and 1200cm-3 with no variation of C abundance as measured by
% C reco lines. Some variation in Te along slit 10'' N of CS.

Given the Galactic latitude and possible evidence for low (O/H), \citet{TorresPeimbert1990} 
suggested NGC~4361
as a Galactic halo PN, so of Population II low-mass progenitor, but given the distance 
from Gaia ($\pi$ = 0.9653 $\pm$ 0.0439 mas, inferred distance 1040\,pc) it may be in the 
thick disc. 

The overall morphology is elliptical with hook-shaped extensions to NE and SW 
without a clear differentiation of a shell. The morphology has been variously 
reported as filamentary and with a halo, but integral field spectroscopy by
\citet{MonrealIbero2006} with the VLT Imaging and Multi-Object Spectrometer (VIMOS)
found no true halo below $\sim$10$^{-2}$ of the central 
surface brightness beyond a nebula extension of $\sim$110$''$.  
From the kinematics, NGC~4361 was initially suggested to be bipolar based on long 
slit observations \citep{Vazquezetal1999}. The more extensive spatial coverage by 
\citet{MuthaAnandarao2001} suggested a double bipolar (quadrupolar) since velocity-split 
line profiles occur along NE-SW and NW-SE axes. The velocity profile separation of upto 
70 km s$^{-1}$ \citep{Vazquezetal1999} is notably high for a PN.

NGC~4361 was included in the MUSE commissioning (c.f., NGC~3132 \citep{MonrealIbero2020}, 
IC~418 \citep{MonrealIbero2022}, IC~4406 \citep{RamosLarios2022}) on
account of its high emission line surface brightness and its size, which overfills 
the MUSE field-of-view (60$\times$60$''$). With their depth, field coverage, spatial 
resolution and spectra over the
optical range, these observations bring important new data to bear on the nature of this
extreme PN, in particular detecting low ionization emission for the first time, but with 
a surprising structure. Section \ref{Sect:ObsRed} describes the MUSE observations and 
the major reduction steps to the data cubes. Section \ref{Sect:Images} presents and 
describes the emission line and line ratio maps and Section \ref{Sect:PhysCond} the 
extinction and line diagnostic images. In Section \ref{Sect:Freckles}  we analyse 
the many compact [\ion{N}{II}] knots, whose striking contrast to the appearance of the 
nebula in all lines of higher ionization and their small size, led us to nickname 
NGC~4361 'The Freckles Nebula'. Section \ref{Sect:Discuss} then presents a discussion 
on these Freckles and their relation to the high ionization nebula. In the outer field, 
a serendipitously discovered galaxy 'shining' through the nebula was identified
\citep{Baconetal2014} and in Section \ref{Sect:Steve} analysis of this target is 
presented. Section \ref{Sect:Concl} draws together the conclusions.  

%__________________________________________________________________

\section{Observations and reductions}
\label{Sect:ObsRed}

\begin{figure}
\centering
\resizebox{\hsize}{!}{
\includegraphics[width=0.45\textwidth,angle=0,clip]{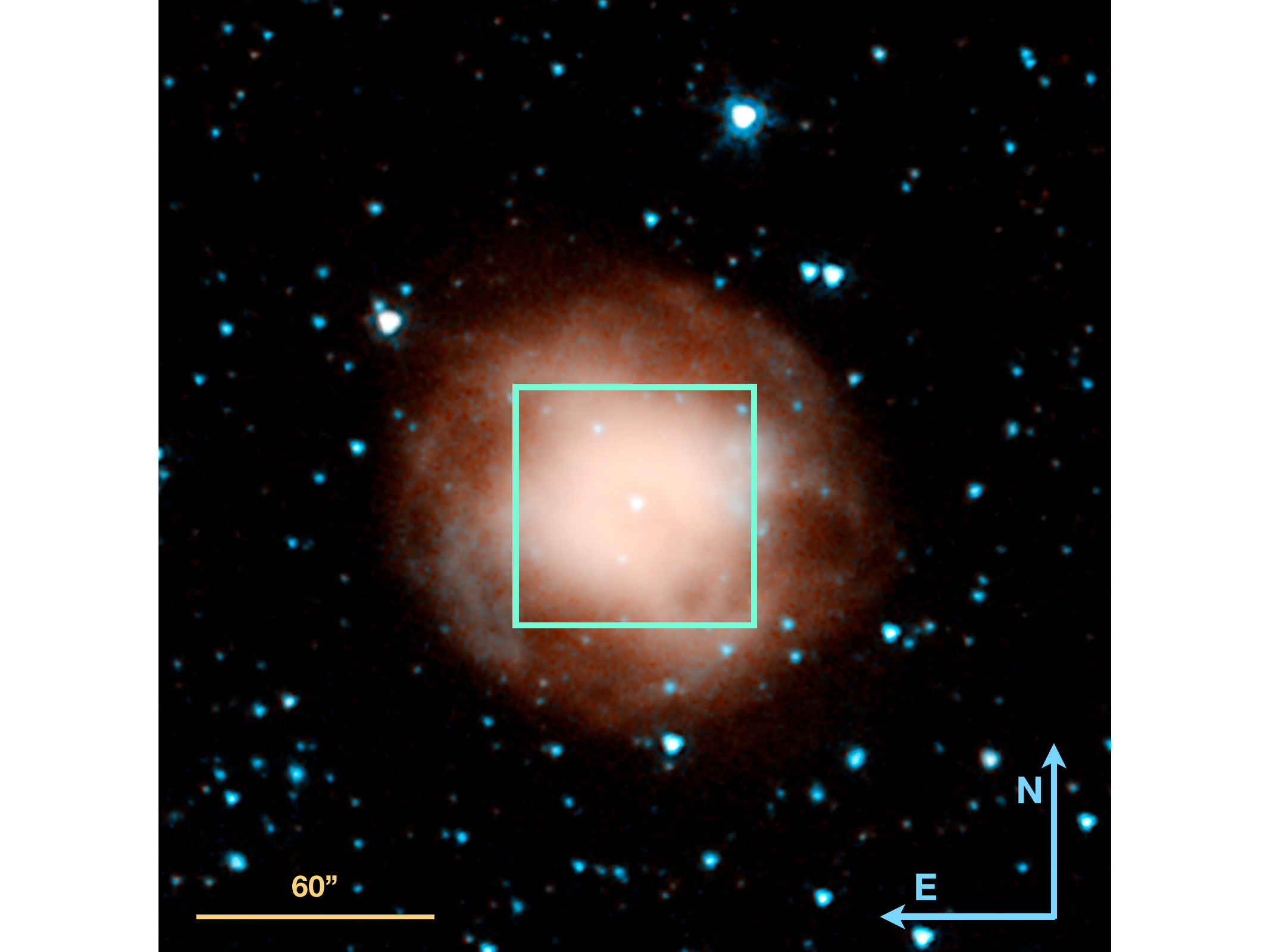}
}

\caption{The location of the central MUSE field is shown, superposed on an image of the
wider field including the nebula extent, from the Spitzer IRAC combined 3.6 and 4.5 
$\mu$m image. The scale and orientation are indicated. \\
Image credit: NASA/JPL-Caltech
}
\label{Fig:MUSEFields} 
\end{figure}

NGC~4361 was observed during the second MUSE commissioning run in May 2014
over 5 nights. MUSE \citep{Baconetal2014} at the Very Large Telescope (VLT) 
was deployed in standard Wide Field Mode (WFM, 60$\times$60$''$ field with 
0.20$''$ spaxels) with the default wavelength range 4750--9300\,\AA\ at 
1.25\,\AA\ sampling (MUSE WFM-NOAO-N mode). A total of 74 exposures 
of 60\,s each were obtained, with 58 centred on the central star of the nebula 
and 16 at an offset position 30$''$ W to sample the extent of the 
nebula. Figure \ref{Fig:MUSEFields} shows the central pointing overlaid on the 
Spitzer Infrared Array Camera (IRAC) 3.5 + 4.5$\mu$m image. Between 
exposures 90$^{\circ}$ rotations (PA's of 0, 90, 180 and 270$^{\circ}$ were 
covered) and dithers of 0.4$''$ offsets were generally applied in order
to average out the pattern of the 24 individual integral field units of
MUSE. On account of poorer seeing and low transparency during some 
exposures, a selection was made based on measured flux in a narrow
passband centred on H$\alpha$ (the passband of the Hubble Space Telescope
Wide Field Camera 3 ultraviolet and visible arm F656N filter was used, 
half power points 6552--6572\AA) during the reduction chain, leading to 
16 (all from 2014-05-03) and 3 exposures being dropped from the centre and 
offset field respectively. Table \ref{Tab:Obs} summarises those observations 
that were selected. The differential image motion monitor (DIMM) seeing 
(where available) is listed in col. 7. 

\begin{table*}
\caption{Log of MUSE observations of NGC~4361}
\centering
\begin{tabular}{lrrrrrrr}
\hline\hline
Target   & RA~~~~          & Dec~~~               & PA       & Date  & Dataset   & Airmass & DIMM   \\
         & ($h$ $m$ $s$)~~ & ($^{\circ}$ $'$ $''$)~~ & ($^{\circ}$) &       & UT(h:m:s) &         & ($''$) \\
\hline
% CS 2014-05-01 Exps. 1-10
NGC~4361 CS & 12 24 30.78 & -18 47 05.6 &   0 & 2014-05-01 & 23:22:25.061 & 1.371 & 0.86 \\
NGC~4361 CS & 12 24 30.78 & -18 47 05.6 &   0 & 2014-05-01 & 23:27:19.740 & 1.347 & 0.75 \\
NGC~4361 CS & 12 24 30.76 & -18 47 05.6 &  90 & 2014-05-01 & 23:29:51.435 & 1.335 & 0.76 \\
NGC~4361 CS & 12 24 30.79 & -18 47 05.7 & 180 & 2014-05-01 & 23:32:23.770 & 1.323 & 0.65 \\
NGC~4361 CS & 12 24 30.81 & -18 47 05.7 & 270 & 2014-05-01 & 23:34:56.477 & 1.312 & 0.72 \\
NGC~4361 CS & 12 24 30.76 & -18 47 05.6 &   0 & 2014-05-01 & 23:37:40.385 & 1.300 & 0.66 \\
NGC~4361 CS & 12 24 30.79 & -18 47 05.7 &  90 & 2014-05-01 & 23:40:11.925 & 1.289 & 0.69 \\
NGC~4361 CS & 12 24 30.81 & -18 47 05.7 & 180 & 2014-05-01 & 23:42:43.771 & 1.279 & 0.66 \\
NGC~4361 CS & 12 24 30.82 & -18 47 05.5 & 270 & 2014-05-01 & 23:45:14.909 & 1.269 & 0.69 \\
NGC~4361 CS & 12 24 30.79 & -18 47 05.7 &   0 & 2014-05-01 & 23:47:56.611 & 1.258 & 0.71 \\
\hline
\end{tabular}
\tablefoot{Only the first 10 entries of this table are shown; the full table is 
contained in the on-line material.
}
\label{Tab:Obs}
\end{table*}

The MUSE observations of NGC~4361 were reduced with $EsoRex$ (version 3.13)
scripts and the MUSE 
instrument pipeline (version 2.8) \citep{Weilbacheretal2014,Weilbacheretal2020}. 
Since these were commissioning observations, the full set of calibration 
files was not necessarily available for each night. So sets of compatible bias frame, 
master flat and wavelength calibration exposures were assembled from the
closest available calibrations to produce master bias, master flat, trace table 
and wavecal tables appropriate for each night. For example, wavelength calibrations 
for the exposures on 2014-05-07 were not available and those for 2014-05-06 
were employed. 
The mean spectral resolving power from the fit of the $muse\_wavecal$ task to the arc 
lamp exposures was 3020$\pm$75.
Line Spread Function profile fits required for the extraction of the 
spectra were also produced using calibration files generated for each night, or for
the closest possible date if not available. A suitable sky flatfield 
was available for 2014-05-06 and was used to produce the necessary skyflat for
all datasets.
The bad pixel and geometry tables developed for commissioning (calib-0.18.2) were 
used in the ($muse\_scibasic$) extraction of the spectra for formation of the 
pixel tables. 

Observations of the white dwarf standard star GD 108 \citep{Bohlinetal2014, Oke1990} 
from 2014-05-06 were reduced similarly to the NGC~4361 observations and a single
response table produced, which was applied to the data for all 5 nights. Since
it was not necessarily known from the nightlogs and the DIMM results which 
exposures were affected by poorer visibility/light cloud, an H$\alpha$ narrowband 
(6552-6572\AA) image was produced for each exposure and the total flux used 
to flag the exposures suffering poorer visibility. Using exposures 
in a V$'$ band filter (square profile, 5350--5650\AA) and centroiding 
(IRAF\footnote{IRAF is distributed by the National Optical Astronomy 
Observatories, which are operated by the Association of Universities for Research
in Astronomy, Inc., under cooperative agreement with the National
Science Foundation.} $imexam$) 
the central star, the offsets between exposures were determined
and used in aligning the individual exposures to produce a single combined
data cube. For the W offset position, the central star was at the extreme edge 
of the field and could not be reliably used for image alignment; since no other 
stars were bright enough for alignment purposes, the default telescope pointing 
enscribed in the headers was used by default to align the multiple exposures. As 
a result image quality and alignment is not as good in the offset field as the 
central pointing.
For the central field, 42 observations over 3 nights were combined (see Tab. 
\ref{Tab:Obs}), and the resulting
V$'$-band (5350--5650\AA) image quality measured in the final combination was 
0.74$''$ (Gaussian FWHM of the central star). 
% For V' band: CS 0.740'', * to S 0.734, * to NW 0.668''
For the offset field the resulting image quality for the combined exposures was 
somewhat worse ($\sim$ 1.0$''$) but from the
signal in the overlap region of the central and offset fields, the W offset field
was found to have only 17\% of the depth of the central field (lower exposure time 
and clouds). 

For the central field, sky subtraction proved to be problematic as there is little
area on the MUSE field occupied by sky uncontaminated by the extended
nebulosity. For each position angle a careful assessment of the minimum area 
to use for sky was made and two small triangles to the SE and NW, consisting of a 
total area of $\sim$1900 spaxels (76 arcsec$^{2}$), 
1.85\% of the MUSE field was chosen (the extents of these two regions 
are indicated by the dark blue shaded areas on the \ion{H}{$\beta$} image, Fig. 
\ref{Fig:FluxImages1}). For the outer field, nebulosity only extends westward over 
about two thirds of the MUSE field, and 8\% of the field was adopted as sky. Some 
weak nebula emission nevertheless occurs over these adopted sky regions for 
the central field, so that the total 
flux of the extended line emission (e.g. H, He) in the MUSE field will not be 
absolute, but since the MUSE field does not cover the whole nebula extent, and 
absolute fluxes are not sought, this effect can be neglected. An indication
of the accuracy of the sky subtraction shows that the extended [\ion{N}{I}]5199,5202\AA\ 
sky line was no longer detectable in the total spectrum. For analysis of
the spatially compact 'Freckles' (Sect. \ref{Sect:Freckles}), small errors in sky subtraction are 
uncritical. 
% Check also [O I] in sky-subtracted cube and in total spectrum.
% In NGC4361/Regions/tab_Neb-Ms.dat, [O I] 6300A has obs 2.23794
% while in whole cube has obs 2.28022
 
The resultant sky-subtracted cubes have dimensions 426 ($\alpha$) by 433 ($\delta$)
[61.2 $\times$ 61.3$''$] by 3640 ($\lambda$) voxels (4750–9300\AA) at the 
default binning of 1.25 \AA. The absolute astrometry was adjusted using coordinates 
of the stars in the Gaia DR3 catalogue \citep{Lindegrenetal2021} which are available 
in the MUSE field-of-view.

%__________________________________________________________________

\section{Results: spectral imaging}
\label{Sect:Images}

The cubes were analysed with a semi-automatic Gaussian line fitter task
already described in \citet{Walshetal2018}, Sect. 3.2. The line list to drive the
fitting for the spectrum in each spaxel was derived from the list of 
emission lines identified by \citet{Aller1979}, supplemented with a manual 
identification of the emission lines present in an integrated spectrum of the
nebula over an area 60$''$ $\times$ 60 $''$ centred on the CS (see Tab. 
\ref{Tab:300x300Spec}). 
The predominance of high ionization lines, and the extreme weakness of low 
ionization lines, as found in earlier studies is confirmed. However the depth 
of the spectra (amounting to 42 $\times$ 60s = 2520s per spaxel) conclusively 
lead to the detection of low ionization lines of [\ion{N}{II}], [\ion{O}{II}] 
and [\ion{O}{I}] for 
the first time. 
% The deep integrated spectrum is analysed in detail in Sect. \ref{Sect:Spect}.

\begin{figure*}
\centering
\resizebox{\hsize}{!}{
\includegraphics[width=0.45\textwidth, angle=0, clip]{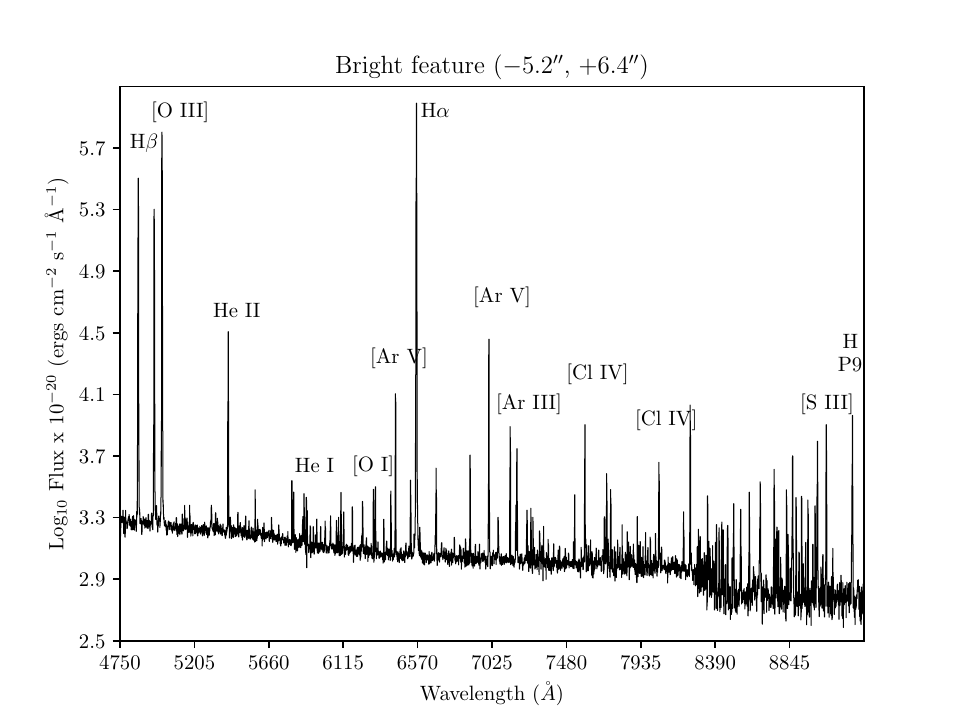}
\vspace{0.1truecm}
}
\resizebox{\hsize}{!}{
\includegraphics[width=0.50\textwidth, angle=0, clip]{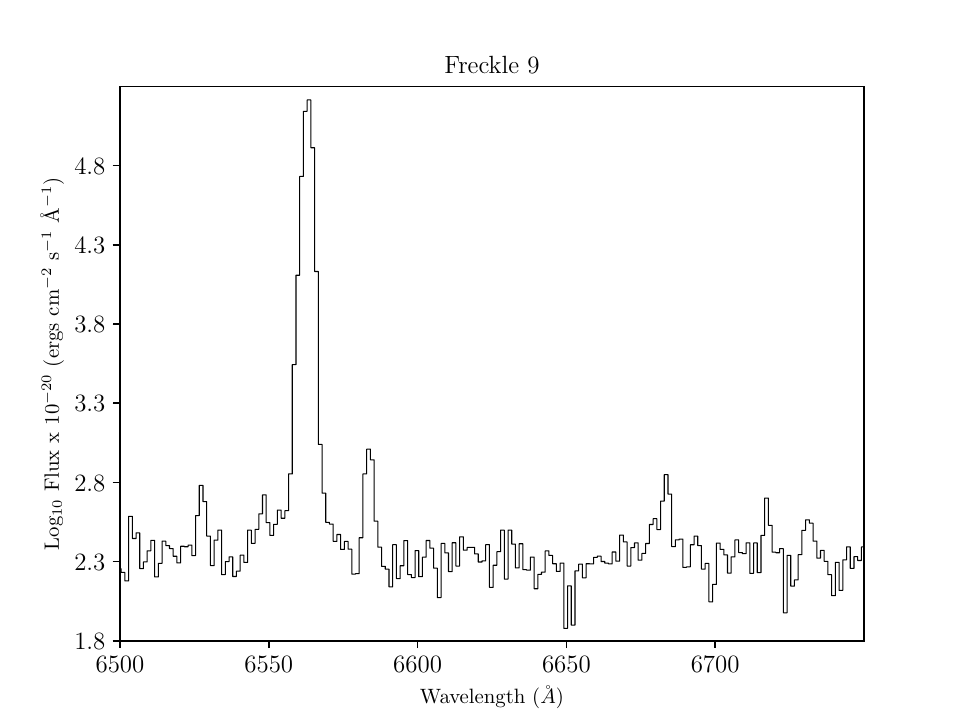}
\hspace{0.1truecm}
\includegraphics[width=0.50\textwidth, angle=0, clip]{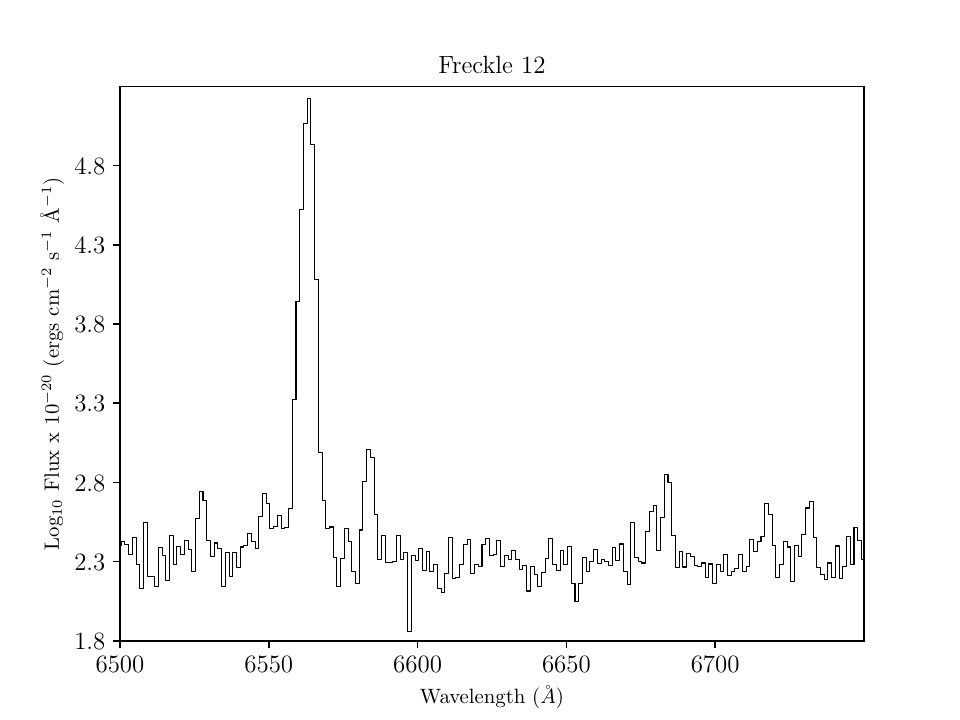}
\hspace{0.1truecm}
\includegraphics[width=0.50\textwidth, angle=0, clip]{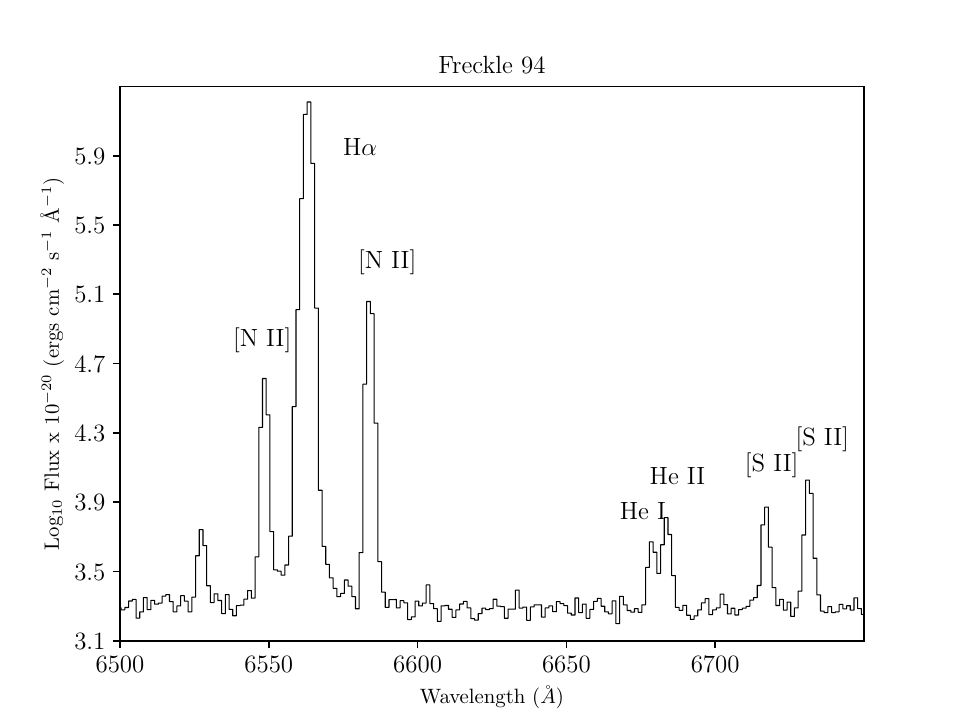}
}
% See NGC4361/Regs_measles/plot_spect_polished.py for the plot of the 1arcsec2 region
% and Regs_measles/plot_spect_Fspecs.py for the plots of the three Freckle spectra
\caption{Some representative spectra (log$_{10}$ flux) of a few selected regions are 
shown. Upper: The full observed spectrum over a bright region in the core of the 
nebula (area 1 arcsec$^{2}$ -- 25 MUSE spaxels) centred at ($\Delta \alpha$ -5.2$''$, 
$\Delta \delta$ +6.4$''$) relative to the position of the CS. \\
Lower: three example spectra over Freckles chosen to sample faint, average and
bright Freckles (based on their extracted H$\alpha$ and [\ion{N}{II}] fluxes; see Sect. 
\ref{SubSec:Freckles2}). \\
Left: spectrum of a faint Freckle (9); centre: spectrum of a Freckle (12) whose
H$\alpha$ and \ion{N}{II} fluxes mark it as of average brightness; spectrum of the brightest
Freckle (94). Note that these spectra of Freckles include the (high ionization) 
background nebular spectrum within the area of the Freckle.
}
\label{Fig:Ex_Specs}
\end{figure*}

Some example spectra of NGC~4361 are shown in Fig. \ref{Fig:Ex_Specs}: the full 
(4750--9300\AA) spectrum, on a log flux scale, of a 1 arcsec$^{2}$ region over 
the bright shell in the core ($\Delta \alpha$ -5.2$''$, $\Delta \delta$ +6.4$''$ 
with respect to the position of the CS); and three red spectra 
(6500--6750\AA) of Freckles (see Sect. \ref{Sect:Freckles} for details). 
Freckle 9 is one of the fainter ones in terms of H$\alpha$ and \ion{N}{II} 6583\AA\ 
% Freckle 009 in class C; Freckle 012 in class B
flux (lower left); Freckle 12 is of average brightness (bottom centre) and Freckle 94 
is the brightest (lower right). Sect. \ref{Sect:Freckles} provides a 
detailed description of the Freckles and Fig. \ref{Fig:N4361+Freckle_vels} shows the position and numbering of all the Freckles. 

\begin{table}
\caption{NGC~4361 integrated spectrum (60$''$ $\times$ 60$''$) - key to line maps}
\centering
\begin{tabular}{lrrl}
\hline\hline
Species & Rest $\lambda$ & Obs. flux       & Fig. \\
        & ~~~~(\AA)      & (H$\beta$=100)  &      \\
 \hline
  
H$\beta$  & 4861.3 & 100.0 & \ref{Fig:FluxImages1} \\
~[O~III]   & 5006.8 & 283.3 & \ref{Fig:FluxImages2} \\
\ion{He}{II}     & 5411.5 &   8.8 & \ref{Fig:FluxImages1} \\
% C~IV      & 5801.4 &   0.6 & \ref{Fig:FluxImages3} \\
% C~IV      & 5812.0 &   0.4 &            ''         \\
\ion{He}{I}      & 5875.7 &   0.7 & \ref{Fig:FluxImages1} \\
H$\alpha$ & 6562.8 & 296.5 & \\
~[\ion{N}{II}]    & 6583.4 &   0.5 & \ref{Fig:FluxImages4} \\
~[\ion{Ar}{III}]  & 7135.8 &   1.5 & \ref{Fig:FluxImages2} \\
~[\ion{Ar}{IV}]   & 7170.6 &   0.4 &  $\ast$                     \\
~[\ion{Ar}{V}]    & 7005.7 &   2.1 & \ref{Fig:FluxImages1} \\
~[\ion{S}{III}]   & 9068.6 &   2.8 & \ref{Fig:FluxImages2} \\
\hline
\end{tabular}
\tablefoot{
log$_{10}$ F(H$\beta$) = -10.777 in 60$''$ $\times$ 60 $''$. \newline
$\ast$ Figure not shown since line is faint.
}
\label{Tab:DispLines}
\end{table}

From the Gaussian fitted line flux per spaxel, emission flux and error 
maps were derived. Over the image of the CS, the continuum is 
strong and displays absorption and emission lines, so subtraction of the stellar
spectrum is necessary to reveal the nebular emission morphology in this area.  
The spectrum of the CS was determined by extracting the star+nebula 
over an area of 530 spaxels (radius 2.6$''$) and then subtracting the 
nebular emission from an annulus over radii 3.0--3.8$''$. The resulting 
stellar spectrum (presented in Appendix \ref{App:CS})
was then subtracted from each spaxel spectrum over the 
central area (5$\times$5 $''$) by scaling the stellar continuum over 
5 windows (4770--4830, 5130--5370, 5450--5780, 6800-7000 and 7350-7510\AA), 
selected as without strong nebula emission lines (but of course still
containing a contribution of some weak lines). This effectively removes 
the stellar contribution over the central area before Gaussians were fitted
to the emission lines, but neglects the very minor nebular continuum 
contribution.   
  
\begin{figure*}
\centering
\resizebox{\hsize}{!}{
\includegraphics[width=0.45\textwidth,angle=0,clip]{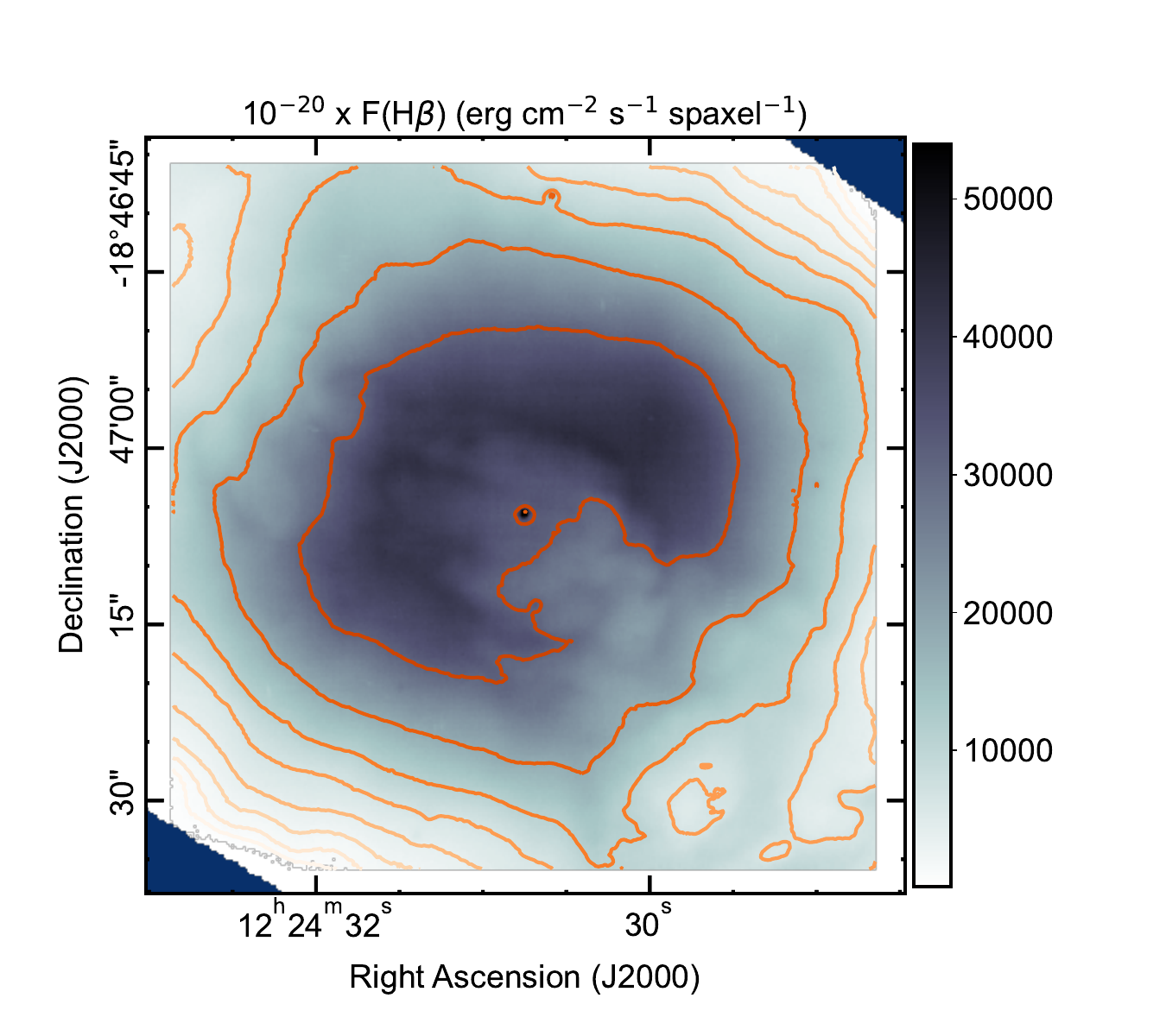}
\hspace{0.2truecm}
\includegraphics[width=0.45\textwidth,angle=0,clip]{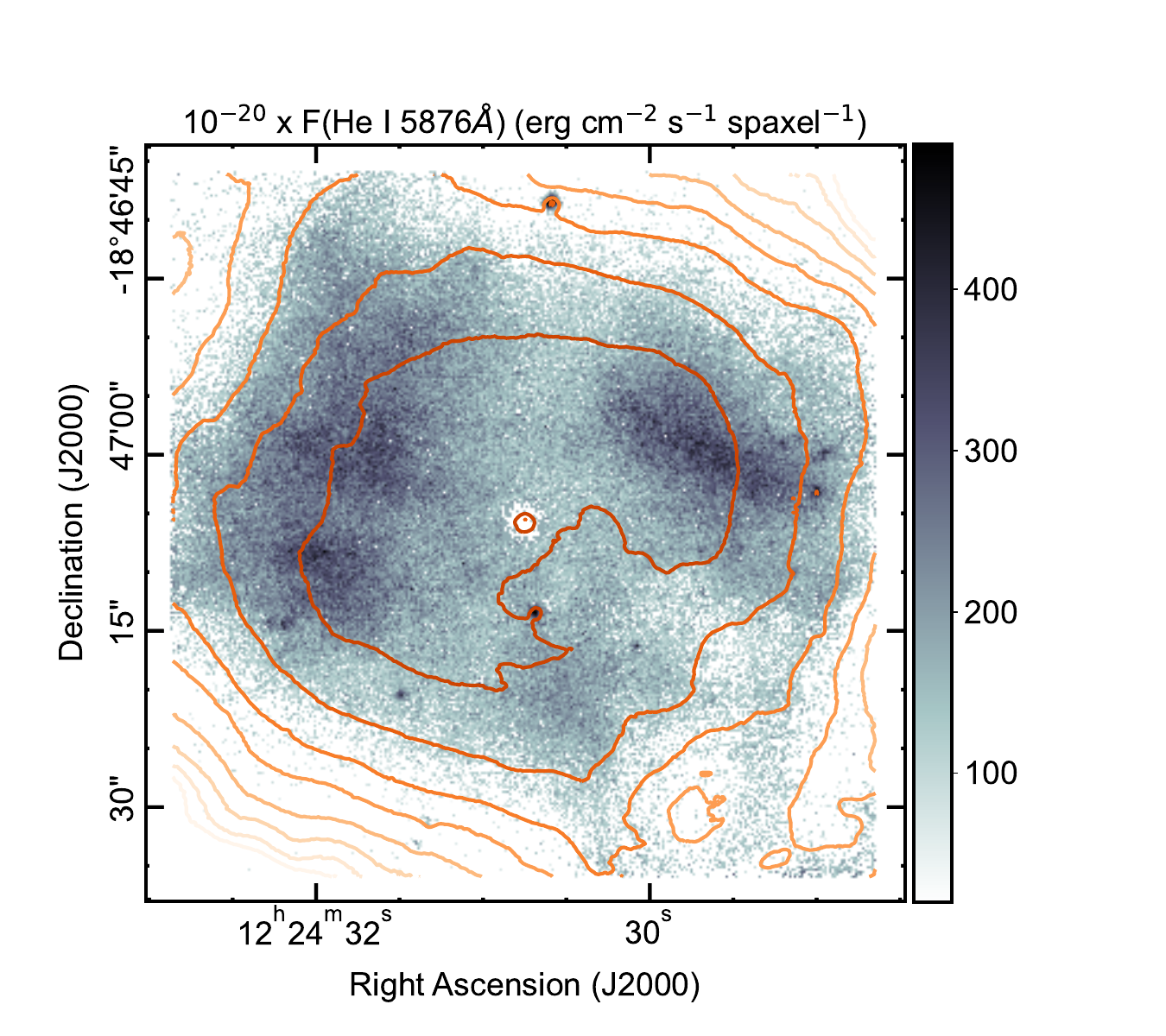}
\hspace{0.2truecm}
\includegraphics[width=0.45\textwidth,angle=0,clip]{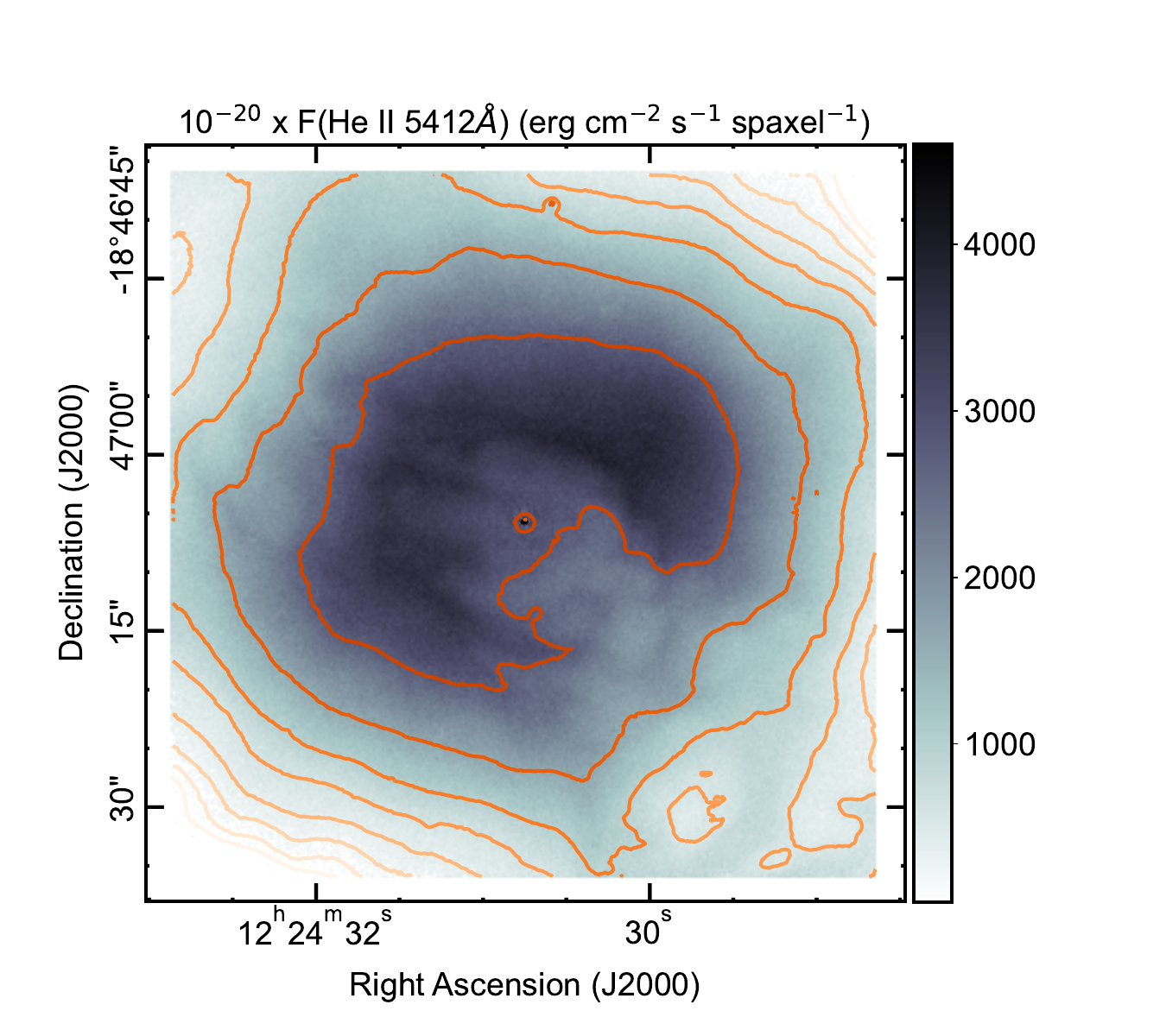}
}
\caption{Images of NGC~4361 in \ion{H}{$\beta$}, \ion{He}{I} 5876\,\AA\  and 
\ion{He}{II} 5412\,\AA\ with linear colour table lookup (see the colour bar 
right of each figure). The two coloured wedges at the SE and NW 
corners of the field shown on the H$\beta$ image represent the regions used for sky
background estimation. The contour map on the images is derived from the 
H$\beta$ image. 
}
\label{Fig:FluxImages1}
\end{figure*}

\begin{figure*}
\centering
\resizebox{\hsize}{!}{
\includegraphics[width=0.45\textwidth,angle=0,clip]{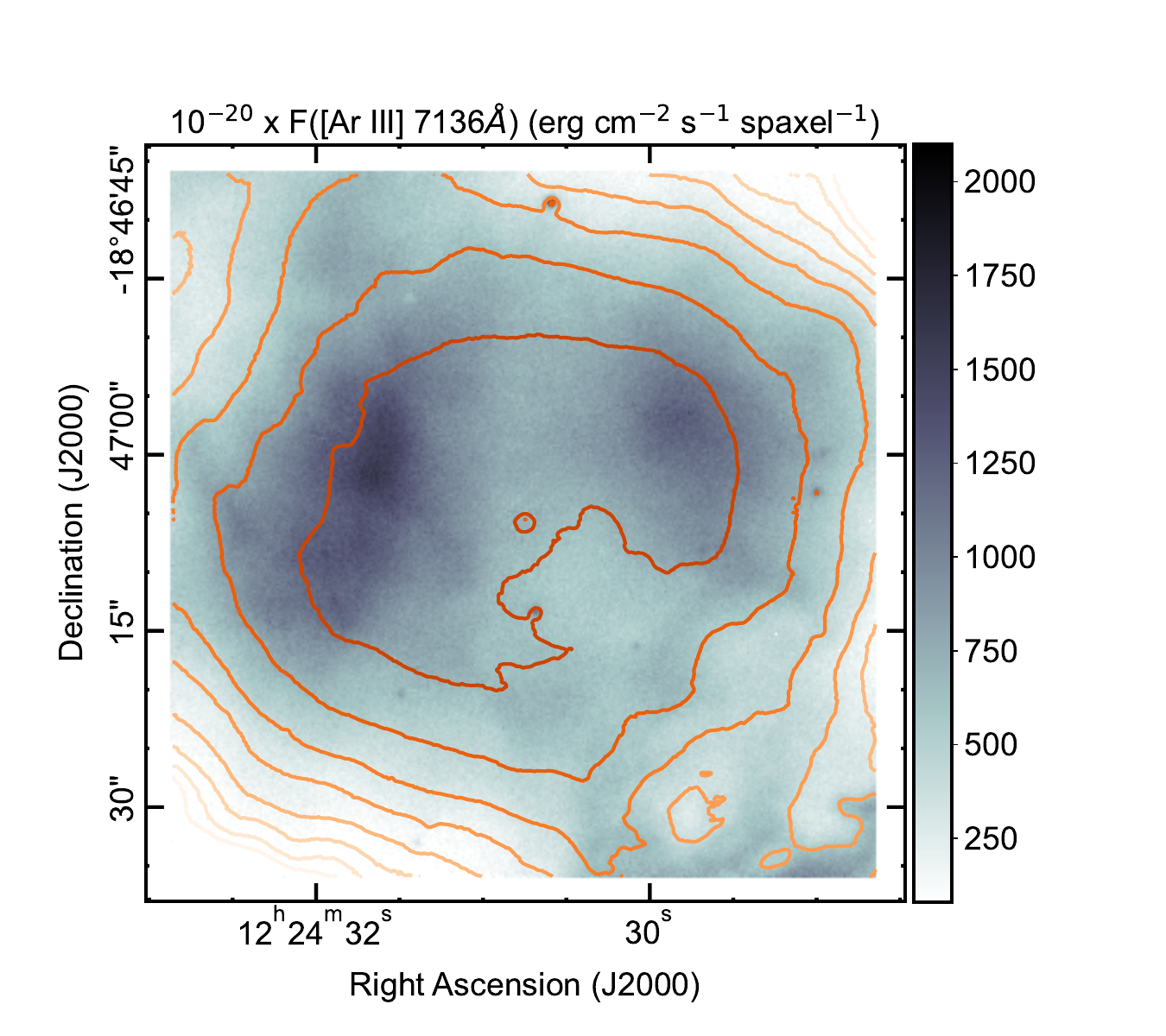}
\hspace{0.2truecm}
\includegraphics[width=0.45\textwidth,angle=0,clip]{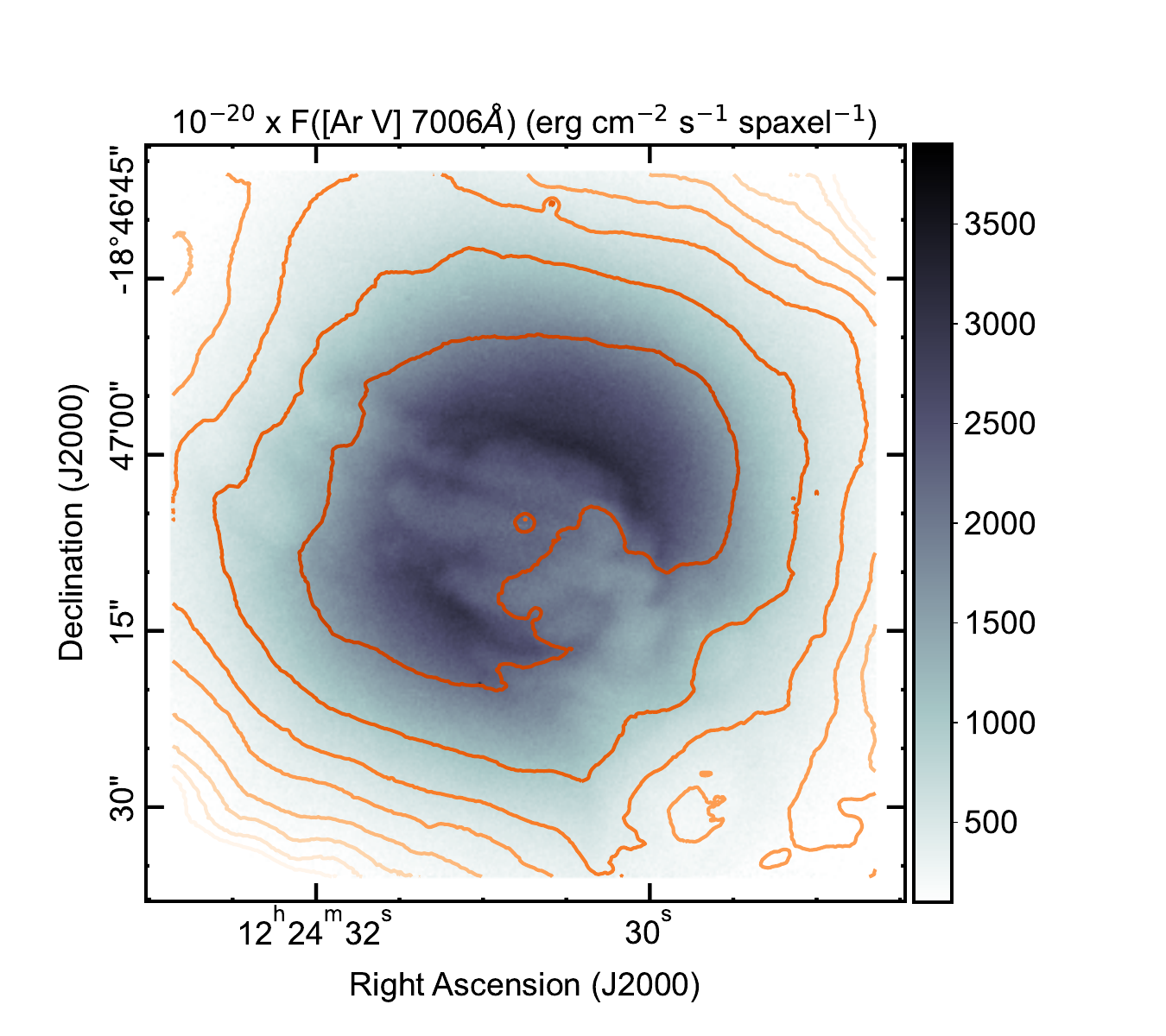}
}
\resizebox{\hsize}{!}{
\includegraphics[width=0.45\textwidth,angle=0,clip]{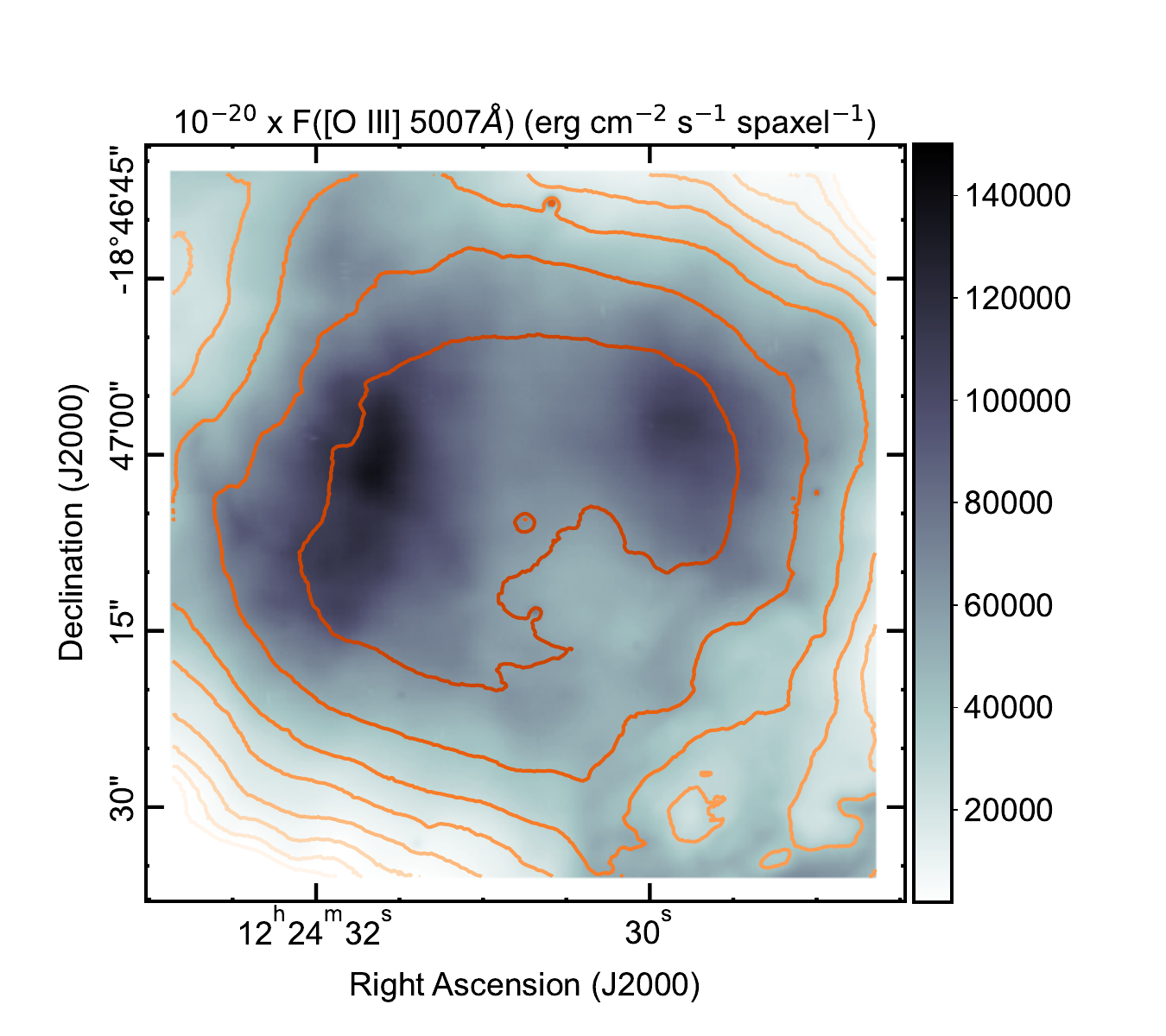}
\hspace{0.2truecm}
\includegraphics[width=0.45\textwidth,angle=0,clip]{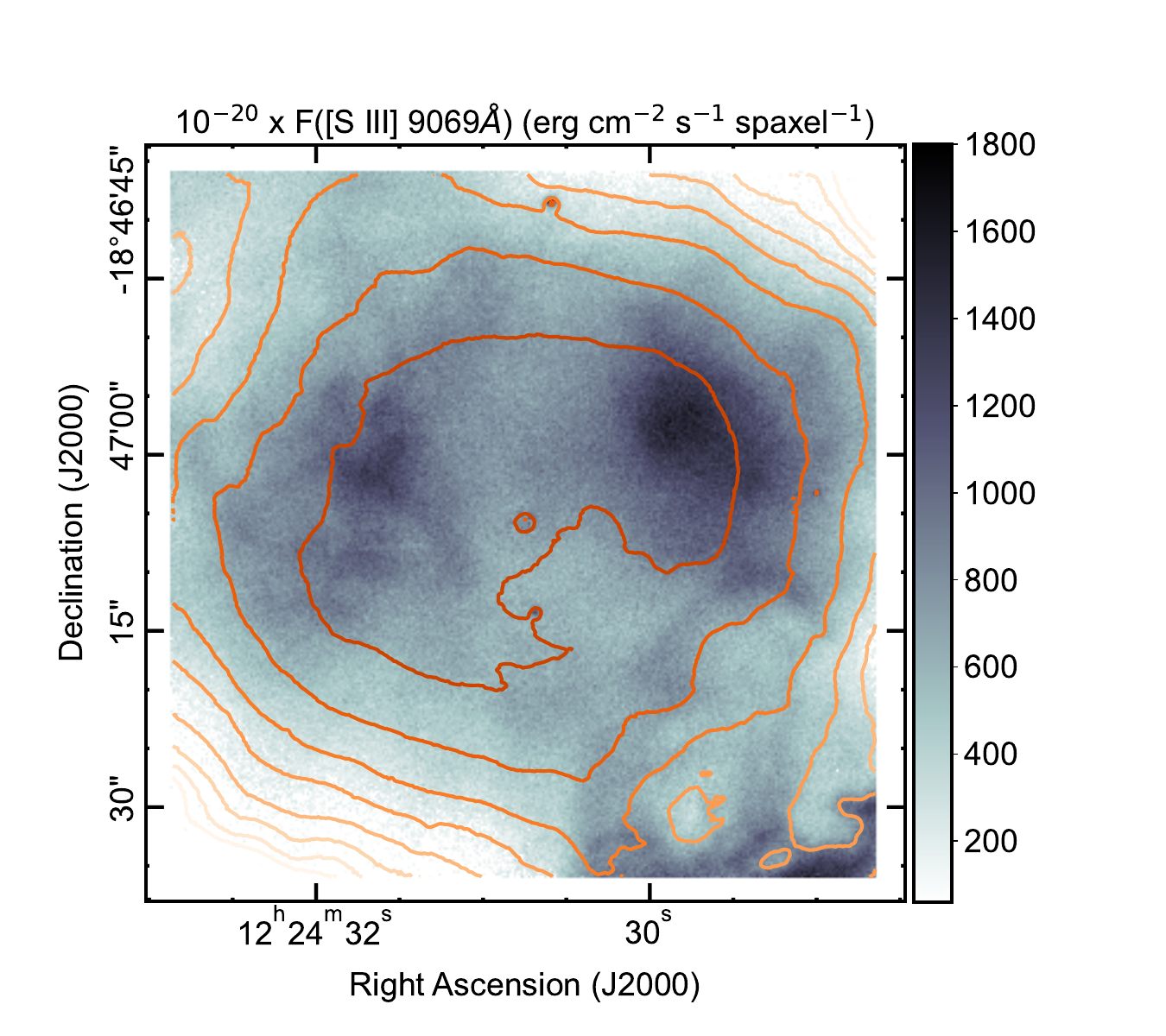}
}
\caption{Upper: Images of NGC~4361 in two Ar ions -- [\ion{Ar}{III}]7136\,\AA\ 
and [\ion{Ar}{V}]7006\,\AA\ -- demonstrating the differing morphology at two 
ionization levels for the same element;
Lower: Images in the two strongest lines of medium ionization species -- 
[\ion{O}{III}]5007\,\AA\ and [\ion{S}{III}]9069\,\AA. The contour lines are 
derived from the \ion{H}{$\beta$} image in Fig. \ref{Fig:FluxImages1}.
}
\label{Fig:FluxImages2}
\end{figure*}

\begin{figure*}
\centering
\resizebox{\hsize}{!}{
\includegraphics[width=0.45\textwidth,angle=0,clip]{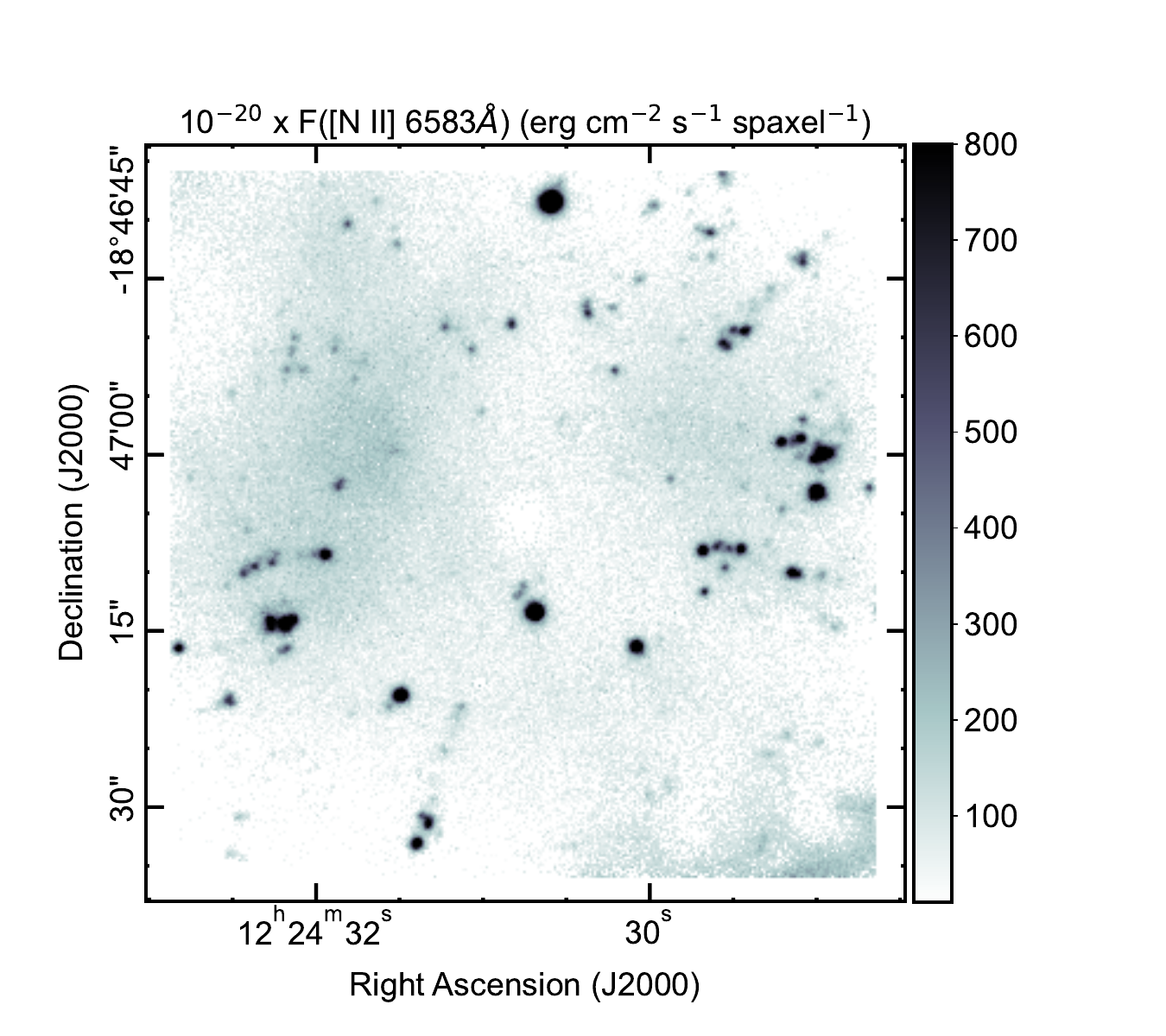}
}
\caption{The strikingly contrasted appearance of NGC~4361 at low ionization
is shown by this image of the brightest optical line [\ion{N}{II}] 6583\,\AA.
}
\label{Fig:FluxImages4}
\end{figure*}

A selection of these flux images is shown in Figs 
\ref{Fig:FluxImages1}, \ref{Fig:FluxImages2} 
% \ref{Fig:FluxImages3} 
and \ref{Fig:FluxImages4}; the (rest) wavelengths and integrated 
fluxes of the selected lines 
displayed are listed in Table \ref{Tab:DispLines} for reference. 
% These images from /mnt/Scientific\ Data/ANAL/MUSE/PNCOMM/NGC4361/Maps/

% \begin{figure}
% \centering
% \resizebox{\hsize}{!}{
% \includegraphics[width=0.45\textwidth,angle=0,clip]{N4361_C4_5801_12flux.pdf}
% }
% \caption{An image of NGC~4361 in C$^{3+}$ from the sum of the permitted \ion{C}{IV} 
% 5812\,\AA\ and 5812\,\AA\ lines.}
% \label{Fig:FluxImages3}
% \end{figure}

Fig. \ref{Fig:FluxImages1} show the images of the \ion{H}{$\beta$} (4681.3\AA), 
\ion{He}{I} 5876.0\,\AA\ and \ion{He}{II} 5411.5\,\AA\ (recombination) line emission. 
The contour map shown overlaid on the line images is derived from 
\ion{H}{$\beta$}. Fig. \ref{Fig:FluxImages2} contrasts the morphology between 
the lines of the same element for differing ionization level, 
[\ion{Ar}{III}]7135.8\,\AA\ (ionization potential, I.P., 40.7\,eV) and 
[\ion{Ar}{V}]7005.7\,\AA\ (I.P. 74.8\,eV), and also shows the bright 
medium ionization level [\ion{O}{III}]5006.8\,\AA\ (I.P. 54.9\,eV) and 
[\ion{S}{III}]9068.6\,\AA\ (I.P. 34.9\,eV) 
line images. 
%
% Decided to drop C IV image
% Of some interest is the \ion{C}{IV} ($3s ^{2}S - 3p ^{2}P_{o}$)  
% 5801.4 and 5812.0\AA\ doublet, which are present in both the central 
% star and the nebula. Fig. \ref{Fig:FluxImages3} shows the resulting 
% image of the sum of both 5801 and 5812\AA\ lines, without subtraction 
% of the stellar spectrum.

Figure \ref{Fig:FluxImages4} shows the striking morphology 
of the low ionization emission as exemplified by the strongest low
ionization line, [\ion{N}{II}]6583.4\,\AA. In contrast to higher ionization species, 
the emission is mostly prominent as compact knots, some in striking radial 
orientations. They were dubbed 'Freckles' on account of their unexpected
appearance and frequency over the face of the nebula, in comparison to the 
other emission line images. 
From the [\ion{N}{II}]6583\,\AA\ image, a total of 102 compact knots were 
counted by eye. There is in addition very faint extended [\ion{N}{II}] emission 
similarly following the morphology of the [\ion{O}{III}] line for example 
(brighter regions to WNW and E). More details on the 'Freckles' are 
provided in Sect. \ref{Sect:Freckles} where the extracted spectra are 
analysed.

\begin{figure*}
\centering
\resizebox{\hsize}{!}{
\includegraphics[width=0.45\textwidth,angle=0,clip]{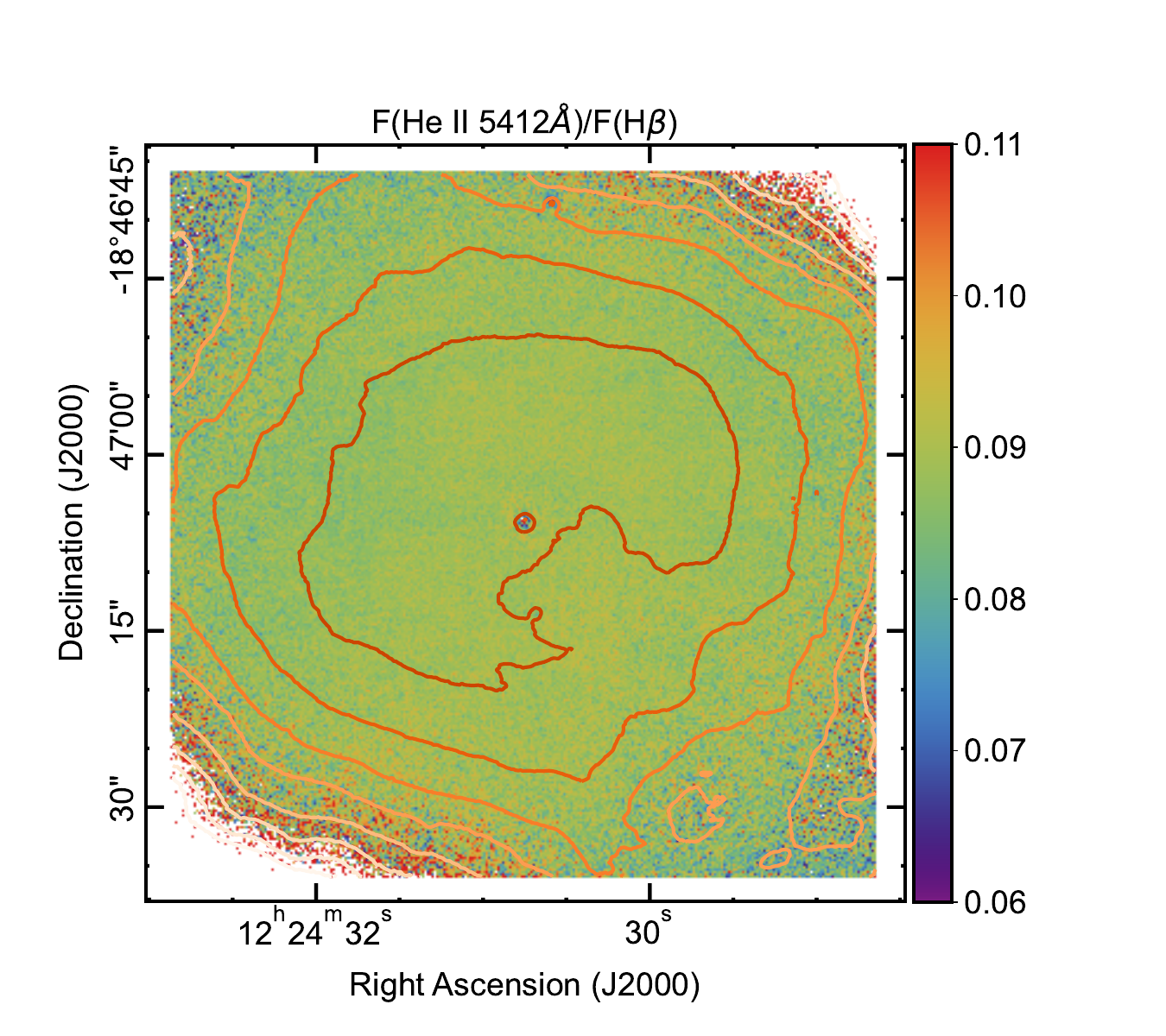}
\hspace{0.2truecm}
\includegraphics[width=0.45\textwidth,angle=0,clip]{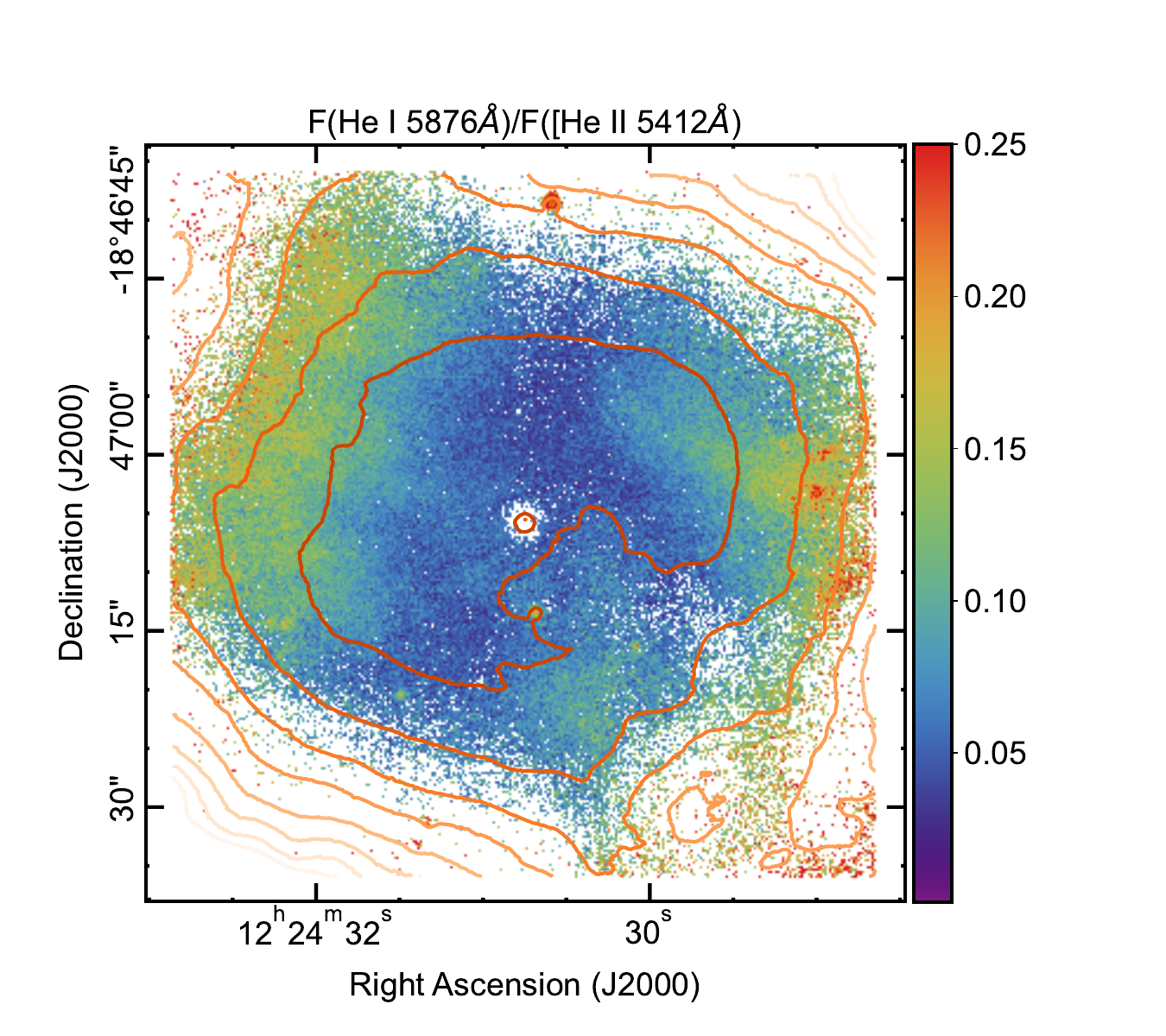}
}
\resizebox{\hsize}{!}{
\includegraphics[width=0.45\textwidth,angle=0,clip]{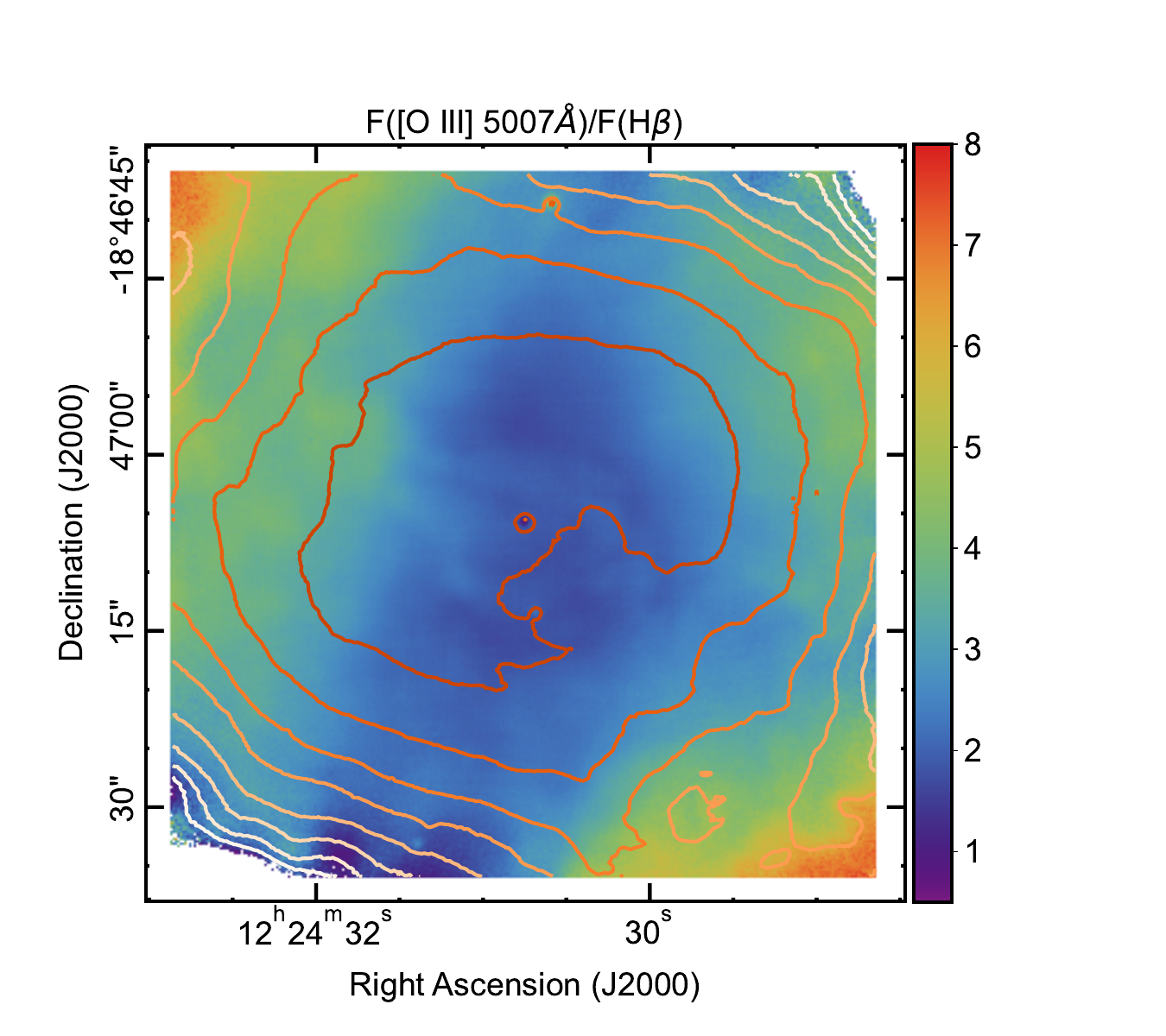}
\hspace{0.2truecm}
\includegraphics[width=0.45\textwidth,angle=0,clip]{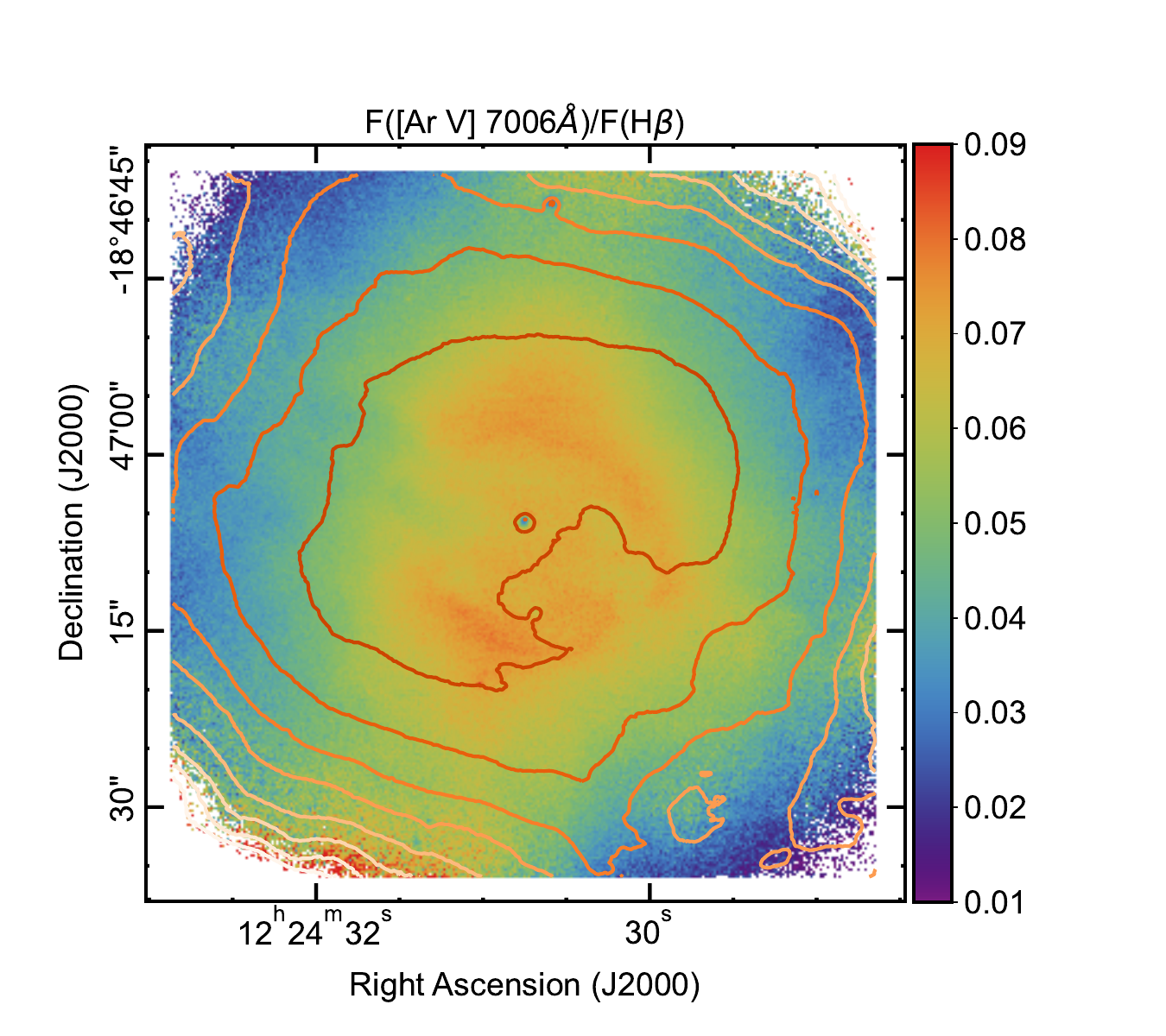}
}
\caption{
Upper: Ratio images of NGC~4361 in He lines: \ion{He}{II} 5412\,\AA\,/ H$\beta$ (left)
and \ion{He}{I} 5876\,\AA\,/ \ion{He}{II} 5412\,\AA\ (right) demonstrating that a very large fraction
of He is fully ionized throughout the nebula;
Lower: Ratio images displaying the ionization zones: [\ion{O}{III}] 5007\,\AA\,/ H$\beta$ (left)
for the medium ionization zone and [\ion{Ar}{V}] 7006\,\AA\,/ H$\beta$ for the highest ionization
(right).
}
\label{Fig:RatioImages1}
\end{figure*}

The morphology is remarkably similar between \ion{H}{$\beta$} and \ion{He}{II} 
5412\,\AA\ as shown in Fig. \ref{Fig:RatioImages1}. The flat appearance of 
\ion{He}{II}/\ion{H}{$\beta$} (ratio 0.089 $\pm$ 0.0073) 
continues into the outer field (not shown on account of lower signal-to-noise) 
without an
% Halpha extends to 56'' W of CS; He 2 detectable to 30'' W
obvious decrease to lower values, reinforcing evidence of 
an optically thin nebula. When plotting the ratio 
\ion{He}{I} 5876\,\AA / \ion{He}{II} 5412\,\AA,
it is clear that He$^{++}$ is beginning to recombine to He$^{+}$ towards
the edges of the field, although the 
ratio is low (0.095 $\pm$ 0.044); several of the low ionization
Freckles appear in this plot since they contain singly ionized He.
That recombination occurs to the outer edges of the MUSE field is
evident from the [\ion{O}{III}]5007\,\AA/\ion{H}{$\beta$} ratio image (Fig. 
\ref{Fig:RatioImages1}), where higher ionization species of O
(O$^{4+}$ and O$^{3+}$) in the central region must be recombining 
to O$^{2+}$ outwards, predominantly to the NE, NW and SW. The column 
of low [\ion{O}{III}]/\ion{H}{$\beta$} (NNW to SSE) is notable. Additional 
evidence for lower ionization is found in the [\ion{N}{II}] image,
where faint extended emission is found at the peaks of [\ion{O}{III}] emission
to the E and NW. The ratio [\ion{Ar}{V}]7006\,\AA/\ion{H}{$\beta$} 
image is quite distinct from any other (Fig. \ref{Fig:RatioImages1}) and 
shows a prominent elliptical ring (minor axis radius 19$''$) with openings 
at the ends of the major axis
to ENE and WSW, aligned with the large scale nebula extension visible in 
Fig. \ref{Fig:MUSEFields}.      

% ratio_C4_Hb.fits looks like a circular distrib of nebula flux (scattering)

%__________________________________________________________________

\section{Extinction and diagnostics of physical conditions}
\label{Sect:PhysCond}

\subsection{Extinction determination}
\label{SubSect.Extin}
From images of the ratios of Balmer and Paschen line fluxes 
compared to the \citet{MenzelBaker} Case B values, a map of the
extinction across the nebula can be constructed. The value of the
electron temperature $T_{\rm e}$ and electron density $N_{\rm e}$ 
must be provided, such as from collisionally excited line (CEL) indicators,
for example [\ion{S}{III}] 6312/9069\,\AA\ and [\ion{Cl}{III}] 5517/5538\,\AA\
respectively. Since the He$^{++}$ lines are strong, then contamination
of Balmer and Paschen line fluxes by nearby (within $\sim$3\AA) \ion{He}{II} 
emission must also be taken into account. The signal-to-noise per spaxel 
in the CEL diagnostic lines was however insufficient to compute the spatial 
distribution of $T_{\rm e}$ and $N_{\rm e}$, so integrated values for the 
whole MUSE field were adopted, $T_{\rm e}$ = 17000\,K and 
$N_{\rm e}$=1500 cm$^{-3}$, see following sub-section.
 
\begin{figure}
\centering
\resizebox{\hsize}{!}{
\includegraphics[width=0.45\textwidth,angle=0,clip]{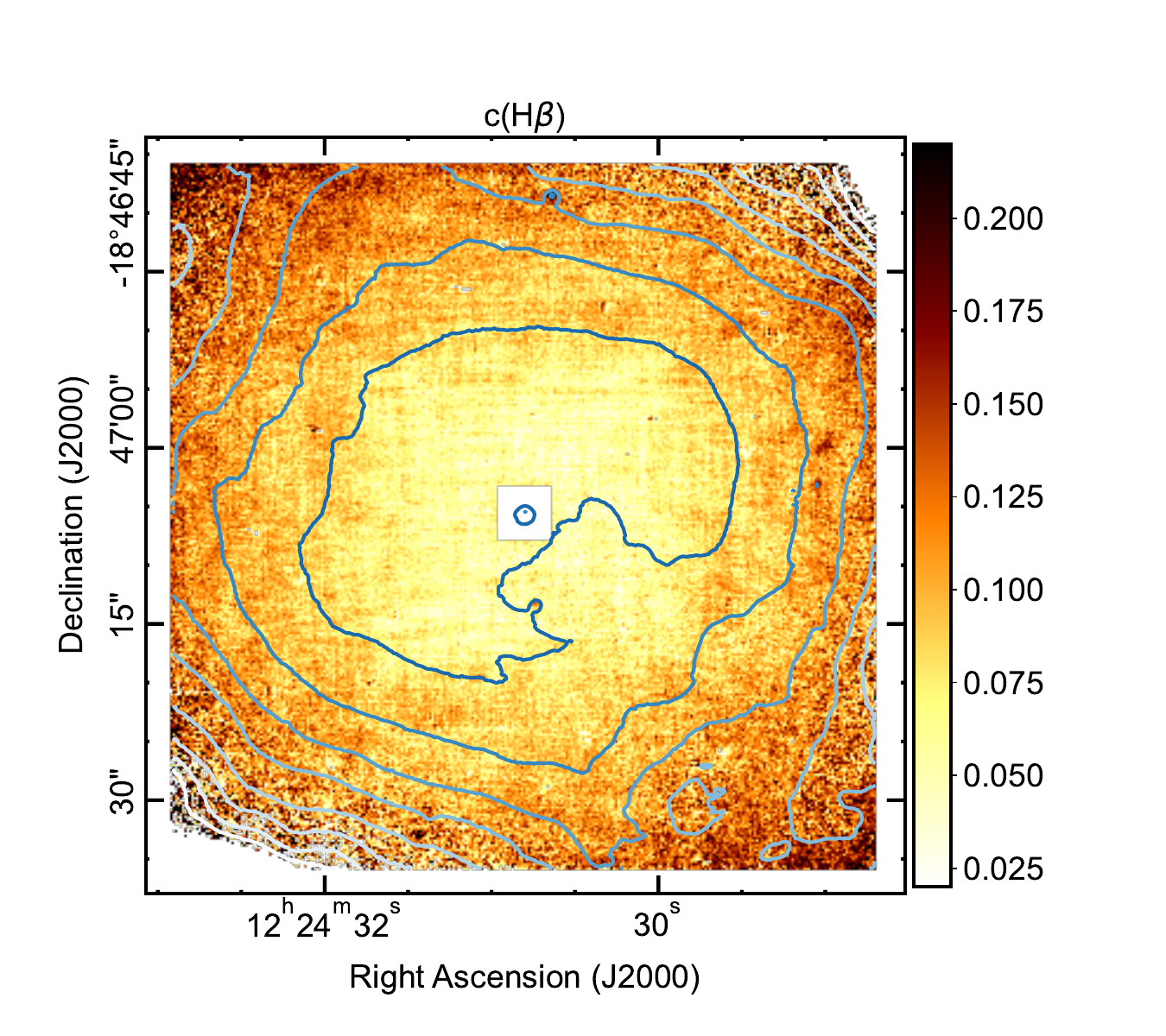}
}
% Maps/N4361_c.fits
\caption{The log extinction at H$\beta$, c, image of NGC~4361 is shown for the central 
MUSE field. The area over the bright central star has been substituted by a constant value
(see text for more detail). The contour lines are for \ion{H}{$\beta$} flux as in Fig. 
\ref{Fig:FluxImages1}.
}
\label{Fig:cmap}
\end{figure}

Figure \ref{Fig:cmap} shows the resulting extinction map of $c$,
the log extinction at \ion{H}{$\beta$}, from the ratio of 
\ion{H}{$\alpha$} / \ion{H}{$\beta$}, correcting \ion{H}{$\beta$}
for the presence of \ion{He}{II}\,4859.3\AA\ n=4--8 
% ($1/2-7/2 4f+_2F* - 1/2-15/2 8g+_2G$) 
and \ion{H}{$\alpha$} for the presence of 
\ion{He}{II}\,6560.10\AA\ n=4--6.  
% ($1/2-7/2 - 1/2-11/2 2F^{o} - 2G$).
The fluxes of the \ion{He}{$^{++}$} lines are given by the 
flux of the \ion{He}{II}\,5412\AA\ (n=4--7) line and the ratios of the
\ion{He}{II} Pickering series from the 
computations of \citet{HummerStorey1987}, adopting Case B, $T_{\rm e}$=17000\,K
and $N_{\rm e}$=1500 cm$^{-3}$. The mean value of $c$ in Fig. \ref{Fig:cmap} 
is 0.096 $\pm$ 0.031 with 3 rounds of 3$\sigma$ rejection (mean 
signal-to-noise 4.7 per spaxel); the mean value is higher by 
0.009 if the corrections for the two \ion{He}{II} lines are neglected. If 
Case A for \ion{He}{II} is adopted, then the mean value of $c$ is higher by 
0.001. Including the Paschen lines into the determination of the $c$
map, has almost negligible effect, and increases the mean $c$ value by 0.001,
while increasing the error on account of the much lower Paschen line 
fluxes. Fig. \ref{Fig:cmap} shows the resulting extinction map, with lower
values over the central 30$''$, increasing slightly to the NE and SW. Over the
area of the central star the strong absorption lines of \ion{H}{I} and \ion{He}{II} hinder
a clean determination of the H Balmer and Paschen line fluxes and the 
\ion{He}{II}\,5412\AA\ line flux, and so the value of c over this area was 
substituted by the mean of the surroundings. The observed emission line flux 
images were all dereddened using the
$c$ map shown in Fig. \ref{Fig:cmap} where \ion{He}{II} line contamination of
the Balmer lines was taken into account.
% File: NebSt_he2cor_c_B.fits

% imstat NebSt*.fits lower=0.01 upper=0.4 nclip=3
% #               IMAGE      NPIX      MEAN    STDDEV       MIN       MAX
%     NebSt_HPas_c.fits     67790   0.09741   0.03085   0.01132    0.1915
%  NebSt_HPas_cerr.fits     61673   0.02869   0.01757      0.01   0.09649
%      NebSt_Hab_c.fits     87086    0.1051   0.03137   0.01061    0.1999
%   NebSt_Hab_cerr.fits     84058   0.02339   0.01009   0.01171    0.0575
%  NebSt_he2cor_c_A.fits     86958   0.09553   0.03121   0.01001    0.1898
%  NebSt_he2cor_c_B.fits     86967   0.09643   0.03121   0.01002    0.1907 USED for FIG.
%  NebSt_he2cor_cerr_A.fits     84062   0.02339   0.01009   0.01171   0.05751
%  NebSt_he2cor_cerr_B.fits     84062    0.0234    0.0101   0.01171   0.05752

% For whole cube Reg_cube300x300s1_11.dat (CS included and no He2 correction)
%    VALUE OF C FROM H-ALPHA/H-BETA =  0.097733 +&-   0.000064
%        VALUE OF C FROM P-9/H-BETA =  0.081634 +&-   0.000428
%       VALUE OF C FROM P-10/H-BETA =  0.038285 +&-   0.000467
%       VALUE OF C FROM P-11/H-BETA =  0.129290 +&-   0.000465
%       VALUE OF C FROM P-12/H-BETA =  0.043912 +&-   0.000265
%       VALUE OF C FROM P-13/H-BETA =  0.095973 +&-   0.000915
%       VALUE OF C FROM P-14/H-BETA =  0.044778 +&-   0.000461
%   VALUE OF C =   8.34961534E-02  AND ERROR =   2.82689501E-02
%  VALUE OF AV =  0.179061010      AND ERROR =   6.06239550E-02
% So P09 and P11 fair agreement, but P10 and P12 poor, P13 okay.
%

\subsection{$T_{\rm e}$ and $N_{\rm e}$ determination}
\label{SubSect.TeNe}

\subsubsection{CEL diagnostics}
The collisionally excited lines (CELs) in NGC~4361 are all weak with respect to H
and He, and therefore the available CEL diagnostics for electron temperature 
and density ([\ion{N}{II}], [\ion{S}{III}] and [\ion{Ar}{III}] for $T_{\rm e}$, 
[\ion{S}{II}] and [\ion{Cl}{III}] for $N_{\rm e}$) have very low 
signal-to-noise per spaxel, effectively precluding making images of 
$T_{\rm e}$ and $N_{\rm e}$ at the original spaxel resolution (0.2$''$).
The summation of spaxels into successively larger bins was attempted
until sufficient signal-to-noise (S/N) was reached to achieve a fully filled image of 
$T_{\rm e}$ or $N_{\rm e}$. For example, this was achieved with $T_{\rm e}$
from [\ion{S}{III}] 6312/9069\,\AA\ with 5$\times$5 spaxel bins (1.0$''$).
However for $T_{\rm e}$ from [\ion{N}{II}] 5755/6583\,\AA\ the bins had 
to be at least 20$\times$20 (4.0$''$) before an image of sufficient S/N could 
be produced to compute an image filled with valid $T_{\rm e}$ values.
For $N_{\rm e}$ from [\ion{S}{II}] 6716/6731\,\AA\ 
and [\ion{Cl}{III}] 5517/5537\,\AA\ the lines are even weaker and even
4.0$''$ bins were not adequate. Thus only for $T_{\rm e}$ from 
[\ion{S}{III}] could an adequately sampled image be produced; for the
other diagnostics a single value was computed based on the integrated 
spectrum of the whole field. All determinations
of H and He and metal CEL line ratios were made with PyNeb 
\citep{Luridiana2015}. The sources for the atomic data
employed for all the PyNeb CEL computations are listed in Appendix
\ref{App:AtomicData}.  

\begin{figure}
\centering
\resizebox{\hsize}{!}{
\includegraphics[width=0.45\textwidth,angle=0,clip]{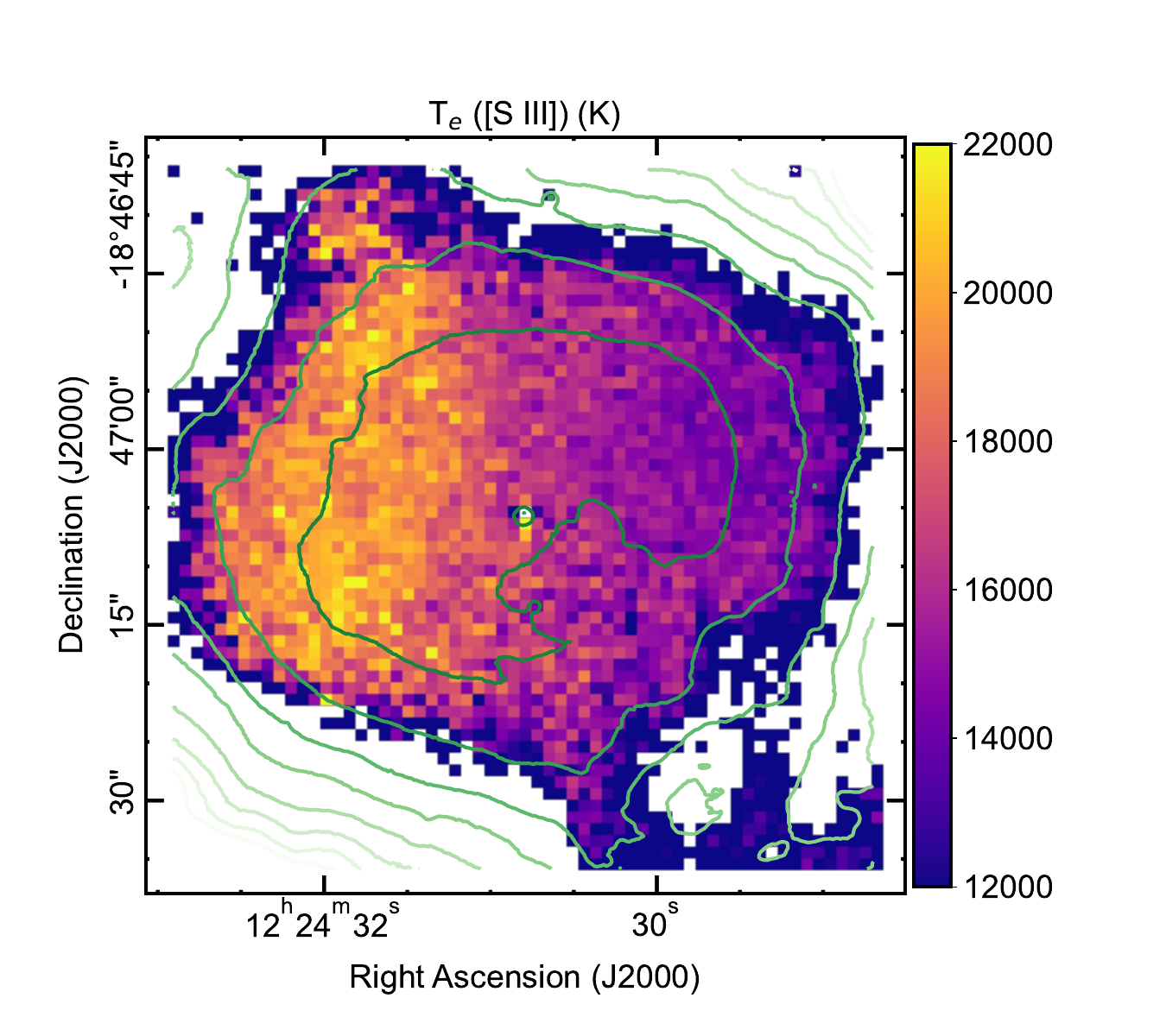}
}
\caption{The $T_{\rm e}$  image from [\ion{S}{III}] 6312/9069\AA. The original
MUSE spaxels were binned 5$\times$5 (1.0$\times$1.0$''$) to increase the 
signal-to-noise in the fainter line (6312\AA).
%The plotted range is 9000 to 22000\,K.   
}
\label{Fig:S3Temap}
\end{figure}

However for [\ion{S}{III}] $T_{\rm e}$, the range of many of the
values of the ratio 6312/9069\,\AA\ (dereddened line fluxes) was 
beyond the feasible range of the ratio of 0.0066 -- 0.2096 (for
$N_{\rm e} \leq$ 2000 cm$^{-3}$). Various factors could cause this
condition, such as telluric absorption of the red [\ion{S}{III}] 
9068.6\,\AA\ or a contaminating line at or close to the wavelength of
the weaker [\ion{S}{III}] 6312.1\,\AA. Indeed there is an \ion{He}{II} 
n = 5 series line at rest wavelength 6310.85 (5-16), close enough to
6312.1\,\AA\ to be included in the integrated flux of the [\ion{S}{III}]
line at the MUSE spectral resolution of $\sim$2.5\,\AA. From the dereddened
flux of \ion{He}{II} 5412\,\AA\ and the Case B line ratios 
from \citet{HummerStorey1987} for an adopted $T_{\rm e}$ of 17000\,K 
and $N_{\rm e}$ of 1500 cm$^{-3}$, 
%   6310.854         He II E1  5-16 
the ratio 6311/5412\,\AA\ = 0.04623. Subtracting the expected
flux of this line lowers the flux of the fitted 6312\,\AA\ flux by 
$\sim$40\%. With this correction the 6312/9069\,\AA\ ratio enters 
feasible values and $T_{\rm e}$ ([\ion{S}{III}]), can be calculated.
Fig. \ref{Fig:S3Temap} shows the resulting image, where the mean 
$T_{\rm e}$ is 14940 $\pm$ 3800\,K, but a very noticeable increase to 
the eastern lobe of the nebula (c.f., the gradient in [\ion{O}{III}]
$T_{\rm e}$ shown by \citet{Liu1998}, Fig. 4a for a NS slit, 
positioned 10$''$ E of the central star).
 
\begin{table*}
\caption{NGC~4361 Large area spectrum 60$''$ $\times$ 60$''$}
% Based on Regions/NEWtab_cube300x300_s1_s11.tex, from Reg_cube300s1_11r.dat
\centering
\begin{tabular}{llrrrr}
\hline\hline
 Rest $\lambda$ & Line Species & Observed &         Error & Dered & Error  \\
 ~~~~~ (\AA)    &              & flux     &         flux  & flux  & flux  \\
\hline
  4785.4 & \ion{C}{IV} 2D--2P$_{o}$          &    0.036 &    0.002 &    0.036 &    0.003 \\
  4861.3 & \ion{H}{I} 2--4                   &  100.000 &    0.000 &  100.000 &    0.000 \\
  4921.9 & \ion{He}{I} 1P$_{o}$--1D          &    0.064 &    0.001 &    0.064 &    0.004 \\
  4931.2 & [\ion{O}{III}] 3P--1D             &    0.050 &    0.001 &    0.050 &    0.003 \\
  4944.0 & \ion{N}{V} 2G--2H$_{o}$           &    0.089 &    0.001 &    0.089 &    0.006 \\ 
  4953.5 & ?                                 &    0.159 &    0.001 &    0.158 &    0.010 \\
  4958.9 & [\ion{O}{III}] 3P--1D             &   93.867 &    0.012 &   93.350 &    6.007 \\
  4971.7 & [\ion{Fe}{VI}] 2P--2F             &    0.166 &    0.001 &    0.165 &    0.010 \\ 
  5006.9 & [\ion{O}{III}] 3P--1D             &  283.283 &    0.029 &  280.978 &   17.976 \\
  5018.4 & \ion{C}{IV} 2P$_{o}$--2S          &    0.096 &    0.013 &    0.095 &    0.015 \\ 
\hline
\end{tabular}
\tablefoot{
Line identifications were primarily from the Atomic Line List \citep{vanHoof2018}, supplemented
by the extensive line lists for NGC~7009 \citep{FangLiu2011} and NGC~7027 \citep{Zhang2007}. If a line is identified, then the rest wavelength is listed; if no identification was found, the wavelength 
is the observed value, corrected for the systemic heliocentric radial velocity ($+$10 km\,s$^{-1}$). \\

Observed \ion{H}{$\beta$} normalising flux: 1.66 $\times$10$^{-11}$ $\pm$ 6.88$\times$10$^{-16}$ ergs cm$^{-2}$ s$^{-1}$ \\
Dereddened \ion{H}{$\beta$} normalising flux: 2.09 $\times$10$^{-11}$ $\pm$ 4.82$\times$10$^{-13}$ ergs cm$^{-2}$ s$^{-1}$ \\
Extinction c = 0.10 $\pm$ 0.01 calculated with $T_{\rm e}$ = 17000 K; $N_{\rm e}$ = 1500 cm$^{-3}$. \\

Only the first 10 entries of this table are shown; the full table is 
contained in the on-line material. \\
} 
\label{Tab:300x300Spec}
\end{table*}

Table \ref{Tab:300x300Spec} presents the observed line fluxes (normalised 
to I(H$\beta$)=100), calculated by Gaussian
fits to the detected emission lines (as in Sect. \ref{Sect:Images}),
for the total spectrum within the MUSE central field (60$''$ $\times$ 60$''$). 
Column 4 presents the error on the line flux 
from the Gaussian fit (with the error on H$\beta$ set to zero and so 
propagated to all other flux errors). The extinction determined from 
H$\alpha$/H$\beta$, with correction for the presence of \ion{He}{II} lines, is 
also listed in Tab. \ref{Tab:300x300Spec}. Column 5 lists the dereddened line 
fluxes and col. 6 the propagated errors. Table \ref{Tab:CELDiags} then presents the 
CEL diagnostics for the whole MUSE field based on the dereddened fluxes 
listed in Tab. \ref{Tab:300x300Spec}. 
The $N_{\rm e}$  and $T_{\rm e}$ error values were determined by Monte Carlo 
based on the (1$\sigma$) flux ratio errors for 1000 trials. 
The value for $T_{\rm e}$ ([\ion{S}{III}]) is included for comparison with the 
value from the binned spaxel determination 
(Fig. \ref{Fig:S3Temap}). On account of the high $T_{\rm e}$ values and the
large errors on the density sensitive [\ion{S}{II}] and [\ion{Cl}{III}] line
ratios, in PyNeb it was found to be more reliable to compute the diagnostics from 
the single line ratio, rather than the $getTemDen$ task, using the appropriate
$T_{\rm e}$  or $N_{\rm e}$ values, after iteration. For the integrated spectrum
of the full MUSE field, the [\ion{Ar}{III}] 5191.8\AA\ is detectable, so the
determination of $T_{\rm e}$ from the 5192/7136\AA\ becomes feasible, and the 
value is listed in Tab. \ref{Tab:CELDiags}.   

\begin{table}
\caption{NGC~4361 CEL diagnostics for MUSE field}
\centering
\begin{tabular}{lllrrll}
\hline\hline
Species & Rest $\lambda$s & Flux  & Error & Diagnostic & Error \\
        & ~~~~(\AA)       & ratio &       &            &       \\
\hline  
~[\ion{S}{III}]  & 6312/9069 & 0.1735 & 0.0023 & 17100\,K & 180  \\
~[\ion{S}{II}]   & 6716/6731 & 0.7506 & 0.0068 & 1600 cm$^{-3}$ & 50 \\
~[\ion{Cl}{III}] & 5517/5537 & 1.1337 & 0.0119 & 1500 cm$^{-3}$ & 90 \\
~[\ion{N}{II}]   & 5755/6312 & 0.0663 & 0.0014 & 24600\,K & 550  \\
~[\ion{Ar}{III}] & 5192/7136 & 0.0165 & 0.0002 & 15800\,K & 100  \\
\hline
\end{tabular}
% \tablefoot{}
\newline
\label{Tab:CELDiags}
\end{table}
          
\subsubsection{ORL diagnostics}
\label{SubSubSect.ORLs}

The ratios of H$^{+}$, He$^{+}$ and He$^{++}$ lines can also be used for 
$T_{\rm e}$, $N_{\rm e}$ determination. In particular: the ratios of 
\ion{He}{I} singlet lines can be used for $T_{\rm e}$ determination
\citep{Zhang2005, Walshetal2018}; the ratio of higher H Paschen lines can 
be used for $N_{\rm e}$ measurement \citep{Zhang2004, Walshetal2018};
and the \ion{H}{I} Paschen jump (8204\AA) is sensitive to $T_{\rm e}$ 
\citep{Zhang2004, Walshetal2018}, as is the \ion{He}{II} n=5 series limit jump at 
5694\AA\ \citep{FangLiu2011}. Given the extreme weakness of \ion{He}{I} lines (Tab. 
\ref{Tab:DispLines} and \ref{Tab:300x300Spec} and Fig. \ref{Fig:FluxImages1}), no attempt 
was made to use the line ratios for $T_{\rm e}$ estimation.

Employing the \citet{HummerStorey1987} Case B emissivities for \ion{He}{II}, an 
attempt was made to use the ratios of brightest n=5 lines (n=5-9 8236.8\AA, 
n=5-10 7592.8\AA, n-5-11 7177.5\AA, n=5-12 6890.9\AA, n=5-13 6683.2\AA, 
n=5-14 6527.1\AA, n=5-15 6406.4\AA\ and n=5-17 6233.8\AA) to n=4-7 5411.5\AA)
to estimate $T_{\rm e}$; however there is too little dependence of the ratios of 
these lines with $T_{\rm e}$ at the available S/N and accuracy of the flux 
calibration, even for integrated spectra, to derive any likely values within
5000--35000\,K. Given the evidence that the nebula is optically thin, Case 
A emissivities are suggested, but the same conclusion results as for Case B. 

$N_{\rm e}$ (\ion{H}{I}) determination from higher Paschen line ratios (P15 - P26,
[8545.4, 8502.5, 8467.3, 8438.0, 8413.3, 8392.4, 8374.8, 8359.0, 8345.5, 
8333.8, 8323.4 and 8314.3 \AA),  c.f. \citet{Walshetal2018}, was also
attempted. However the quality of the flux calibration, and the proximity of
some lines to nearby strong \ion{He}{II} lines of the n=6 series, affects the 
line fitting of the H Paschen lines, and overall did not allow a useful estimate 
of $N_{\rm e}$ to be determined without large uncertainties.

The \ion{H}{I} Paschen jump (PJ) at 8204\AA\ is prominent in the spectra across the MUSE
field and the Paschen jump was calculated in the same way as in \citet{Walshetal2018}.
However on account of the strength of the \ion{He}{II} lines, the regions selected
to be free of all but very weak emission lines for measurement of the 
continuum on both sides of the jump were altered from those given 
in the Appendix A of \citet{Walshetal2018}. To the blue of the jump, 
the region 8100--8180\AA\ was chosen and to the red of the jump, 
regions 9115--9140 and 9170--9215\AA. An attempt to form an image of the 
jump, normalised by the Paschen 11 (8862.8\AA) emission line flux,
(called PJ/P11), required binning of the spaxels to reach an image not 
dominated by large value fluctuations, at the level of at least 
10$\times$10 spaxels (2.0$''$) but showing little structure not related 
to the flux level of the nebular continuum (c.f., H$\beta$ image, 
Fig. \ref{Fig:FluxImages1}). However the region east of the CS, 
which displayed elevated $T_{\rm e}$ ([\ion{S}{III}]) (see Fig. \ref{Fig:S3Temap}), 
showed slightly higher PJ/P11 (lower PJ $T_{\rm e}$), while
the region west of the CS with lower $T_{\rm e}$ ([\ion{S}{III}]) had 
a lower PJ/P11 (higher $T_{\rm e}$). However these PJ/P11 values 
suggested very high temperatures ($>$ 25000\,K), close to the cut-off value 
of 30000\,K (the \citet{HummerStorey1987} H line emissivity tabulations do not go 
beyond this value) and the PJ $T_{\rm e}$ errors (estimated by Monte Carlo 
based on 100 trials with the 1$\sigma$ flux errors delivered by the MUSE pipeline) 
are of similar magnitude to the $T_{\rm e}$ values.

Therefore, given the generally low S/N of the Paschen jump even in binned 
spectra, the Paschen jump was computed only in large integrated regions: the 
full MUSE field, with a region of radius 1.8$''$ around the central star 
excluded; the region of high $T_{\rm e}$ ([\ion{S}{III}]) (650 
arcsec$^{2}$), see Fig.\ref{Fig:S3Temap}; the region of lower 
$T_{\rm e}$ ([\ion{S}{III}]) (702 arcsec$^{2}$). Since the Freckles are 
distinct regions and, as shown in the following section, of lower ionization and
electron temperature than the bulk nebula, a separate region of the
whole field, excluding the central star and the area of the emission
of all the Freckles (189 arcsec$^{2}$) was formed. Table \ref{Tab:PJTes} 
lists the PJ/P11 values determined from the continuum sections defined.
Column 4 lists the derived Paschen jump $T_{\rm PJ}$ values, using the same 
methodology described in \citet{Walshetal2018} to compare the observed PJ/P11 
to the theoretical values with the modified blue and red continuum extents. 
A value of 0.099 was calculated for the He/H abundance (He$^{+}$/H$^{+}$ = 0.0041,
He$^{++}$/H$^{+}$ = 0.0951) based on the strengths of the
\ion{He}{I} 5876\AA\ and \ion{He}{II} 5412\AA\ lines for a single
value of $T_{\rm e}$ of 17000\,K and $N_{\rm e}$ of 1500 cm$^{-3}$.

\begin{figure}
\centering
\resizebox{\hsize}{!}{
\includegraphics[width=0.45\textwidth,angle=0,clip]{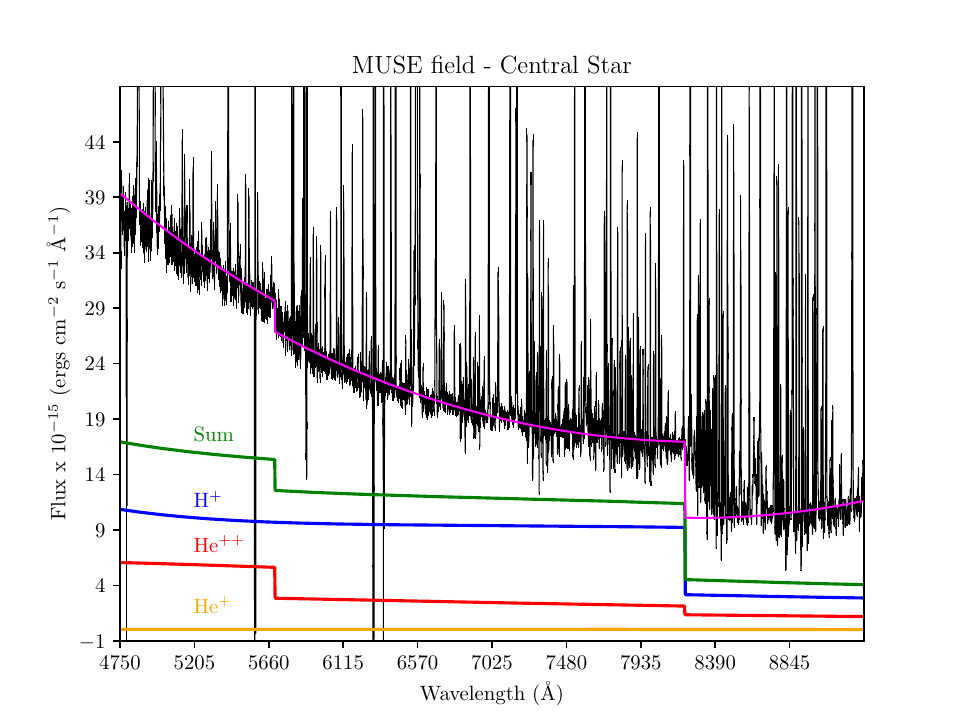}
}
\caption{The dereddened spectrum of the full MUSE field of NGC~4361, with the area of
the central star subtracted, is shown with the fit of the H$^{+}$, He$^{+}$ and 
He$^{++}$ nebular continua (in blue, yellow and red respectively). The sum of the
three sets of continua is shown in green. The fit to the spectrum is shown in
magenta and includes a scattered light contribution modelled
as a 3rd order polynomial continuum to match the sum of the nebular continua
to the measured nebular spectrum. See text for further details
and Tab. \ref{Tab:PJTes}, cols. 5 \& 6 for the fitted jump temperatures 
(\ion{H}{I} Paschen jump and \ion{He}{II} n=5 jump).
}
\label{Fig:PJplot_Reg_Neb-star}
\end{figure}

However, when a graphical comparison of the dereddened spectra of
the regions listed in Tab. \ref{Tab:PJTes} with the computed nebular
continuum (including \ion{H}{I} bound-free (bf) transitions, free-free
and 2-photon (2$\nu$) , \ion{He}{I} bf \& ff and \ion{He}{II} bf, ff \& 2$\nu$) 
was made, the computed \ion{He}{II} n=5 continuum jump at 5694\AA\ was 
too large with respect to the local continuum in this vicinity.
The implication is that the $T_{\rm e}$ (\ion{He}{II}) is larger than for 
\ion{H}{I}. This was not unforeseen given previous determinations of
the \ion{He}{II} jump $T_{\rm e}$, c.f. \citet{LiuDanziger1993}, 
\citet{FangLiu2011} which showed elevated $T_{\rm e}$ (\ion{He}{II})
compared to $T_{\rm e}$ (\ion{H}{I}) from PJ/P11. Given the 
high abundance of He$^{++}$, amounting to 95\% of the He abundance,
then the contribution of the \ion{He}{II} (n=6) jump to the \ion{H}{I} 
Paschen jump is significant (18\%) so that the assignment of the
$T_{\rm e}$ (\ion{He}{II}) plays an important role in determining  
$T_{\rm e}$ (\ion{H}{I}) from PJ/P11. The values of $T_{\rm e}$ (\ion{H}{I}) from PJ
and \ion{He}{II} from n=5 jump were thus calculated iteratively. From
the value of PJ $T_{\rm e}$ listed in Tab. \ref{Tab:PJTes}, the 
\ion{H}{I} and \ion{He}{I} continua were computed and subtracted from the 
dereddened spectrum of the NGC~4361 integrated regions, leaving
the \ion{He}{II} continuum to be included \footnote{Summing the \ion{H}{I}, \ion{He}{I}
and \ion{He}{II} continua with a single value of $T_{\rm e}$ and
appropriately scaling by the dereddened H$\beta$ flux, produced a 
continuum lower than the observed one. The source of this continuum
is assumed to be some instrumental scattered light, as a similar 
effect was found for the NGC~7009 MUSE data \citep{Walshetal2018},
Fig. A1. The spectrum of this scattered light was smooth but did not 
show a similar spectral shape to the central star (either observed
or dereddened); it is assumed that it arises from locally scattered 
nebular light within the MUSE IFU's. In order to match the theoretical nebular 
continuum spectrum to the dereddened target spectrum, the two were 
subtracted and a smooth (3rd order) continuum was fitted to the 
difference. See the caption to Fig. \ref{Fig:PJplot_Reg_Neb-star} 
for details.}. The \ion{He}{II} jump at 5694\AA\ was then fitted matching
the jump graphically (an approach using continuum bands to the blue
and red of the jump as for the Paschen jump was not successful on
account of the difficulty of finding line-free continuum regions 
in this wavelength range). This value was then used to compute the 
Paschen jump allowing the $T_{\rm e}$ (\ion{H}{I}) to be a variable. The
resulting \ion{H}{I} PJ $T_{\rm e}$ was then lower than the initial value 
(assuming a single value of $T_{\rm e}$ for H$^{+}$ and He$^{++}$) on 
account of the higher $T_{\rm e}$ (\ion{He}{II}), which lowered the He$^{++}$
contribution to the Paschen jump. Table \ref{Tab:PJTes} lists the
resulting \ion{H}{I} and $T_{\rm e}$ (\ion{He}{II}) values. No error evaluation 
as such was performed but typical errors of $\pm$1000\,K are indicated for
the $T_{\rm e}$ (\ion{H}{I}) and $\pm$2000\,K for \ion{He}{II} on account of the
much weaker break.

\begin{table*}
\caption{NGC~4361 PJ/P11 $T_{\rm e}$ measurements}
\centering
\begin{tabular}{lrrrrr}
\hline\hline
Region  & PJ/P11           & Error          & T$_{PJ}$ & T$_{PJ}$ (\ion{H}{I}) & $T_{\rm e}$ (\ion{He}{II}) \\
        & (\AA$^{-1}$)~~   & (\AA$^{-1}$)~~ & K~~    & K~~ & K~~  \\
 \hline  
 Field - CS                      & 0.02453 & 0.00057 & 10400 & 7500 & 15000 \\
 Field - CS - Freckles           & 0.02442 & 0.00062 & 10600 & 7500 & 15000 \\
 High $T_{\rm e}$ ([\ion{S}{III}])  & 0.02213 & 0.00056 & 15900 & 9000 & 18000 \\
 Low $T_{\rm e}$ ([\ion{S}{III}])   & 0.02192 & 0.00055 & 16900 & 9000 & 17000 \\
\hline
\end{tabular}
\tablefoot{T$_{PJ}$ listed in col. 4 is computed from the PJ/P11 value based on 
calibration of the flux ratios in continuum windows (see text for details);
\ion{H}{I} T$_{PJ}$ in col. 5 is computed from the total fit to the spectrum, including 
the contribution of He$^{++}$ nebular continuum (bf, ff and 2$\nu$) at the temperature
listed in col. 6.
\newline
}
\label{Tab:PJTes}
\end{table*}

%__________________________________________________________________

\section{Census of the NGC~4361 Freckles}
\label{Sect:Freckles}

\subsection{Identification and location}
\label{SubSect:Freckles1}

\begin{figure*}
\centering
\resizebox{\hsize}{!}{
\includegraphics[width=0.45\textwidth,angle=0,clip]{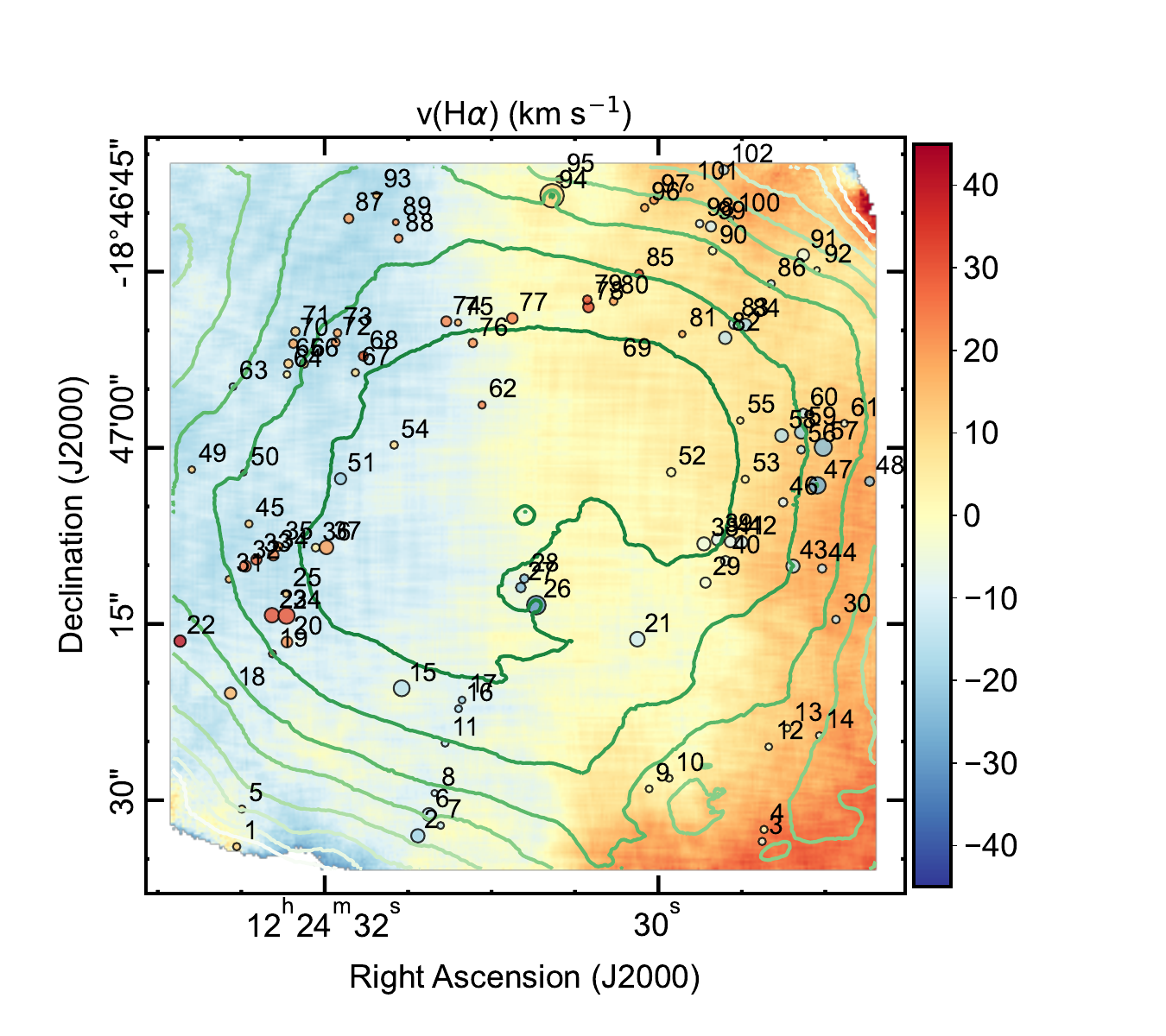}
}
\caption{The velocity field for the central MUSE field NGC~4361 determined 
from H$\alpha$ is shown together with the numbering for the Freckles, with a
circle for each proportional in size to the [\ion{N}{II}]6583\AA\ flux.
The H$\alpha$ and individual Freckle offset velocities 
([\ion{N}{II}] $-$ local H$\alpha$), in km\,s$^{-1}$, are colour coded as 
shown by the colour bar, at right.
}
\label{Fig:N4361+Freckle_vels}
\end{figure*}

The Freckles are the most obvious features on the [\ion{N}{II}] image 
on account of their compact emission superposed on a very low 
background (Fig. \ref{Fig:FluxImages4}). However
in the strong emission lines of H and higher ionization CELs, the features
are difficult to distinguish. The brightest Freckle (94, $\Delta \alpha$ -2$''$,
$\Delta \delta$ +27$''$, see below) has a contrast in the H$\alpha$ image
of $\sim$40\% with respect to the surroundings, the 2nd brightest Freckle a 
contrast of 14\%, and the rest of the Freckles having an H$\alpha$ contrast 
$<$10\%. 102 separate Freckles were identified on the [\ion{N}{II}] image by eye 
and Table \ref{Tab:FrecklesatN2} presents their designation, centroid positions 
as $\Delta \alpha$, $\Delta \delta$ with respect to the position of the central 
star (Gaia DR3 coordinate: 12$^{h}$ 24$^{m}$ 30.75$^{s}$, 
$-$18$^{\deg}$ 47$'$ 5.73$''$), their area as delineated (to $\sim$3$\sigma$ 
above the background emission) in arcsec$^{2}$ and the 
[\ion{N}{II}] 6583\AA\ flux (but including the background contribution). The 
faintest Freckle is 0.3\% of the brightest. By excluding the area of all the 
Freckles, an integrated spectrum of the high ionization nebular component can be
formed, as referenced in Tab. \ref{Tab:PJTes}. The designation of the 
Freckles is shown on the [\ion{N}{II}]\,6583\AA\ image (Fig. 
\ref{Fig:N4361+Freckle_vels}).

% Table of Freckle positions areas, [N II] flux , velocity and velocity - local Ha velocity

% \input{Table_Freckle_params_SHORT.tex}

\begin{table*}
\caption{NGC~4361 Freckles: morphological summary, [\ion{N}{II}] flux and velocity}
\centering
\begin{tabular}{lrrrrrr}
\hline\hline
 No. & $\Delta \alpha$ & $\Delta \delta$ & Area      & ~[\ion{N}{II}] & Vel. & $\Delta$ V([\ion{N}{II}]-H$\alpha$) \\
     & ~~~~($''$)      & ~~~~($''$)     & ($''^{2}$) & Flux $\times$ 10$^{20}$ & (km\,s$^{-1}$) & (km\,s$^{-1}$) \\
\hline
% Now with NEW numbering and corrected N2 fluxes
 1 & 24.6 & -28.4 & 2.20 & 3513 & 11 & 13 \\ 
 2 & 9.2 & -27.4 & 3.84 & 34800 & -25 & -20 \\ 
 3 & -20.2 & -27.8 & 0.60 & 3200 & 21 & -6 \\ 
 4 & -20.4 & -26.8 & 0.72 & 3311 & 27 & 1 \\ 
 5 & 24.2 & -25.2 & 2.00 & 3467 & 4 & 8 \\ 
 6 & 8.2 & -25.6 & 4.24 & 29434 & -20 & -15 \\ 
 7 & 7.2 & -26.4 & 1.04 & 2571 & -16 & -15 \\ 
 8 & 7.8 & -23.8 & 0.80 & 1791 & -17 & -12 \\ 
 9 & -10.6 & -23.4 & 0.88 & 2873 & 7 & 1 \\ 
 10 & -12.2 & -22.4 & 1.00 & 2895 & 5 & 0 \\ 
\hline
\end{tabular}
\tablefoot{The units of flux are ergs cm$^{-2}$ s$^{-1}$ $\times$ 10$^{-20}$. \\
Only the first 10 entries of this table are shown; the full table is 
contained in the on-line material.
} 
\label{Tab:FrecklesatN2}
\end{table*}

Column 6 of Tab. \ref{Tab:FrecklesatN2} lists the [\ion{N}{II}] velocity
and col. 7 the difference in velocity between the [\ion{N}{II}] and
H$\alpha$ emission (fitted by single Gaussians) over the extent of
each Freckle. On account of the relative strength of the H$\alpha$ and
[\ion{N}{II}] emission over the Freckles compared to the background,
the [\ion{N}{II}] velocity measured that of the Freckle, while the 
H$\alpha$ velocity that of the local bulk nebular emission. At the MUSE
spectral resolution ($\sim$ 130 km\,s$^{-1}$ at H$\alpha$), it was not 
feasible to deconvolve the H$\alpha$ velocities of the Freckle and the 
local bulk ionized gas. The minimum and maximum [\ion{N}{II}] - H$\alpha$ 
velocity differences are -34 and +45 km\,s$^{-1}$, and the mean  
+3.4 km\,s$^{-1}$, to be compared to the systemic (heliocentric) velocity of 
the nebula at H$\alpha$ of +5 km\,s$^{-1}$ for the full MUSE central field. 
The typical error on a single velocity determination is at least 10 km\,s$^{-1}$
for a high line flux, so there is no evidence that the system of Freckles 
has a bulk displacement from that of the overall nebula. Fig. 
\ref{Fig:N4361+Freckle_vels} displays the large scale H$\alpha$ velocity 
field for the central MUSE field and the numbering of the Freckles; the size of
the Freckle marker is proportional to the [N~II] flux and the 
offset velocities of the Freckles from the background emission 
(Table \ref{Tab:FrecklesatN2}, col. 7) are colour coded.
  
Examination of the spatial distribution of [\ion{N}{II}] Freckles shows
that the SW quadrant has the lowest number whilst the other three quadrants 
are fairly similarly populated. There are some obvious close groups of 
Freckles, such as 31-37, 38-41 and 54-60. Examination of Tab. \ref{Tab:FrecklesatN2}
shows that 31-35, 37 have similar $+$ve velocities with respect to
that of the surroundings, 38-41 also with low velocities, and
55-59 have similar $-$ve velocities. It is clear here, and from
correlation of positions with velocities in Tab. \ref{Tab:FrecklesatN2},
that Freckles projected close on the sky may not share the same velocity 
(and presumably line-of sight position, assuming for example that a $-$ve 
velocity implies a Freckle foreground to the local emission).
There are also a few obvious linear alignments: 
to the SSE, 2, 6, 8, 11, 16 17 (all with $-$ve velocities in range 
-17 to -26 km\,s$^{-1}$);
to the NNW, 80, 96, 97; ($+$ve velocities in range 20 -- 36 km\,s$^{-1}$); 
and to the NW, 82, 83, 84, 86, 91 (velocities in range -9 to +5 km\,s$^{-1}$).
The SSE alignment points back to the central star within about 1$''$,
as does the NNW one; the NW one however does not and is tangent to the 
SW of the CS by 8$''$. None of these alignments show convincing evidence of
a coherent velocity gradient, as might be indicative of an ejection / 
acceleration sequence, although the MUSE velocity resolution is 
rather low. 

% (' Average [N II] flux ', 2803.422252374537) so 20.2% of #92
% (' Median [N II] flux ', 1595.4299945831299) so 11.5% of #92
%  #92 has [N II] flux 13901

% Chosen stats on Freckles velocities - see Regs_Freckles/Stats_Mvels.lis
%(' Mean Del Ra, Dec ', -1.0147058823529407, 3.0803921568627457)
%(' Min, Max, Mean Havel ', -15.0, 27.0, 1.8333333333333333, 133.18627450980392)
%(' Min, Max, Mean N2vel ', -36.0, 43.0, 5.2745098039215685, 126.08823529411765)
%(' Min, Max, Mean N2-Ha vel ', -34.0, 45.0, 3.4411764705882355)

\subsection{Spectra}
\label{SubSec:Freckles2}

The properties of the Freckles, as distinct from the bulk emission of
the nebula, such as their ionization, extinction and $T_{\rm e}$, $N_{\rm e}$  
diagnostics can be revealed by emission line spectroscopy. However
as explained in Sec. \ref{SubSect:Freckles1}, their observed 
properties as faint compact sources on a high and structured nebula 
background introduce severe difficulties in extracting their intrinsic 
spectra. A number of approaches were trialed in order to reach the
highest number of non-zero emission line measurements among the 102
Freckles:
\begin{itemize}
\item For each wavelength slice in the MUSE cube, successively fit a 
2D gaussian-smoothed surface to the whole field excluding the areas of 
the Freckles, then extract the spectrum of the background subtracted 
emission for each Freckle. This approach performed poorly on account of 
small scale structure in the background and non-uniformities in 
individual wavelength slices;
\item For each wavelength slice in the cube, perform aperture photometry 
with a circular aperture for the background (excluding spaxels occurring 
over any other Freckles), subtract from the signal over the extent of the
Freckle and form the integrated spectrum. The approach also performed poorly 
and gradients in the background emission often compromised the estimation of 
a single background value over the Freckle area;
\item Given the velocity of the Freckle from [\ion{N}{II}] and the velocity
of the local background from the strong H$\alpha$ emission, simulate the 
total emission line over the Freckle by a sum of Gaussians at the background 
and Freckle velocities, and fit for the fluxes. Performance as judged by the quality
of the fits to bright lines, such as H$\alpha$ or [\ion{O}{III}], was 
generally unsatisfactory on account of the low velocity resolution of
MUSE, the generally small velocity separation of Freckle and background 
and small deviations from true Gaussians for the line profiles;
\item A simpler approach was to apply local background subtraction 
to the emission line maps already produced as described in Sect.
\ref{Sect:Images} and shown in Figures \ref{Fig:FluxImages1},  
\ref{Fig:FluxImages2} 
% DROPPED \ref{Fig:FluxImages3}, 
and \ref{Fig:FluxImages4}.
Rather than circular background apertures the shape and size of the
aperture was tuned to the location of nearby Freckles and the direction of
slope of the background emission. As estimators, both the median value 
over the area of the background external to the Freckle and a 
bilinear fit, resulting in removal of a sloping background, were
tested. Inspection of the subtracted background, showed that the latter 
produced the better fidelity background removal, although the median 
value produced more cases of feasible H$\alpha$/H$\beta$ and 
[\ion{O}{III}]/H$\beta$ ratios, simply because more background 
emission was included in the area of the Freckle.
\end{itemize}

Tab. \ref{Tab:FreckleSpecs} presents the set of line fluxes
for all the Freckles (designation as in Tab. \ref{Tab:FrecklesatN2}), 
determined from the last method (with bilinear background estimate). 
For all the Freckles, fluxes for a set of lines (\ion{H}{I}, \ion{He}{I},
\ion{He}{II}, [\ion{O}{I}], [\ion{O}{II}], [\ion{O}{III}], [\ion{N}{II}], 
[\ion{S}{II}] and [\ion{S}{III}]) are tabulated, but for multiple Freckles 
some or many of these lines clearly have large errors as indicated by 
infeasible values, or are undetected. The factors that 
affect the quality of these spectra range simply from low fluxes for fainter
Freckles (see H$\alpha$ flux listed in Tab. \ref{Tab:FreckleSpecs}, col. 2),
to the strength of the background emission (affecting particularly Freckles
closer to the CS), the proximity of neighbouring Freckles (c.f. the tight groups
39-42 and 57-61) and the local gradient and complexity of the background
nebular emission. Whilst detection of [\ion{N}{II}] 6583\AA\ is simple (strong line
on a weak background), fluxes for lines with strong background contribution, 
such as H$\alpha$, H$\beta$ and [\ion{O}{III}] could be subject to large 
systematics, and, for example, line ratios (H$\alpha$/H$\beta$ and 
[\ion{O}{III}] 4959/5007\AA) were measured outside allowed ranges. We 
therefore decided to rank the quality of the Freckle spectra into three quality
bands:
\begin{description}
\item A -- with high H$\alpha$ flux ($>$ 1.5$\times$10$^{-16}$ ergs cm$^{-2}$ s$^{-1}$), 
most lines detected, 2.5 $\leq$ R([\ion{O}{III}] 5007/4959\AA) $\leq$ 3.5,
and H$\alpha$/$\beta$ $>$ 2.8;
\item B -- with many lines detected and one of 2.0 $\leq$ R([\ion{O}{III}] 5007/4959\AA) 
$\leq$ 4.0 and 2.4 $\leq$ H$\alpha$/$\beta$ $\leq$ 6;
\item C -- neither the R([\ion{O}{III}] 5007/4959\AA) nor H$\alpha$/H$\beta$ within
the bounds of band B.
\end{description}
For category A spectra, extinction can be determined by comparison of 
H$\alpha$/H$\beta$ to the Case B value, and usually $N_{\rm e}$ from 
[\ion{S}{II}] 6716/6731\AA\ and $T_{\rm e}$ from [\ion{N}{II}] 5755/6583\AA), and 
[\ion{S}{III}] 6312/9069\AA.
For category B spectra extinction can also be determined, and often $N_{\rm e}$ from 
[\ion{S}{II}] 6716/6731\AA\ and occasionally $T_{\rm e}$ from [\ion{N}{II}] 5755/6583\AA), and 
[\ion{S}{III}] 6312/9069\AA.
For C spectra, extinction cannot be determined.
The total number of spectra in each quality band were: A 11; B 31; C 58;
these quality bands are listed in the last column of Tab. \ref{Tab:FreckleSpecs}.

Only those lines that were detectable in the Freckles are listed - notably lines 
with ionization energy $\lesssim$45\,eV (corresponding to detection of [\ion{O}{III}] 
and [\ion{Ar}{III}]). In many Freckle spectra, there is weak \ion{He}{II} 5412\AA\ 
detected; however only for Freckle 94 is this line clearly detectable on the 
\ion{He}{II} 5412\AA\ image (Fig. \ref{Fig:FluxImages1}, right) and its dereddened 
flux is only 1.7 (I(H$\alpha$) = 100). The spectra have, unusually, been normalised to 
% Whether to comment that He II residual from subtraction error.
H$\alpha$, since it is $\sim$3$\times$ brighter than H$\beta$ for both 
background and Freckle, so the contrast Freckle:background is greater and 
the errors lower for H$\alpha$ than H$\beta$, resulting in detection 
of a larger number of Freckles.

\begin{table*}
% Version with split parts to encompass ultra-long line
% Without [O I] 6363A 
\caption{Spectra of NGC~4361 Freckles}
\centering
% 11 cols including 6364A
\begin{tabular}{lrrrrrrrrrr}
\hline\hline
 No. & H$\alpha$ & I(H$\beta$)  & I([\ion{O}{III}]) & I([\ion{O}{III}]) & I(\ion{He}{II}) & I([\ion{N}{II}]) & I(He~I) & I(([\ion{O}{I}]) & I([\ion{S}{III}]) & I([\ion{O}{I}]) \\
     & (erg cm$^{-2}$ s$^{-1}$) & 4861\AA & 4959\AA & 5007\AA & 5412\AA & 5755\AA & 5876\AA & 6300\AA & 6312\AA & 6364\AA \\
\hline
% M  H-alpha &  H-beta & 4959 &  5007 &  5412 & 5755 &  5876 &  6300 &  6312 & 6364 \\
% Now with NEW numbering
%
1 & 3.64e-17 & 85.2 & 7.8 & 2.0 & 2.6 & 2.5 & 0.0 & 111.4 & 0.0 & 29.5 \\ 
2 & 8.19e-16 & 24.0 & 46.6 & 132.1 & 7.2 & 0.0 & 1.7 & 21.9 & 0.1 & 7.2 \\ 
3 & 2.52e-17 & 16.7 & 76.8 & 151.4 & 3.7 & 0.0 & 21.2 & 14.2 & 28.3 & 13.0 \\ 
4 & 3.27e-17 & 28.1 & 105.7 & 292.0 & 7.3 & 0.0 & 8.8 & 28.8 & 2.4 & 4.6 \\ 
5 & 7.06e-17 & 33.2 & 33.5 & 26.5 & 11.9 & 0.0 & 0.0 & 50.8 & 0.0 & 22.4 \\ 
6 & 8.00e-16 & 26.2 & 26.0 & 69.9 & 3.1 & 0.1 & 2.0 & 30.8 & 1.0 & 10.9 \\ 
7 & 4.06e-17 & 47.0 & 25.0 & 42.5 & 1.7 & 0.0 & 1.6 & 24.4 & 5.5 & 20.7 \\ 
8 & 1.23e-18 & 405.2 & 533.9 & 1445.6 & 163.3 & 0.0 & 0.0 & 1162.9 & 55.2 & 23.8 \\ 
% Decided to drop 9 as Ha flux only 14
% 9 & 0.0 & 0.0 & 0.0 & 0.00 & 00.0 & 0.0 & 0.0 & 0.0 & 0.0 
10 & 1.43e-17 & 66.8 & 89.4 & 104.5 & 28.7 & 4.7 & 16.9 & 93.1 & 12.7 & 11.8 \\ 
11 & 5.87e-17 & 29.1 & 40.3 & 37.7 & 12.9 & 1.2 & 3.8 & 42.4 & 25.5 & 20.2 \\ 
\hline
\end{tabular}
\vskip 0.5truecm
% 12 cols

\begin{tabular}{lrrrrrrrrrrr}
\hline
 No. & I([\ion{N}{II}]) & I([\ion{S}{II}]) & I([\ion{S}{II}]) & I(He~I) & I([\ion{Ar}{III}]) & I([\ion{O}{II}]) & I([\ion{O}{II}]) & I([\ion{Ar}{III}]) & I([\ion{S}{III}]) & I(H P9) & Grade \\
     &  6583 & 6716\AA & 6731\AA & 7065\AA & 7136\AA & 7319\AA & 7330\AA & 7751\AA & 9069\AA & 9229\AA &  \\
\hline
% M  &  6583 & 6716 &  6731 &  7065 &  7136 &  7320 &  7330 &  7751 &  9069 &  9229 & \\
% Now with NEW numbering
%
1 & 82.9 & 2.5 & 0.0 & 6.6 & 0.0 & 0.0 & 0.0 & 3.5 & 0.0 & 2.6 & C \\ 
2 & 34.2 & 0.9 & 2.1 & 5.0 & 4.0 & 2.8 & 2.1 & 0.8 & 4.0 & 7.2 & A \\ 
3 & 24.4 & 5.3 & 5.8 & 3.4 & 2.7 & 0.0 & 2.7 & 4.3 & 19.5 & 5.3 & B \\ 
4 & 18.1 & 5.1 & 8.9 & 7.7 & 5.5 & 3.3 & 0.0 & 9.1 & 14.7 & 17.0 & B \\ 
5 & 38.9 & 2.1 & 2.4 & 0.7 & 19.6 & 1.5 & 5.7 & 12.8 & 6.7 & 3.1 & B \\ 
6 & 23.0 & 0.5 & 0.5 & 6.8 & 4.3 & 3.8 & 2.3 & 1.6 & 7.3 & 4.6 & A \\ 
7 & 32.0 & 4.4 & 0.4 & 9.7 & 8.5 & 1.6 & 5.8 & 0.0 & 25.8 & 6.9 & C \\ 
8 & 325.5 & 32.8 & 101.5 & 79.1 & 107.3 & 0.0 & 204.7 & 129.6 & 544.0 & 72.7 & C \\
% 9 &  
% 0.0 & 0.0 & 0.0 & 0.0 & 0.0 & 0.0 & 0.0 & 0.0 & 0.0 & 0.0 & 0.0 & \\ 
10 & 42.9 & 10.9 & 11.5 & 23.9 & 1.8 & 15.3 & 6.0 & 44.5 & 16.1 & 27.5 & C \\ 
11 & 19.7 & 2.0 & 3.5 & 11.2 & 2.7 & 5.9 & 4.5 & 23.4 & 6.2 & 2.1 & B \\ 
\hline
\end{tabular}

\tablefoot{The position and area of each Freckle is listed in Tab. \ref{Tab:FrecklesatN2}. The
spectra are tabulated for all Freckles with detectable H$\alpha$ flux and the quality assessment 
of each spectrum is designated by A (highest), B or C (lowest) in the final column. All spectra 
are normalised to I(H$\alpha$) = 100.0. See text for details of the quality 
assignment of the Freckle spectra. \\
Only the first 10 entries of this table are shown; the full table is 
contained in the on-line material.} 
\label{Tab:FreckleSpecs}
\end{table*}

\begin{figure}
\centering
\resizebox{\hsize}{!}{
\includegraphics[width=0.45\textwidth,angle=0,clip]{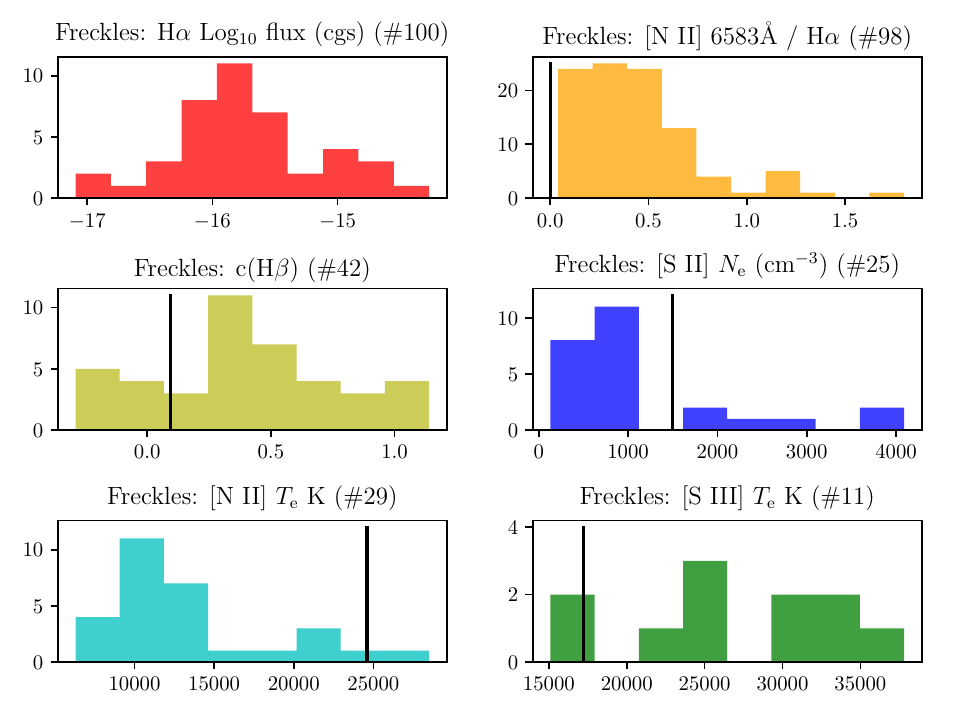}
}
\caption{Histograms of the properties of the Freckles: \newline
the observed H$\alpha$ flux (log erg cm$^{-2}$ s$^{-1}$) (top left);
the [\ion{N}{II}]6583\AA / H$\alpha$ ratio (upper right);
log extinction at H$\beta$, c from H$\alpha$/H$\beta$ (centre left); 
$N_{\rm e}$ ([\ion{S}{II}]) (cm$^{-3}$ (centre right);
$T_{\rm e}$ ([\ion{N}{II}]) K (bottom left); 
$T_{\rm e}$ ([\ion{S}{III}]) K (bottom right); \\
The vertical bar in the plots shows the corresponding value for the 
large scale nebular emission of NGC~4361 from the MUSE central field,
but excluding the area of the Freckles. \newline
Note that each histogram only plots the available number of values of
each parameter (listed in the title to each plot as (\#)\,); for the 
extinction, $N_{\rm e}$ and $T_{\rm e}$ histograms only the values from 
the A and B quality spectra are shown.
}
\label{Fig:FreckleHistos}
\end{figure}

The extinction for each Freckle was determined from the H$\alpha$/H$\beta$
ratio compared to Case B for assumed values of $T_{\rm e}$ = 11000\,K and 
$N_{\rm e}$ = 1500 cm$^{-3}$ (mean of $T_{\rm e}$ ([\ion{N}{II}]) and
$N_{\rm e}$ ([\ion{S}{II}]), see Tab. \ref{Tab:FreckleSumDiags}). 
Tab. \ref{Tab:FreckleDiags} then presents the 
[\ion{N}{II}] 6583\AA\ / H$\alpha$ ratio, and, for A and B quality spectra, the 
extinction, and those physical diagnostics, where the line ratios are available. 
Fig. \ref{Fig:FreckleHistos} shows histograms of the integrated (observed)
H$\alpha$ flux in the Freckles, the [\ion{N}{II}] 65683\AA/H$\alpha$ 
ratio, the extinction coefficient c, $N_{\rm e}$ ([\ion{S}{II}]) from
6716/6731\AA\ ratio, the $T_{\rm e}$ ([\ion{N}{II}]) from 5755/6583\AA\ and
the $T_{\rm e}$ ([\ion{S}{III}]) from 6311/9069\AA\ ratio. 
Since some of these diagnostics were not available for all Freckles, the 
number included in the histogram is listed in the title.
Shown on all the plots except for H$\alpha$ flux, are the equivalent values for
the integrated nebula (see Sect. \ref{Sect:PhysCond}). 
  
% Table of Freckle, c, Ne([S II]), Te([N II]), Te([S III])
% \input{Table_FreckleDiags_SHORT.tex}

\begin{table*}
\caption{NGC~4361 Freckle diagnostics}
\centering
\begin{tabular}{lrrrrr}
\hline\hline
 No. & [\ion{N}{II}]\,6583\AA & c(H$\beta$) & $N_{\rm e}$([\ion{S}{II}]) & $T_{\rm e}$ ([\ion{N}{II}]) &  $T_{\rm e}$ ([\ion{S}{III}]) \\
     &      /H$\alpha$        &             &               (cm$^{-3}$)~ &                       K~~~ &                       K~~~ \\
\hline
% Now with NEW numbering and [N II]/Ha 
% Revised to include only A and B spectra (excluding 79)
% Now with Te=11000K, Ne=1500 cm-3
% Rounded the c, Ne, Te values
%
% 1 & 0.00 & & 0.000 & - & - & - \\ 
2 & 0.34 & 0.52 & 47900 & - & - \\  
3 & 0.24 & 1.01 & 760 & - & - \\  
4 & 0.18 & 0.31 & 4090 & - & 17400 \\  
5 & 0.39 & 0.08 & 790 & - & - \\  
6 & 0.23 & 0.40 & 370 & 6570 & 15100 \\  
% 7 & 0.00 & & 0.000 & - & - & - \\ 
% 8 & 0.00 & & 0.000 & - & - & - \\ 
% 9 & 0.00 & & 0.000 & - & - & - \\ 
% 10 & 0.00 & & 0.000 & - & - & - \\ 
11 & 0.20 & 0.26 & 3600 & 24500 & - \\  
% 12 & 0.00 & & 0.000 & - & - & - \\ 
% 13 & 0.00 & & 0.000 & - & - & - \\ 
% 14 & 0.00 & & 0.000 & - & - & - \\ 
15 & 0.52 & -0.05 & 2320 & 10300 & - \\  
16 & 0.29 & 0.08 & - & - & - \\  
% 17 & 0.00 & & 0.000 & - & - & - \\ 
% 18 & 0.00 & & 0.000 & - & - & - \\ 
% 19 & 0.00 & & 0.000 & - & - & - \\ 
% 20 & 0.00 & & 0.000 & - & - & - \\ 
21 & 1.34 & -0.16 & 390 & 15500 & - \\  
22 & 0.40 & -0.01 & 390 & 6900 & - \\    
\hline
\end{tabular}
\tablefoot{Only Freckle spectra with quality bands A and B are included in this table.
Where one of the lines in the line ratio diagnostic was undetected 
and the value of $N_{\rm e}$ or $T_{\rm e}$ is undetermined, the value is indicated by - \\
Only the first 10 entries of this table are shown; the full table is 
contained in the on-line material.}
\label{Tab:FreckleDiags}
\end{table*}

A thorough error analysis of the Freckle fluxes (Tab. \ref{Tab:FreckleSpecs}) was not 
attempted but the errors over the Freckle areas and the subtraction of the 
local background (based on the flux errors resulting from the Gaussian fitting
of the emission lines, Sect. \ref{Sect:Images}) were propagated. For the brightest
Freckle (94) the errors range from $<$0.5\% for [\ion{N}{II}] 6583\AA\ (bright 
line on a low background) to 25\% for the faint line \ion{H}{I} P9 9229\AA. For 
H$\beta$, a bright line but on a high background, the propagated error was 2.5\%. 
For a Freckle whose flux is at the peak of the histogram (Fig. 
\ref{Fig:FreckleHistos}), e.g., 71, 
% Dont forget old 70 !
H$\alpha$ flux 1.4$\times$10$^{-16}$ ergs 
cm$^{-2}$ s$^{-1}$, the errors range from 11\% on [\ion{N}{II}] 6583\AA\ to
106\% for [\ion{Ar}{III}] 7751\AA. For fainter lines these values are typically
exceeded, depending on the position of the Freckle in the nebula -- i.e., the
background nebular emission and the close proximity of other Freckles. However
it must be considered that systematic errors arising from the fitting to the 
background around the Freckle and its subtraction may be much larger. For
example, even for the brightest Freckle, the difference between the 
total flux resulting from median and bilinear fitted backgrounds is 8\% for
the [\ion{N}{II}] 6583\AA\ line, indicating the level of uncertainty.

The total spectrum of all Freckles is presented in Tab. \ref{Tab:FrecklesTOTSpec},
together with the totals for the A and A+B quality spectra. 
The \ion{He}{II} 5412\AA\ line is listed for the total Freckles spectrum, but 
may be a systematic error from inadequate subtraction of the weak 
Freckle flux from the bright \ion{He}{II} background emission (except for 94).
The summed line fluxes can also be computed including those
above some cut-off. Table \ref{Tab:FreckleSumDiags} summarises the cut-off 
H$\alpha$ flux used in computing the sum, the resulting number of Freckles summed, 
the extinction from Case B (assuming $T_{\rm e}$ = 11000\,K and 
$N_{\rm e}$ = 1500 cm$^{-3}$), $N_{\rm e}$ ([\ion{S}{II}]), $T_{\rm e}$ ([\ion{N}{II}]) 
and $T_{\rm e}$ ([\ion{S}{III}]) (with the flux of [\ion{S}{III}] 6312\AA\ 
corrected for contamination by \ion{He}{II} 6311\AA).
It is clear from this table that the extinction increases as the Freckle 
H$\alpha$ flux increases (and for [\ion{N}{II}] 6583\AA\ flux, given that 
[\ion{N}{II}] 6583\AA / H$\alpha$ is almost independent of Freckle 
brightness). Also the $N_{\rm e}$ ([\ion{S}{II}]) increases with brightness,
while the $T_{\rm e}$ ([\ion{N}{II}]) seems to be independent of 
Freckle brightness. The $T_{\rm e}$ ([\ion{S}{III}]) shows indications 
of a decrease in $T_{\rm e}$ with brightness but given the very high
$T_{\rm e}$ values where this line ratio is not sensitive, this result may be
an effect of increasing S/N as few higher flux Freckles are summed. 
The alternative suggestion to very high $T_{\rm e}$ is of very high $N_{\rm e}$; 
for example a value of $N_{\rm e}$ $\gtrsim$ 2.7$\times$10$^{5}$ cm$^{-3}$ is 
needed to be consistent with $T_{\rm e}$ ([\ion{S}{III}]) of $<$ 20000\,K. However
there is no evidence for such high values from $N_{\rm e}$ ([\ion{S}{II}]).  

% Table of the spectrum of all 100 Freckles added (excluding 9 and 50)
% \input{Table_TOTFreckles.tex}

\begin{table*}
% Version with split parts to encompass ultra-long line
\caption{Summed spectrum of NGC~4361 Freckles}
\centering
% Redo extinction for T=11000K, Ne=1500
\begin{tabular}{lrrrrrrrrrr}
\hline\hline
% See Regs_measles/image_fitt_bilinearNEW/MTOTred.dat and various runs in Qual/

Spec. & H$\alpha$ & I(H$\beta$) & I([\ion{O}{III}])& I([\ion{O}{III}]) & I([\ion{He}{II}]) & I([\ion{N}{II}]) & I(He~I) & I(([\ion{O}{I}]) & I([\ion{S}{III}]) & I([\ion{O}{I}]) \\
     & (erg cm$^{-2}$ s$^{-1}$) & 4861\AA & 4959\AA & 5007\AA & 5412\AA & 5755\AA & 5876\AA & 6300\AA & 6312\AA & 6363\AA \\
\hline
A Obs.   & 1.50$\times$10$^{-14}$ &  26.3 &  52.0 & 150.4 & 2.5 & 0.8 &  3.7 &  23.6 &  1.3 &   7.1 \\
A+B Obs. & 2.03$\times$10$^{-14}$ &  26.6 &  45.2 & 129.6 & 3.2 & 0.8 &  4.3 &  27.7 &  2.0 &  8.9 \\
All Obs. & 2.40$\times$10$^{-14}$ &  31.2 &  45.2 &  124.6 & 4.2 & 0.9 &  5.1 &  32.8 &  2.7 &  10.5 \\
        &                        &       &       &       &     &     &      &       &       &     \\     
% Now Te=11000, Ne=1500
%
% Note that dered Ha for all Freckles less than for the A+B groups - because the extinction for 100 Freckles is lower!!
A Ext.   & 2.78$\times$10$^{-14}$ & 35.2 & 68.0 & 194.6 & 2.9 & 0.8 &  4.0 &  24.5 &  1.3 & 7.3 \\
% OLD T=1e4, Ne=1000 A Ext.   & 2.74$\times$10$^{-14}$ & 35.0 & 67.6 & 193.6 & 2.9 & 0.8 &  4.0 &  24.5 &  1.3 & 7.3 \\
A+B Ext. & 3.69$\times$10$^{-14}$ & 35.2 & 58.6  & 166.4 & 3.8 & 0.9 &  4.7 &  28.6 & 2.1 & 9.1 \\
All Ext. & 3.10$\times$10$^{-14}$ & 35.2 & 50.5 & 138.7 & 4.5 & 0.9 &  5.3 &  33.2 &  2.7 &  10.6  \\
\hline
\end{tabular}
\vskip 0.5truecm
\begin{tabular}{lrrrrrrrrrr}
\hline\hline
% See Regs_measles/image_fitt_bilinearNEW/QualAB/MTOTlins.dat and MTOTred.dat

Spec. & I([\ion{N}{II}]) & I([\ion{S}{II}]) & I([\ion{S}{II}]) & I(He~I) & I([\ion{Ar}{III}]) & I([\ion{O}{II}]) & I([\ion{O}{II}]) & I([\ion{Ar}{III}]) & I([\ion{S}{III}]) & I(H P9) \\
      & 6583\AA & 6716\AA & 6731\AA & 7065\AA & 7136\AA & 7319\AA & 7330\AA & 7751\AA & 9069\AA & 9229\AA \\
\hline
A Obs.   & 43.7 & 2.4 &  3.3 &  4.6 &  4.0 &  6.3 &  5.1 &  2.3 &  4.3 &  2.3 \\
A+B Obs. & 44.1 & 2.8 &  3.6 &  5.6 &  4.3 &  6.8 &  5.7 &  3.3 &  4.9 &  3.4 \\
% All Obs without 009
All Obs. & 44.9 & 3.3 &  4.6 &  6.3 &  4.7 &  7.3 &  6.5 &  4.4 &  5.7 &  4.6 \\
      &     &      &      &      &      &      &      &      &      &  \\
% OLD T=1e4, Ne=1000 A Ext.   & 43.6 &  2.2 & 3.3 &  4.4 &  3.7 &  5.7 &  4.7 &  2.0 &  3.4 &  1.8 \\
A Ext.   & 43.6 &  2.3 & 3.3 &  4.3 &  3.7 &  5.7 &  4.7 &  2.0 &  3.4 &  1.8 \\
A+B Ext. & 44.0 &  2.8 & 3.6 &  5.3 &  4.1 &  6.2 &  5.2 &  2.9 &  3.9 & 2.6 \\
% Now Te=11000, Ne=1500
All Ext. & 44.8 &  3.3 &  4.0 &  6.1 &  4.6 &  7.1 &  6.3 &  4.2 &  5.1 &  4.1 \\ 
\hline
\end{tabular}
\tablefoot{All spectra normalised to I(H$\alpha$) = 100.0. \newline
The extinction value applied (from H$\alpha$/H$\beta$, assuming $T_{\rm e}$ 
of 11000K, $N_{\rm e}$ of 1500 cm$^{-3}$) for the 12 A quality spectra summed 
is c=0.394, for the 42 A+B quality spectra summed c = 0.382 and for all 100 
spectra summed c = 0.164. See Tab. \ref{Tab:FreckleSpecs} for the quality 
assignment of the Freckle spectra.}
\label{Tab:FrecklesTOTSpec}
\end{table*}

\begin{table*}
\caption{NGC~4361 Freckle diagnostics - summed properties}
\centering
% Check these with Te=11000, Ne=1000
\begin{tabular}{lrrrrr}
\hline\hline
 H$\alpha$ cutoff & No. & c(H$\beta$) & $N_{\rm e}$ ([\ion{S}{II}]) & $T_{\rm e}$ ([\ion{N}{II}]) &  $T_{\rm e}$ ([\ion{S}{III}]) \\
( ergs cm$^{-2}$ s$^{-1}$)  &     &               & (cm$^{-3}$)                & K                          & K \\
\hline
% Values for 6311/5412 = 0.04623 (Te=17000K, Ne=1500cm-3) used to correct 6312A
% Revised to T=11000, Ne=1500
$>$ 1.0$\times$10$^{-18}$ & 100 & 0.164 & 1060 & 11340 & 61400 \\
$>$ 1.0$\times$10$^{-16}$ &  42 & 0.235 & 1200 & 11320 & 49600 \\
$>$ 2.0$\times$10$^{-16}$ &  21 & 0.306 & 1350 & 11170 & 47000 \\
$>$ 3.0$\times$10$^{-16}$ &  15 & 0.355 & 1590 & 10870 & 42100 \\
$>$ 4.0$\times$10$^{-16}$ &  10 & 0.426 & 1880 & 11220 & 37500 \\
$>$ 5.0$\times$10$^{-16}$ &   8 & 0.415 & 2140 & 11290 & 33900 \\
$>$ 1.0$\times$10$^{-15}$ &   5 & 0.386 & 1800 & 11380 & 35200 \\
                          &     &       &      &       &  \\
% A now only 11 spectra summed. Revised to T=11000K, Ne=1500cm-3
A summed                  &  11 & 0.394 & 1640 & 11000 & 33700 \\
A+B summed                &  42 & 0.382 & 1240 & 11400 & 60300 \\
\hline
\end{tabular}
% \tablefoot{} 
\label{Tab:FreckleSumDiags}
\end{table*}

%__________________________________________________________________

\section{Discussion}
\label{Sect:Discuss}

\subsection{Optically thin / optically thick}
\label{SubSect.ThinThick}

There is abundant evidence that NGC~4361 is optically thin to Lyman continuum
radiation, but with the detection of low ionization species, in particular the 
'Freckles' strong in [\ion{N}{II}], the question is now raised whether the nebula is
density bounded at its outer extremities. The qualitative evidence that a PN
is optically thin typically derives from its lack of extended low ionization
line emission, particularly in the outer emission zone. The quantitative evidence 
that a nebula is optically thin usually derives from a comparison of the \ion{H}{I} and 
\ion{He}{II} Zanstra temperatures. The Zanstra temperature \citep{Zanstra1931} is 
derived by comparing the ionizing flux required to produce a given 
emission line flux (either an \ion{H}{I} or an \ion{He}{II} line) 
compared to the flux in the central star continuum at some measured wavelength 
(typically in the optical range close in wavelength to the emission line in 
question).  The relation between the stellar flux measured in the optical and
the far-UV ionizing flux required to produce the line emission can be provided by
either a black body (BB) assumption or by a model atmosphere. Several authors have
determined \ion{H}{I} and \ion{He}{II} Zanstra temperatures for NGC~4361 -- \citet{Phillips2003}
(41\footnote{Note that \citet{Phillips2003} actually derives \ion{He}{I} Zanstra temperature}
and 93kK, BB assumption), \citet{Mendez1992} (43 and 99kK, non local thermodynamic 
equilibrium (NLTE) model atmosphere).
A more fundamental approach to determining the optical depth of a nebula is
to construct a photoionization model that matches the emission and general 
appearance of the nebula with a gas shell of given composition and a compatible
central star. The photoionization model of \citet{Aller1979} finds H$^{+}$ 
$>$99\% ionized throughout the nebula and the model of \citet{TorresPeimbert1990} 
(their model B) indicates that only 7\% of the Lyman continuum is absorbed by 
the nebula. From the spectra of the central star and NLTE model atmosphere fits, 
\citet{Mendez1992} determined that 11\% of the H Lyman continuum is absorbed by 
the nebula. 

None of these earlier studies had significantly detected low ionization species, 
such as [\ion{O}{II}] or [\ion{N}{II}], but while the [\ion{N}{II}] image in Fig. 
\ref{Fig:FluxImages4} shows some extended low ionization emission, there is no 
definitive evidence of a classical Str\"o{}mgren sphere expected for an ionization
bounded nebula. The \ion{He}{I} and \ion{He}{II} images (Fig. \ref{Fig:RatioImages1}, upper 
right) show that \ion{He}{I} is very weak throughout the nebula and He$^{+}$/H$^{+}$ 
integrated over the whole nebula is only 4.5\% (Sect. \ref{SubSubSect.ORLs}) 
% (He$^{+}$/H$^{+}$ = 0.0045, He$^{++}$/H$^{+}$ = 0.0956)
of the total He/H$^{+}$. This
raises the question whether NGC~4361 is unusual in also being, at least partially,
optically thin in the \ion{He}{II} ionizing continuum (shortward of 228\AA). The ratio
map \ion{He}{I}/\ion{He}{II} (Fig. \ref{Fig:RatioImages1}) shows marginal radial 
enhancement, and for the offset field (NGC~4361 W, Tab. \ref{Tab:Obs}), in which 
emission is detected to radii of $\gtrsim$40$''$, \ion{He}{I}
emission remains very weak. From the integrated emission of all the Freckles,
the He$^{+}$/H$^{+}$ ratio is 0.11 (see following section), whilst for the
extended nebula (excluding the Freckles) the He$^{+}$/H$^{+}$ = 0.004 and 
He$^{++}$/H$^{+}$ = 0.095, so that the total He/H$^{+}$ for the nebula
is close to the value for the Freckles, implying that there is little margin for
extra He$^{+}$/H$^{+}$ (see Tab. \ref{Tab:Abunds} and discussion in the following 
section).

Does the optical thinness arise from overall low density or an inhomogeneous 
structure with large low density opening angles, or a combination of both? 
The spatially extended (i.e., excluding the Freckles) $N_{\rm e}$ values from [
\ion{S}{II}] 6716/6731\AA\ and [\ion{Cl}{III}] 5517/5537 \AA\ ratios indicate a 
mean value (weighted by inverse errors) of 1580 cm$^{-3}$. 
\citet{TorresPeimbert1990} quote a total dereddened H$\beta$ flux 
(corrected for contamination by the \ion{He}{II}\,4859.3\AA\ n=4--8 line)
of 3.63$\times$10$^{-11}$ ergs cm$^{-2}$ s$^{-1}$ in an 81$''$ diameter aperture;
compared to the dereddened [c=0.10] integrated H$\beta$ flux in the central 
MUSE field (diameter 68$''$) of 1.97 $\times$10$^{-11}$ ergs cm$^{-2}$ s$^{-1}$ 
(also corrected for \ion{He}{II}\,4859\AA), thus confirming that flux was lost
through lowered atmospheric transparency and the non-optimal sky subtraction
(Sect. \ref{Sect:ObsRed}), in addition to the smaller aperture. For an assumed 
spherical nebula of diameter of 80 arcsec\footnote{On the Digital Sky Survey 
(DSS2) red image, the area is 3.3 arcmin$^{2}$ and the full radius 62$''$.}, 
% See NGC4361/process_dss.cl
and using the \citet{HummerStorey1987} H$\beta$ emissivity for  
the \ion{H}{I} T$_{PJ}$ of 7\,500\,K (Tab. \ref{Tab:PJTes}, col. 5), the root 
mean square density is 190 cm$^{-3}$, implying a filling factor 
(\citet{TorresPeimbert1990}, Sect. 6) of 0.015,
for an electron yield per atom of 1.25 (since 95\% of He is in the form of 
He$^{++}$). The very low filling factor suggests that the line of sight 
is filled by only a small volume of gas whose density is that determined
from the CEL diagnostic ratios (viz. [\ion{S}{II}] and [\ion{Cl}{III}]). From the
appearance of the nebula, a possible morphology could be a barrel whose 
rim has the density given by the diagnostic CEL lines but whose centre has
a much lower density and whose ends (to the NE and SW where the 'ears'
are seen) are open to H Lyman continuum radiation escape. The kinematic 
mapping of \citet{MuthaAnandarao2001}, interpreted as a bipolar (or even 
quadrupolar) form, would at least suggest a structure whose NE lobe is 
tilted towards the line-of-sight, and SW one away from the line-of-sight, 
as also implied by the large-scale trend in the H$\alpha$ velocity image (Fig. 
\ref{Fig:N4361+Freckle_vels}). 

The optical colour temperature of the central star indicates a BB temperature
$\sim$ 89\,kK, whilst fits to \citet{Rauch2003} model atmospheres indicate values
$\sim$ 80\,kK (Appendix \ref{App:CS}). These values are close to, and certainly not 
exceeding, the \ion{He}{II} Zanstra temperature estimates, indicating no strong evidence 
for escape of \ion{He}{II} ionizing photons and hence underestimate of the number of 
photons between 13.6 and 54.4eV. Nevertheless the extreme \ion{He}{II}/H$\beta$ line 
ratio of NGC~4361 among PNe, the very high ionization level of the bright 
emission lines and the weakness of the \ion{He}{I} emission (except in the Freckles),
suggest that the nebula could be optically thin in the He$^{+}$ ionizing 
continuum, at least in some directions. Only a detailed photoionization model 
with a tuned CS model atmosphere together with deeper spectroscopy of the outer 
extent of the nebula, could resolve this question.

\subsection{The properties of the NGC~4361 Freckles}
\label{SubSect. FreckleProps}

The histograms presented in Fig. \ref{Fig:FreckleHistos} summarise some of the
essential properties of the Freckles:
% Now included new c and extinction corrected Te, Ne values for Te=11000K, Ne=1500cm-3
\begin{description}
\item the [\ion{N}{II}]/H$\alpha$ ratio (mean value 0.45) far exceeds the 
ratio for the integrated high ionization nebula (0.0013), which led to 
their detection on the [\ion{N}{II}] image (Fig. \ref{Fig:FluxImages4});
% See Regs_Freckles/image_fitt_bilinearNEW/Freckles_N2Ha.dat (mean value 0.514 for 0.0 and <3.5)

\item the extinction is generally larger than for the large scale-nebula (mean value
0.39);
% See Regs_Freckles/image_fitt_bilinearNEW/Freckle_c.dat excluding c=0.0, 46 values

\item the electron density is more often lower than the value determined for the 
nebula overall, also the mean value ($N_{\rm e}$ = 1160 cm$^{-3}$); 
% See Regs_Freckles/image_fitt_bilinearNEW/Freckle_Ne.dat (55 values 20<Ne<1e4)

\item the $T_{\rm e}$ ([\ion{N}{II}]) is much lower (mean value 13200\,K)
than the equivalent value for the total nebula (excluding the Freckles) of 27700\,K; 
% See Regs_Freckles/image_fitt_bilinearNEW/Freckle_N2Te.datfor 46 values, excluding 5000<Te<30000)

\item the $T_{\rm e}$ ([\ion{S}{III}]) is much larger (mean value 27100\,K) than the 
equivalent value for the total nebula of 17100\,K. 
\end{description}

Concerning the sizes of the Freckles, they are not spatially resolved. The apparent 
size of the Freckles in the [\ion{N}{II}] 6583\AA\ image (Fig. 
\ref{Fig:FluxImages4}) does not differ from the point spread function, as measured 
for the central star at the same wavelength based on 2D Gaussian fits to nine well 
isolated and obviously single Freckles (full width at half maximum, 0.69$\pm$0.04$''$, 
compared to the PSF of the CS, 0.73$''$, at 6600\AA). Thus the dimension of the 
Freckles is $<$10$^{16}$ cm. 

A comparison of the abundances of the light elements (He, N \& O) between the 
Freckles and the bulk nebula could strengthen indications that the Freckles
represent a distinct component. Based on the $T_{\rm e}$ and $N_{\rm e}$ values
presented in Tab. \ref{Tab:FreckleSumDiags} for the summed Freckle (A+B) spectra 
(adopting $T_{\rm e}$ ([\ion{N}{II}]) 11000\,K and $N_{\rm e}$ ([\ion{S}{II}]) 1500 cm$^{-3}$) 
and values of $T_{\rm e}$ 17000\,K and $N_{\rm e}$ ([\ion{S}{II}]) 1500cm$^{-3}$ from
Tab. \ref{Tab:CELDiags} for the high ionization medium (Sec. \ref{SubSect.ThinThick}), 
Tab. \ref{Tab:Abunds} presents He$^{+}$, He$^{++}$, N$^{+}$, and O$^{+}$, O$^{++}$, 
S$^{+}$, S$^{++}$, and total He/H$^{+}$, N/H$^{+}$, O/H$^{+}$, 
S/H$^{+}$ and N/O abundances. The total abundances for N, O and S, were calculated 
with the ionization correction factors (ICFs) of \citet{KingBarlow1994} and
\citet{Delgado-Inglada2014}, but it should be borne in mind that both scales are
developed on lower ionization PNe and so the ICF's will not be accurate for
the integrated high ionization component of NGC~4361 (for example, the 
ICF for N is 329 based on \citet{KingBarlow1994}). The value of N/O should be
more reliable than N/H$^{+}$, and shows that the Freckles
have a higher value than the bulk nebula gas; the values of He/H$^{+}$ 
are not significantly different between both components, given that the 
error on He$^{+}$/H$^{+}$ for the Freckles is $\sim$10\%. However given that
N/O derived for the high ionization component is very low compared to
typical Galactic Disk PNe \citep{KingBarlow1994}, while the value for the 
Freckles is closer to the current Solar value \citep{Amarsietal2021}, suggests 
problems in determining N/O for such an extreme ionization medium. 
The O/H$^{+}$ value is also larger for the summed Freckles than 
for the bulk nebula by about a factor 2 (depending on choice of ICF); this 
increase hints at the O abundance of the Freckles being larger, but again 
must be treated with caution given that the ICF's are being applied beyond 
their recommended range (in He$^{++}$/He). S/H$^{+}$ is also higher for the 
Freckles than the bulk nebula with the ICF from \citet{KingBarlow1994}, but 
lower with the ICF from \citet{Delgado-Inglada2014}.
 
\begin{table}
\caption{NGC~4361 Comparison of abundances for extended nebula and Freckles}
\centering
\begin{tabular}{lrrrlr}
\hline\hline
Ionic ratio         & Nebula  & Freckles \\
\hline
% He+/H+ and He++/H+ recomputed with Te=17000, Ne=1500 (see He_abunds_5876.py in Regions/PasJump
% For Freckles see Regions/Abunds and using A+B spectra and Te=11000K, Ne=1500cm-3
% see calc_ON_REV.cl and calc_S.cl and ICFs.cl. Summary is icfs.txt
He$^{+}$/H$^{+}$    & 0.0038  & 0.0918  \\
He$^{++}$/H$^{+}$   & 0.0950  & $<$0.02 \\
He/H$^{+}$          & 0.0988  & 0.111    \\
\hline
N$^{+}$/H$^{+}$     & 2.38$\times$10$^{-8}$ & 2.04$\times$10$^{-5}$ \\

N/H$^{+}$ $\ast$    & 7.84$\times$10$^{-6}$ & 4.06$\times$10 $^{-5}$ \\
N/H$^{+}$ $\dagger$ & 6.68$\times$10$^{-6}$ & 3.73$\times$10 $^{-5}$ \\
\hline
O$^{+}$/H$^{+}$     & 6.06$\times$10$^{-7}$ & 2.52$\times$10 $^{-4}$ \\ 
O$^{++}$/H$^{+}$    & 2.22$\times$10$^{-5}$ & 1.21$\times$10 $^{-4}$ \\

O/H$^{+}$ $\ast$    & 1.99$\times$10$^{-4}$ & 5.03$\times$10 $^{-4}$ \\
O/H$^{+}$ $\dagger$ & 2.42$\times$10$^{-4}$ & 4.94$\times$10 $^{-4}$ \\
\hline
N$^{+}$/O$^{+}$     & 0.0393                 & 0.0807 \\

N/O $\ast$          & 0.0393                 & 0.0807 \\
N/O  $\dagger$      & 0.0276                 & 0.0755 \\
\hline
S$^{+}$/H$^{+}$     & 3.13$\times$10$^{-9}$ & 4.41$\times$10$^{-7}$ \\ 
S$^{++}$/H$^{+}$    & 1.69$\times$10$^{-7}$ & 1.41$\times$10$^{-6}$ \\

S/H$^{+}$ $\ast$    & 8.25$\times$10$^{-7}$ & 1.94$\times$10$^{-6}$ \\
S/H$^{+}$ $\dagger$ & 3.93$\times$10$^{-6}$ & 2.40$\times$10$^{-6}$ \\
\hline
\end{tabular}
\tablefoot{ $\ast$ calculated with Kingsburgh \& Barlow (1994) ICF; \\
$\dagger$ calculated with Delgado-Inglada et al. (2014) ICF.
}
\label{Tab:Abunds}
\end{table}

Another aspect of the Freckles that distinguishes them from the extended high
ionization nebula, is their radial velocity. Figure \ref{Fig:N4361+Freckle_vels}
shows the Freckle velocities on the large scale H$\alpha$ velocity field, whose
main feature is a gradient from E to W ($\sim$ -12 to +20 km\,s$^{-1}$ with respect 
to systemic velocity, or -15 to +24 km\,s$^{-1}$ NE to SW). Most Freckles show a 
positive or negative velocity offset
with respect to H$\alpha$ (Tab. \ref{Tab:FrecklesatN2}, col. 7). Examining the
image of the velocity difference ([\ion{N}{II}] - H$\alpha$) reveals that the 
Freckles segregate between $+$ve offset to the NE and $-$ve offset to SW, with
only 12/102 within $\pm$5 km\,s$^{-1}$ of the zero velocity (= local H$\alpha$ 
velocity). The line demarcating
the $-$ve from $+$ve Freckle velocity offsets is at PA$\sim$ 136 $^{\circ}$, thus 
approximately perpendicular to the large-scale ellipticity of the PN morphology
(PA $\sim$ 44 $^{\circ}$). This strongly suggests that the system of Freckles forms
a thick equatorial structure, perhaps a disc, around the more diffuse high ionization 
core, with the NE portion titled away from the observer ($+$ve Freckle velocities) and
the SW towards the observer into the plane of the sky ($-$ve velocities). 

However this picture is complicated by the faint extended [\ion{N}{II}] structure 
which is visible in the 6583\AA\ image (Fig. \ref{Fig:FluxImages4}) composed of a
(mirror inverted) comma-shaped cloud to the E \& NE and a more elliptical one to the NW. 
On the ([\ion{N}{II}] - H$\alpha$) velocity difference image, the eastern cloud is 
velocity offset $-$ve ($\sim$ -8 km s$^{-1}$) and the NW cloud offset in $+$ve
velocity ($\sim$ +9 km s$^{-1}$). This trend is in the opposite velocity sense to 
the Freckles, suggesting it is a different structure, related by similarity to 
the large scale nebula, shown in the H$\alpha$ velocity field in Fig.
\ref{Fig:N4361+Freckle_vels}. It is plausible that these regions are lower ionization
layers outside the high ionization zone, and indeed these features are
seen on higher ionization CEL images, such as [\ion{O}{III}] (Fig. 
\ref{Fig:FluxImages2}, lower left) and the \ion{He}{I} 5876\,\AA / \ion{He}{II} 5412\,\AA\  
(Fig. \ref{Fig:RatioImages1}, upper right). Partial recombination of He$^{++}$ is
occurring in these regions implying increased optical depth to He$^{+}$ ionizing
photons in some parts of the nebula, perhaps in outer extremities of the structure.
However the putative disc of Freckles does not show up as extended emission and
as mentioned has a distinctly different velocity displacement. North and South of the CS,
where there is increased spatial density of Freckles, extended low ionization 
emission is very weak. 

\subsection{The nature of the NGC~4361 Freckles}
\label{SubSect:Nature}

The Freckles present a radically dichotomous nebular phase to the bulk 
very high ionization medium, and as a result their nature is mystifying;
the most extreme difference is seen in the factor $\sim$340 for 
[\ion{N}{II}] 6583\AA/H$\beta$ ratio between Freckles and the
integrated high ionization emission. 
% N. B.
They also present a distinct component to the extended low ionization 
emission regions to the E, NW and SW (Fig. \ref{Fig:FluxImages4}) which also 
have similar morphology to the brighter mid-ionization (30--60\,eV) CEL 
emission (Fig. \ref{Fig:FluxImages2}). 
Their distinction from the other nebula components raises the question    
whether the Freckles could be remoter from the CS than the extended CEL 
layers. They may represent late ejections from the star or  
the remains of a distinct component of the asymptotic giant 
branch (AGB) envelope, such as a disk, perhaps even from an 
earlier (pre-AGB) phase, such as a circumstellar disk with the 
Freckles as planets/asteroids. Their spatial
distribution and trend in velocity from $-$ve SW of the CS to $+$ve to
the NW, suggests a distribution in a thick disc oriented almost 
perpendicular to the extension of the high ionization optically thin 
component. This orientation also coincides with 'Reflection axis B' in the
quadrupolar interpretation of \citet{MuthaAnandarao2001}. However the 
velocity tilt of the disc along a NE-SW axis is distinct from the
velocity trend mapped by \citet{MuthaAnandarao2001}, suggesting
a sextupolar structure in their picture. 

It is tempting to interpret the radial orientation of some knots as 
indication of a collimated ejection origin, but only a minority
of the Freckles can be fitted into this picture -- so either
the Freckles are a mix of density peaks within a larger (disc) 
structure or later ejection knots, with the eye led to over-interpret 
almost linear associations as connected features (particularly since
linear trends of knots do not display distinct velocity gradients). 
Higher velocity resolution spectra and an investigation of any proper 
motions (a long term goal given that the Freckles have only been detected 
in the epoch 2014 observations) would be required to advance these suggestions. 
The increased N/H$^{+}$ for the Freckles, and more uncertainly He/H$^{+}$, 
would tend to favour a distinct ejection, perhaps associated
with a later evolutionary phase, but the uncertainty in N/O determination
for the high ionization gas weakens this claim. 

The Freckles are suggested to be condensations within a larger equatorial disc
structure, remnants of dense regions which have been photo-evaporated
and are currently ionized. However from the lack of high $N_{\rm e}$ for
the Freckles, it is not clear that they are high density 
condensations, but the increased extinction, the low ionization and 
presence of neutral emission ([\ion{O}{I}]) points in the direction
of optically thick clumps; then the 
low [\ion{S}{II}] $N_{\rm e}$ could be a measure only of the 
ionized flow from a dense globule. Also the trend of higher
$T_{\rm e}$ ([\ion{S}{III}]) than $T_{\rm e}$ ([\ion{N}{II}]) could
suggest photo-evaporated gas which becomes more highly ionized
(and excited) further from the condensation, approaching the higher ionization 
optically thin conditions. 

From their properties with respect to the bulk nebular emission 
(low ionization, elevated extinction, lack of high $N_{\rm e}$, 
similar abundances), the Freckles resemble the Low Ionization Structures (LIS) 
investigated by \cite{AkrasGoncalves2016} and \cite{Belenetal2023}. However 
the Freckles show a much higher spatial frequency
than, for example, LIS's in NGC~7009 and NGC~6543 \citep{Belenetal2023}, are 
more compact, show enhanced $T_{\rm e}$ ([\ion{N}{II}]) and occur in an 
older PN\footnote{On the basis of the observed V mag of 13.26 \citep{Frew2016} 
(dereddened 13.04 mag. for c = 0.10) and T$_{eff}$ of 89\,000K 
(Appendix \ref{App:CS}), the BB luminosity is 1200L$_{\odot}$. The CS thus 
appears to be on the knee of the low-mass track \citep{MillerBertolami2016}, 
with an age of $\sim$10$^{4}$yr.}

%
% Extract from Weidmann et al. 2016
% Designat    Log g                             Log Teff                   Log L                            V mag 
% 294.1+43.6   5.50             MK1992  eff      5.100             GM2019b  3.540   GM2019b  v   13.26           FP2016
% My little prog for computing L from mag and Teff
% ~/MOCASSIN/PyMOCASSIN/MakeDenMod/Star/BBstarLT4.e
%

It is therefore more probable that the Freckles are analogous to
the cometary knots in the Helix Nebula (NGC~7293) and the extinction
knots in NGC~6720 \citep{Wessonetal2024}, rather than to LIS's in
younger PN. While the extinction to the 
Freckles is generally higher than the surroundings, extinction 
cores to the Freckles are not detected in the 
[\ion{O}{III}] images (or for higher ionization lines) even
for Freckles with $-$ve velocities (presumably in the foreground), as 
in the  case of the cometary knots in the Helix Nebula, NGC~7293 
\citep{Meaburn1992, ODell2000}, or similar extinction knots in 
NGC~6720 \citep{Odell2013}. NGC~4361 is however 
about five times more distant than the NGC~7293. Again higher velocity 
resolution observations could reveal more details of the ionized 
structure of the knots and also imaging at higher resolution could 
test if the knots show structural similarities to those in NGC~7293
and NGC~6720.
 
The Freckles disc could be remoter from the CS than the extended CEL 
layers hinting at an earlier evolutionary status, perhaps related to 
the preceding AGB phase. If the low ionization of the Freckles were a 
result of their being very remote from the CS, in a zone where the 
nebula is optically thick, it is tempting to suggest they could be similar
to planets/asteroids. For Freckle 94, 26.9'' from the CS, a projected distance
$\gtrsim$ 0.14\,pc, this would rather be equivalent to an Oort Cloud body 
(2$\times$10$^{3}$ -- 2$\times$10$^{5}$ AU) rather than a planet in the 
Solar System context.
Given the largest systemic radial velocity of the Freckles is $\sim$ 40
km\,s$^{-1}$, then a lower limit on the age for a late phase ejection 
is $\gtrsim$ 2700 yr, generally less than the kinematic ages derived for 
various velocity components by \citet{Vazquezetal1999, MuthaAnandarao2001}.
However if the measured radial velocities of the Freckles correspond to an 
ionized flow from a dense neutral condensation, and not of the globule itself, 
then any age estimate is biased.

%__________________________________________________________________

\section{NGC~4361's galaxy in hiding}
\label{Sect:Steve}

\begin{figure*}
\centering
\resizebox{\hsize}{!}{
\includegraphics[width=0.5\textwidth, trim={0 110 0 110}, angle=0, clip]{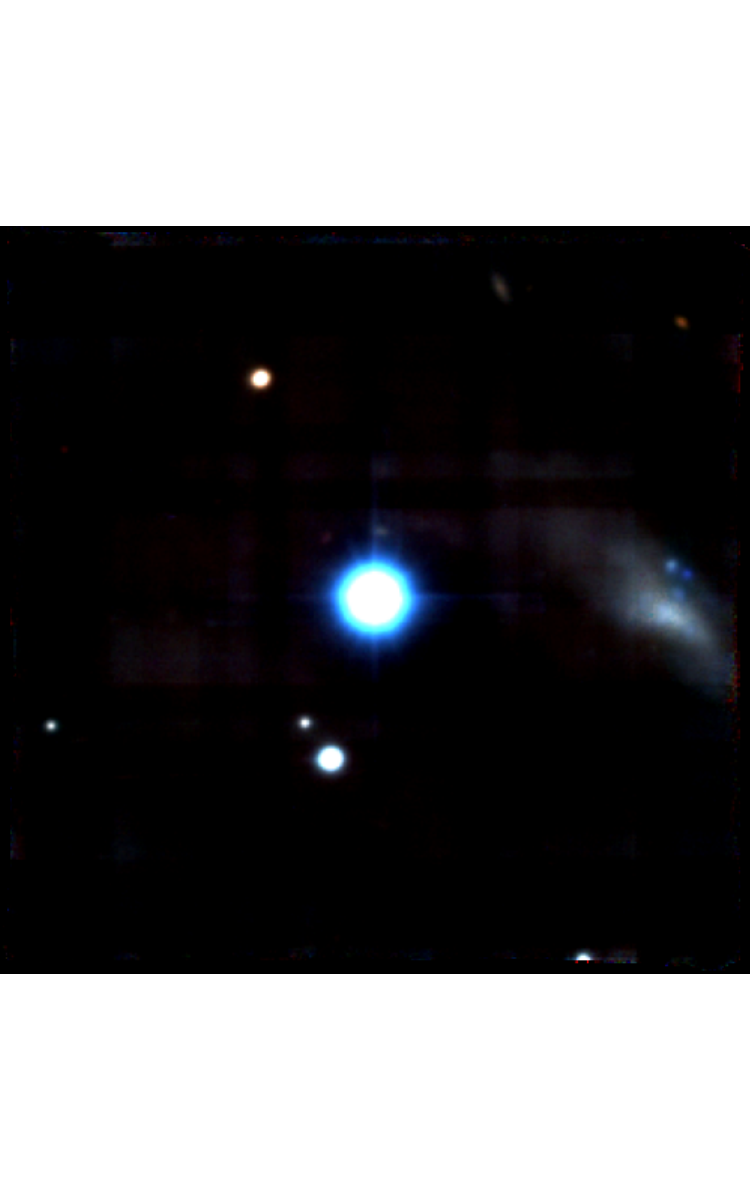}
\hspace{0.2truecm}
\includegraphics[width=0.63\textwidth, trim={0 10 0 10}, angle=0, clip]{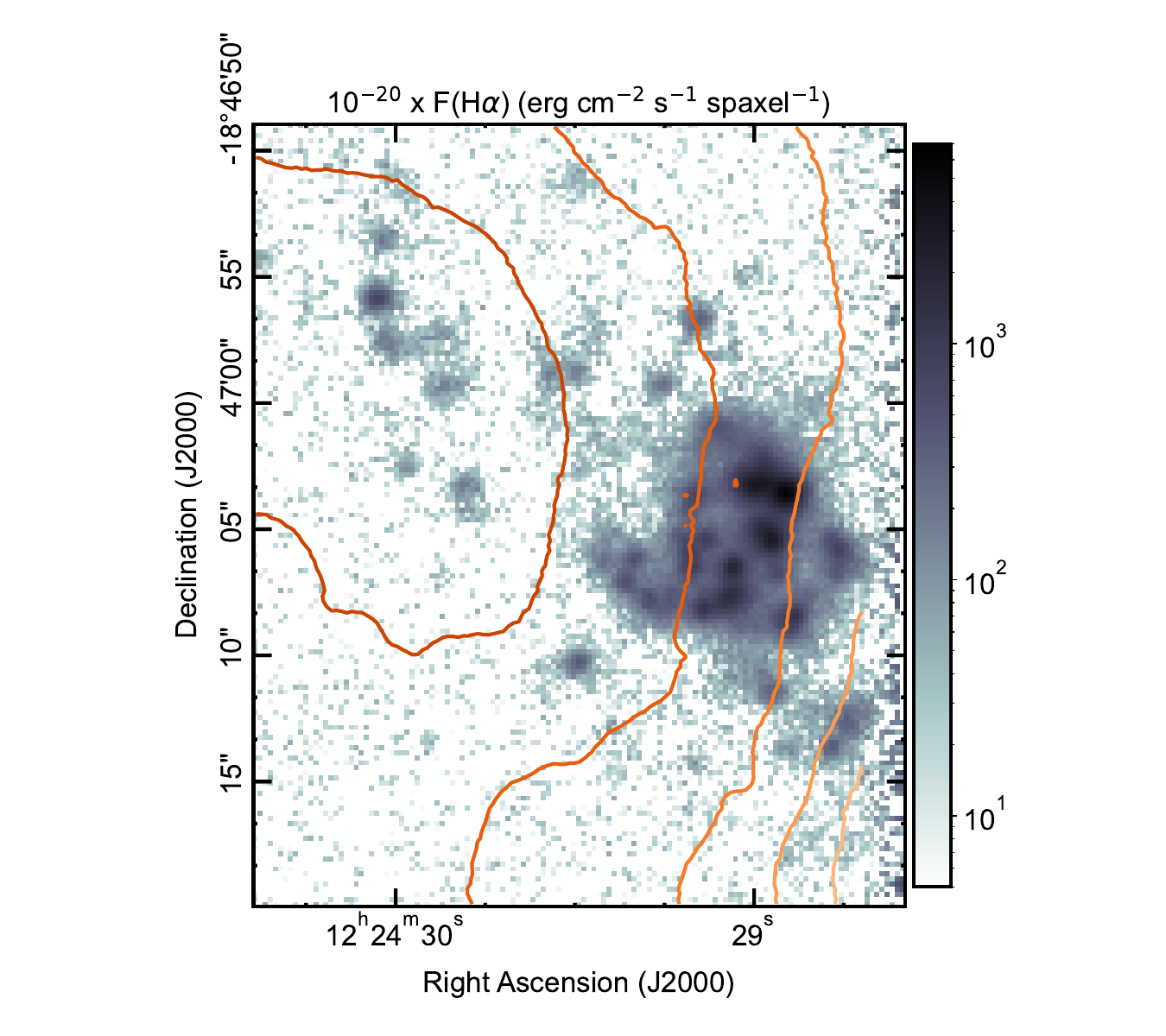}
\hspace{0.2truecm}
\includegraphics[width=0.63\textwidth, trim={0 10 0 10}, angle=0, clip]{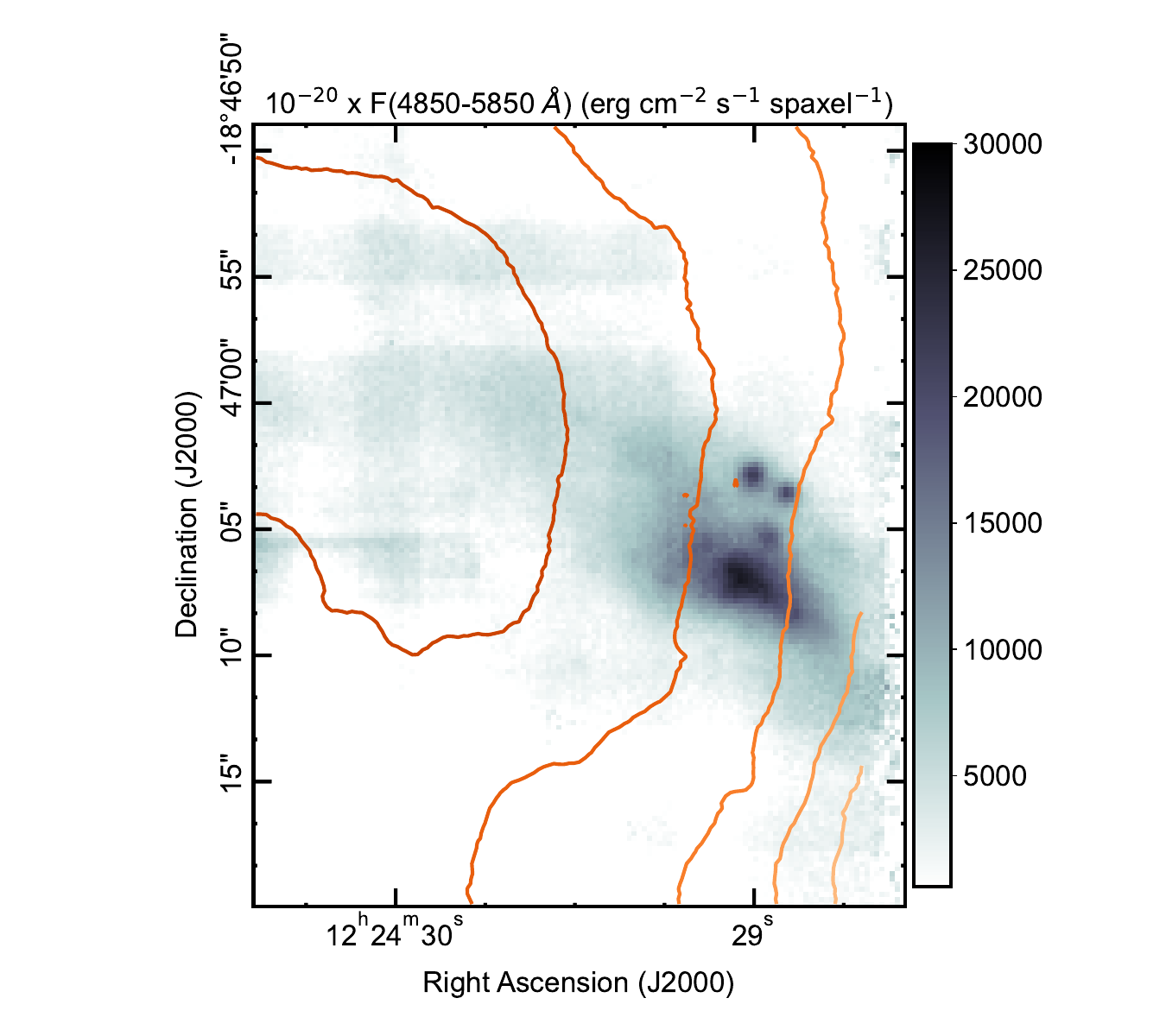}
% Now with coordinates fixed
}
\caption{Left: RGB image (R: 7500 - 8500\AA; G: 5600 - 6600\AA; B: 4850 - 5850\AA) of the 
whole MUSE field-of-view after NGC~4361 nebular continuum subtraction. The continuum emission of the background galaxy is clearly seen to the west. \\
Middle: H$\alpha$ image of the $z$ = 0.021 background galaxy to NGC~4361. The contours of NGC~4361 H$\beta$ emission are as in Fig. \ref{Fig:FluxImages1}.\\
Right: Broadband 'blue' (4850-5850\AA) image of the galaxy after NGC~4361 nebular continuum subtraction and foreground extinction correction, again with NGC~4361 H$\beta$ contours.
}
\label{Fig:SteveHaCon}
\end{figure*}

When panning through all the wavelength slices of the MUSE cube, a 
small region showing an emission line at 6703\AA\ was noticed, 
which does not match any bright PN nebular line. Other lines were 
found in the same area and their separations clearly showed a set 
of typical nebular emission lines (H$\alpha$, H$\beta$,
[\ion{O}{III}] 4959, 5007\AA, [\ion{N}{II}] 6548, 6563\AA\
and [\ion{S}{II}] 6717, 6731\AA) with H$\alpha$ as the brightest. 
There is thus a background emission line galaxy at this position 
(25.6$''$ offset from NGC~4361 CS at PA 267$^{\circ}$) 'shining' 
through the emitting gas from the PN. Subsequent to this 
finding, it was noticed that in the Spitzer IRAC two colour image 
shown in Fig. \ref{Fig:MUSEFields} there
is a hint of a bluer area directly W of the central star at 
a similar position to the galaxy's line emission. We investigated 
the nature of this previously hidden galaxy, which is designated as
{\tt NGC4361-BgGal1224290-184707}\footnote{The investigation of this
'galaxy in hiding' was completed as part of the Bachelor thesis by two 
of the authors; during this work the galaxy was nicknamed 'Steve'.}. 
While the observed emission lines from this galaxy do not coincide with
the NGC~4361 nebular lines, the strong continuum (bf, ff and 2-$\nu$) 
poses a challenge to extracting its spectrum. The nebular continuum is 
similarly strong to the galaxy continuum, so must be removed to investigate 
its stellar properties, such as luminosity, radial profile and 
stellar population. The spatial and wavelength behaviour of the 
nebular continuum distribution was fitted by a bicubic in both
$\lambda$ and offset from the central star and scaled by the normalised 
H$\alpha$ image of NGC~4361. Only the median continuum flux in steps of 
100\AA\ was fitted over the full cube and regions around the CS and the 
galaxy itself were excluded by a mask. Fig. \ref{Fig:SteveHaCon}, left 
image, shows the resulting red-green-blue (RGB) tagged image of the galaxy 
after this foreground continuum removal. Johnson V and Cousins I 
photometry was performed on the cleaned spectra of the galaxy using the
\texttt{mpdaf}\footnote{https://mpdaf.readthedocs.io/en/latest/index.html} 
package \citep{Piquerasetal2019} and the radial surface
brightness, with the area of the H\,II regions excluded, was fitted 
by a Sersic profile \citep{CiottiBertin1999}. Fig. \ref{Fig:SteveHaCon}, 
right image, shows the expanded view of the 'blue' (4850--5850\AA) 
continuum of the galaxy. 

From the H$\alpha$ image, the H\,II regions were identified by using
hierarchical clustering (dendrogram) models with the Python 
\texttt{astrodendro}\footnote{https://dendrograms.readthedocs.io/en/stable/}
package. 39 H\,II regions were thus identified, rejecting any with equivalent
circular diameters below the seeing limit (0.7$''$, or 16 spaxels area).
Fig. \ref{Fig:SteveHaCon}, middle image, shows the H$\alpha$ image of the
galaxy. The ionized gas properties of the H\,II regions were determined by 
Gaussian fitting the cube over the region of the galaxy; most lines were
well separated from the (much brighter) NGC~4361 lines with the
exception of H$\beta$ which is within 6\AA\ of [\ion{O}{III}] 4959\AA.
Table \ref{Tab:HIIRegions} lists the H\,II region 
positions (as offsets from the galaxy centre -- see Tab. \ref{Tab:Steve}), 
fitted radii from \texttt{astrodendro}, H$\alpha$ luminosities (for
a Hubble distance of 87.3 Mpc) and radial velocities, with repect to the 
systemic velocity of the galaxy of 6416 kms$^{-1}$ (Tab. \ref{Tab:Steve}).
Table \ref{Tab:HIIRegionSpectra} then lists the fluxes of the brightest
detected lines ([\ion{O}{III}] 5007\AA, H$\alpha$ and [\ion{N}{II}] 6583\AA\
and [\ion{S}{II}] 6716\AA) from Gaussian fits to the integrated spectrum 
of each H\,II region with errors. The line fluxes have been dereddened by 
the foreground extinction of NGC~4361. The metallicity, $Z$, was determined 
from the O3N2 calibration of \citep{Marinoetal2013} for those (14) H\,II regions 
where the lines required for the O3N2 calibration ([\ion{O}{III}]5007\AA\, 
H$\beta$\footnote{Since flux measurement of H$\beta$ was affected by the presence 
of the nearby strong [\ion{O}{III}] 4959\AA\ line of NGC~4361, the H$\beta$ flux 
was assumed from the H$\alpha$ flux given Case B for $T_{\rm e}$ 10\,000K and 
$N_{\rm e}$ 100 cm$^{-3}$).}, [\ion{N}{II}]6583\AA\ and H$\alpha$) were detected.

% Tab_HIIregions.tex
\begin{table}
\caption{Parameters of {\tt NGC4361-BgGalJ1224290-184707} H\,II Regions}
\centering
\begin{tabular}{lrrrrr}
\hline\hline
ID & $\Delta \alpha$ & $\Delta \delta$ &  Radius &  L(H$\alpha$) &   V~~ \\
   &      ($''$) &      ($''$) & ($''$)~ & (L$_{\odot}$) & (km\,s$^{-1}$) \\ 
\hline
% Error in HII_region_details.csv with RA =-dec and dec=ra fixed
% 40 and 56 without L(Halpha) dropped
 1 &   -5.5 &   -7.3 &  0.68 &  11.0 &        \\ 
 2 &    2.1 &   -7.5 &  0.58 &   4.2 &    180 \\ 
 3 &   -3.5 &   -6.2 &  1.50 &  94.9 &    550 \\ 
 4 &   -1.5 &   -7.0 &  0.51 &   5.8 &     23 \\ 
 5 &    7.0 &   -3.8 &  1.29 &  34.6 &    -16 \\ 
 6 &   -0.6 &   -4.4 &  0.81 &  33.1 &     30 \\ 
 7 &   -5.5 &   -4.6 &  0.56 &  12.2 &     55 \\ 
 8 &   -1.6 &   -1.8 &  0.84 &   120 &     43 \\ 
 9 &    0.9 &   -0.7 &  1.07 &   299 &      1 \\ 
10 &    3.5 &   -1.2 &  0.62 &  42.4 &    -25 \\ 
\hline
\end{tabular}
\tablefoot{
Coordinates with respect to the $V$ band centre of the galaxy 
(12$^{h}$ 24$^{m}$ 29.0$^{s}$, $-$18$^{\deg}$ 47$'$ 07$''$). \\
The H$\alpha$ luminosity is based on a Hubble distance of 87.3 Mpc. \\
The radial velocity is given in the frame of the systemic velocity
of the galaxy, 6416 km\,s$^{-1}$. \newline
Only the first 10 entries of this table are shown; the full table is 
contained in the on-line material.
}
\label{Tab:HIIRegions}
\end{table}

% Tab_HIIspectra.tex
\begin{table*}
\caption{Line fluxes and metallicity of {\tt NGC4361-BgGalJ1224290-184707} H\,II Regions}
\centering
\begin{tabular}{lrrcrcrcrcr}
\hline\hline
ID & F(H$\alpha$) & Error & I([\ion{O}{III}])~~~  & Error & I(\ion{N}{II}])~~~   & Error & I[\ion{S}{II}])~~~   & Error & $Z$(O3N2) & Error \\
   &  (cgs)~~     &       &  5007\AA/H$\alpha$ &       & 6583\AA/H$\alpha$ &       & 6716\AA/H$\alpha$ &       &           &        \\       
\hline
% 40 and 56 without line fluxes not included
 1 &    4.6 &    1.2 &  111.5 &   57.1 &   18.9 &   25.5 &   12.9 &   23.1 &      &      \\ 
 2 &    1.8 &    0.2 &  112.0 &   19.0 &   41.7 &   12.1 &   48.1 &   15.4 & 8.44 & 0.04 \\ 
 3 &   40.0 &    1.3 &   75.6 &    4.6 &    4.0 &    2.1 &   15.5 &    2.6 & 8.21 & 0.07 \\ 
 4 &    2.4 &    0.2 &   29.1 &    7.9 &        &        &   15.0 &    5.0 &      &      \\ 
 5 &   14.6 &    0.9 &   67.0 &    6.1 &        &        &   32.2 &    7.3 &      &      \\ 
 6 &   14.0 &    0.6 &   32.5 &    3.5 &    5.0 &    3.4 &   15.3 &    3.3 & 8.34 & 0.08 \\ 
 7 &    5.2 &    0.8 &   43.0 &   39.2 &   14.5 &    9.2 &   52.0 &   22.2 & 8.43 & 0.13 \\ 
 8 &   50.6 &    1.8 &   65.2 &    3.5 &   10.4 &    1.1 &   14.7 &    8.4 & 8.34 & 0.03 \\ 
 9 &  126.1 &    3.8 &   59.2 &    2.8 &    9.5 &    0.9 &   13.3 &    2.1 & 8.34 & 0.02 \\ 
10 &   17.9 &    0.5 &   58.3 &    2.8 &    6.8 &    1.7 &   12.2 &    2.0 & 8.30 & 0.04 \\ 
\hline
\end{tabular}
\tablefoot{Total flux of H$\alpha$ (F(H$\alpha$) in units of 10$^{-17}$ ergs cm$^{-2}$ 
s$^{-1}$ corrected for the extinction to NGC~4361. Relative intensities of other lines 
with respect to H$\alpha$ = 100.0.
\newline
Metallicity, $Z$, from O3N2 calibration \citep{Marinoetal2013} for those H\,II regions where 
H$\alpha$, [\ion{O}{III}] and [\ion{N}{II}] line fluxes measured. 
\newline
Only the first 10 entries of this table are shown; the full table is 
contained in the on-line material.
}
\label{Tab:HIIRegionSpectra}
\end{table*}

Subtracting the mean PN extinction (see Sect. \ref{SubSect.Extin}), 
the extinction to the galaxy was determined (assuming minimal Galactic 
extinction beyond the PN);
for most H\,II regions the value is compatible with zero but for the
brighter central region (ID. 17 in Tab. \ref{Tab:HIIRegions})
% HII region 17 = Old 35
$E_{B-V}$ = 0.20 (c = 0.14) was found.
In some H\,II regions the [\ion{S}{II}]6716/6731\AA\ ratio
allowed $N_{\rm e}$ to be measured, but errors are large and the results are
compatible with the low density limit. The radial velocity of the galaxy was 
then determined as the mean of the velocities of these H\,II regions (listed 
in Tab. \ref{Tab:Steve}), and their galactocentric radial velocities indicate a 
rotation amplitude of $\sim$ 140 km\,s$^{-1}$ with the SW side receding.

Table \ref{Tab:Steve} lists the properties of this previously hidden 
galaxy. {\tt NGC4361-BgGalJ1224290-184707} has a projected size of $\sim$14 kpc 
making it a similar size to NGC~300 (but note that the whole galaxy is not 
covered in the deep central MUSE field), while it is comparable in luminosity 
to the Small Magellanic Cloud ($M_{V}$ -17.1, \cite{deVaucouleursetal1991}) and mass 
\citep{Besla2015}. The bulge appears to be much brighter
than the disk, however, suggesting that it is likely an Sa or Sb
galaxy (barred or unbarred), as these have more prominent bulges.
In summary, it is a low-mass spiral or a Magellanic irregular 
at $\sim$ 87 Mpc serendipitously viewed through the nebular shell of NGC~4361.

\begin{table}
\caption{{\tt NGC4361-BgGalJ1224290-184707} properties}
\centering
\begin{tabular}{lr}
\hline\hline
Parameter & Value \\
\hline
$\alpha$, $\delta$ (J2000) & 12$^{h}$24$^{m}$29.0$^{s}$, $-$18$^{\circ}$47$'$07$''$ \\
$v$ (km\,s$^{-1}$ & 6416 $\pm$ 3 \\
$z$ & 0.021401 $\pm$ 0.000010 \\
D (Mpc) & 87.3 \\
L(H$\alpha$) & 1.2 $\times$ 10 $^{40}$ erg s$^{-1}$ \\
% log(SFR) & $-$1.03 M$_{\odot}$ yr$^{-1}$ \\
%$Z$ & 8.39 $\pm$ 0.11 \\
$Z$ & 8.35 $\pm$ 0.10 \\
% Mean Z and error on mean
$m_{V}$ & 17.38 mag. \\
$M_{V}$ & $-$17.33 mag.\\
($V-I$) & +0.66 mag. \\
Sersic Index (V) & 0.92 \\
% Effective Radius (V) & 2.70 kpc \\
Stellar Mass & 3.4 $\times$10$^{8}$ M$_{\odot}$ \\
\hline
\end{tabular}
\tablefoot{
Distance, D, for H$_{0}$ = 72.5 km s$^{-1}$ Mpc$^{-1}$ \citep{Riessetal2022}. \\
Mean metallicity, $Z$, based on the O3N2 calibration \citep{Marinoetal2013} 
for the 14 H\,II region listed in Tab. \ref{Tab:HIIRegionSpectra}. \\
Stellar mass estimated from $M_{V}$ and ($V-I$) according to \citet{BelldeJong2001}.
}
\label{Tab:Steve}
\end{table}

%__________________________________________________________________

\section{Conclusions}
\label{Sect:Concl}
% Now shortened from conclusions.txt -> conclusions2.txt

NGC~4361 was observed during MUSE Commissioning in 2014. In total 76 MUSE
exposures were obtained (amounting to 4560s) and preliminary examination revealed
that the quality of most of the data was good. The reduced data have opened up new 
perspectives on this enigmatic PN.

\begin{itemize}
\item NGC~4361 shows an exceptionally high level of ionization as demonstrated 
by the \ion{He}{II}/\ion{H}{I} line ratios, and had been classified as an optically thin 
(density bounded) PN. Prior to this study no low ionization CEL species had been detected.
New evidence that the nebula is optically thin in the H ionizing continuum
comes from the very low \ion{He}{I} emission. The lack of decrease in 
He$^{++}$/H$^{+}$ with radius raises the question if the medium is, at least
partially, optically thin in the He$^{++}$ ionizing continuum also. 
Perhaps deeper imaging in the outer regions could
provide indications that \ion{He}{II} recombines at the extremities. 

\item This study is the first to find the expected, but very weak, low ionization
emission, both as extended emission in some regions of the nebula and, most 
spectacularly as a family of compact (unresolved at 0.73$''$ resolution) knots,
dubbed 'Freckles'. The Freckles are most prominently detected in [\ion{N}{II}], 
but also in [\ion{O}{II}] and some in [\ion{O}{I}]. 102 were detected in this study,
but deeper imaging may reveal many more. 

\item A map of $T_{\rm e}$ from the [\ion{S}{III}] line ratio (Fig. \ref{Fig:S3Temap}) 
shows the regions to the W and NE, which display mid and low ionization extended 
emission, have differing $T_{\rm e}$ with a differences up to $\sim$4000\,K, actually
larger than found by \citet{Liu1998} from a long slit placed 10$''$ E of the CS. 
This large-scale temperature difference in a PN is very 
distinct from the usual temperature differences, such as measured by 
different diagnostic ratios, and the temperature fluctuation effects observed 
in many PNe (such as between CEL and ORL diagnostics, see e.g., \citet{Liu2006}). 
$N_{\rm e}$ was determined from [\ion{Cl}{III}] and [\ion{S}{II}] line ratios and shows 
values around 1500 cm$^{-3}$ (Tab. \ref{Tab:CELDiags}).
  
\item The Freckles were studied in some depth - they behave much like a typical
mid-ionization medium with $T_{\rm e}$ $\sim$~11000\,K (from the diagnostic 
[\ion{N}{II}] ratio) and $N_{\rm e}$ $\sim$1500 cm$^{-3}$. They generally show larger 
extinction  than the high ionization nebula, indicating a distinct
dust component, but not significantly higher $N_{\rm e}$. 
Sect. \ref{SubSec:Freckles2} and Tab. \ref{Tab:FreckleSpecs} present
the extracted spectra of the Freckles. The lower ionization of these knots is very 
striking and, for example, the He/H abundance of $\sim$0.1 is wholly due to He$^{+}$, 
whilst in the high ionization gas the majority is He$^{++}$. There is no compelling
evidence that the Freckles are enriched in either He or N, with respect to the high
ionization medium; [O/H] for the Freckles is higher than values found 
for the high ionization medium from earlier studies. 
Comparison to low ionization knots in other PNe, such as the cometary globules in the
Helix Nebula (NGC~7293) do not show similar structure (compact extinction core and 
ionized head and tail), but the $\sim$5$\times$ greater distance of NGC~4361 does not allow
further resolution of their structure from these data.

\item The spatial distribution of the Freckles with several groups of almost
linear alignments invites the suggestion that
they are ejecta from the CS, but no convincing alignments back to the CS
are found. The low spectral resolution of MUSE only allows limited data on
the velocity structure of the Freckles but there is a clear trend between
those Freckles to the W of an axis at PA $\sim$140$^{\circ}$ (blue-shifted with
respect to the bulk emission) and those to the East (red shifted). 
The picture is advanced that 
the Freckles are aligned in a thick disk, perhaps of larger radius than the large-scale 
high ionization structure (NE-SW orientation) and with perpendicular orientation. 
The Freckles could be part of an older structure, such as from the AGB phase, especially as they 
are dusty, or from an earlier epoch. High spectral resolution would help to elucidate their 
velocity structure and relation to the bulk high ionization gas.
 
\item In the western lobe of the nebula an emission line galaxy viewed through the
nebula emission was found on panning through the MUSE cube. Further investigation 
showed this to be a typical H\,II region galaxy and the redshift was 0.0214. 
Carefully extracting the galaxy 
spectrum from the bright nebula emission, a low-luminosity ($M_{V}$ -17.3 mag.) 
disk galaxy with bright H\,II regions was revealed. The gas phase metallicity is 
[O/H] $\cong$ 8.4 and it is suggested as a Magellanic irregular or low-mass spiral
galaxy at 87 Mpc.
\end{itemize}

%______________________________________________________________

\begin{acknowledgements}
We thank the MUSE Team for boldly taking the observations of this unusual
and less-studied PN during the instrument commissioning time. 

% Gaia acknowledgement:
This work has made use of data from the European Space Agency (ESA) mission
Gaia (\url{https://www.cosmos.esa.int/gaia}), processed by the Gaia
Data Processing and Analysis Consortium (DPAC,
\url{https://www.cosmos.esa.int/web/gaia/dpac/consortium}). Funding for the DPAC
has been provided by national institutions, in particular the institutions
participating in the Gaia Multilateral Agreement. \\

%PyNeb
This research has greatly benefited from the use of PyNeb \citep{Luridiana2015}, 
which has vastly homogenized calculations of nebular physical conditions and
abundances. We are also grateful to the communities who have developed the many 
Python packages used in this research, such as MPDAF \citep{Piquerasetal2019}, 
Astropy \citep{AstropyCollaboration13, AstropyCollaboration18,AstropyCollaboration22}, 
numpy \citep{Walt11}, scipy \citep{Jones01} and matplotlib \citep{Hunter07}.

\end{acknowledgements}

%______________________________________________________________

\bibliographystyle{aa} % style aa.bst
\bibliography{N4361}

%______________________________________________________________

\newpage

\begin{appendix}

\section{Spectrum of the central star of NGC~4361}
\label{App:CS}

\begin{figure*}
%
% Old for one single plot
% \centering
% \resizebox{\hsize}{!}{
% \includegraphics[width=0.45\textwidth,angle=0,clip]{NGC4361_CSspect.pdf}
%
\centering
\resizebox{\hsize}{!}{
\includegraphics[width=0.45\textwidth, angle=0, clip]{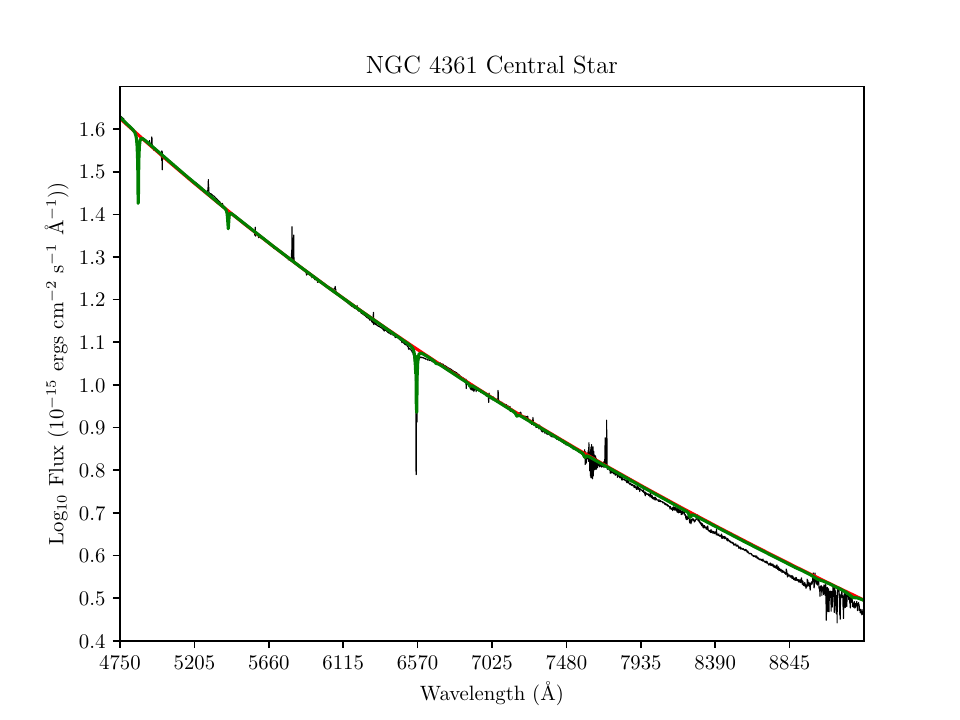}
\vspace{0.1truecm}
}
\resizebox{\hsize}{!}{
\includegraphics[width=0.50\textwidth, angle=0, clip]{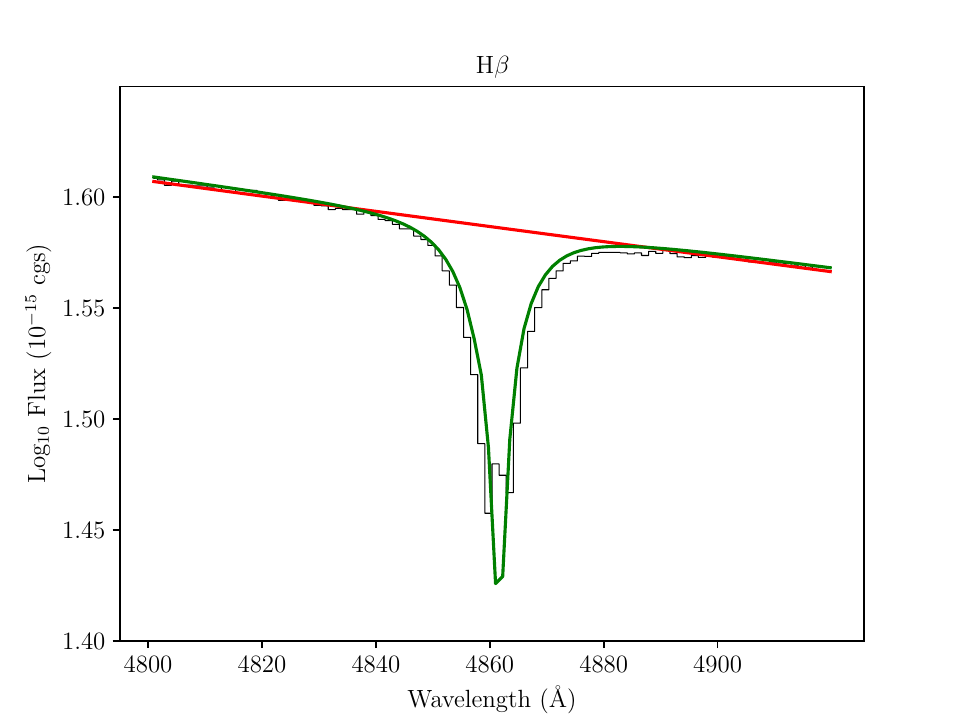}
\hspace{0.1truecm}
\includegraphics[width=0.50\textwidth, angle=0, clip]{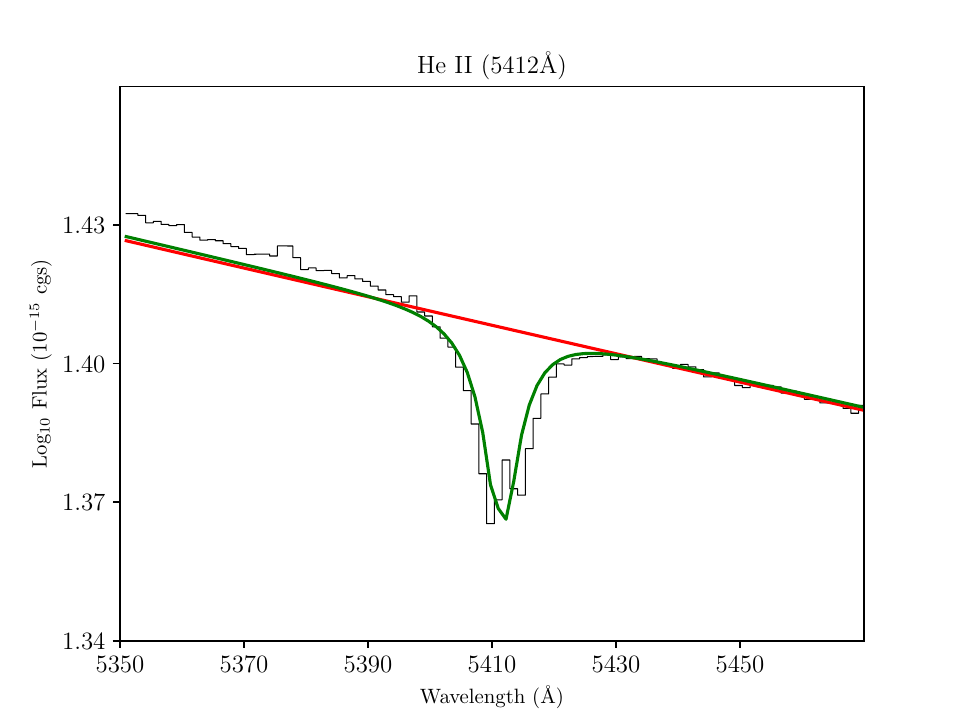}
\hspace{0.1truecm}
\includegraphics[width=0.50\textwidth, angle=0, clip]{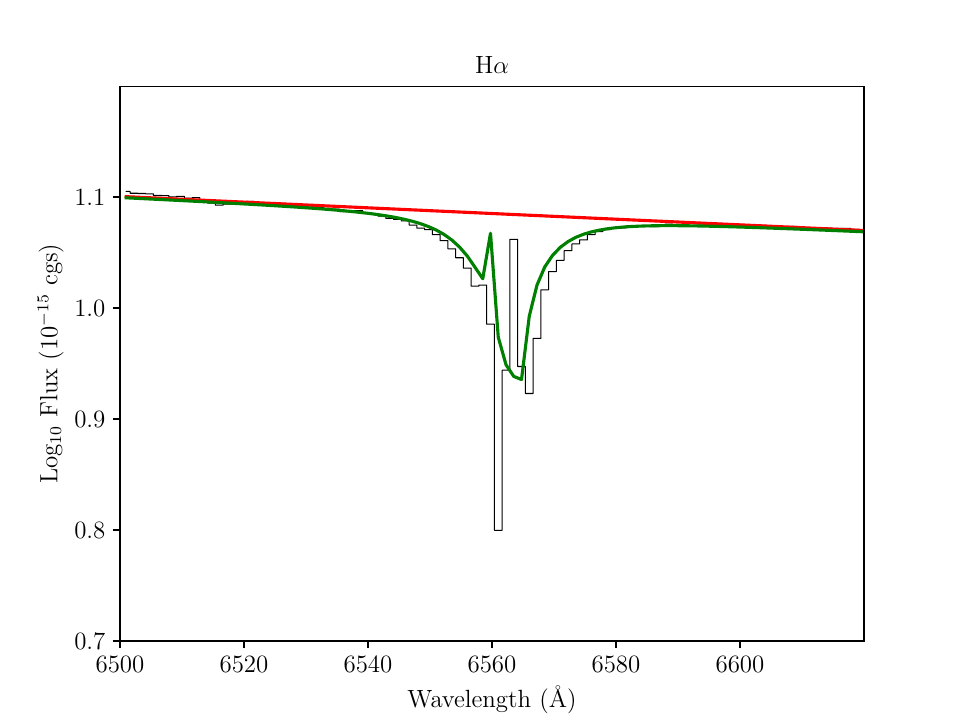}
}
% See NGC4361/Regions/CS for plotting scripts (plot_CS.py and CS+3detail+Hb.py,
% plot_CS+3detail+He2.py and plot_CS+3detail+Ha.py), and pdfs
\caption{Upper: The dereddened spectrum (c=0.10) of the central star (CS) of NGC~4361
is shown, together with a black body fit (T = 89\,000\,K, in red) and a \citet{Rauch2003} 
model atmosphere for T$_{eff}$ 80\,000\,K, log $g$ = 5.5 and H:He = 0.9:0.1 in green. \\
Lower: Details of the matches of the BB fit and the model atmosphere match for H$\beta$
(left), He~II 5412\AA\ (middle) and H$\alpha$ (right).} 
\label{Fig:SpecCS}
\end{figure*}

The spectrum of the central star of NGC~4361 was extracted over an 
area of 42.2 arcsec$^{2}$ (effective radius 3.7$''$) chosen 
to include $\sim$91\% of the V-band stellar flux (since the local 
nebula background varies spatially, the actual flux loss is uncertain).
Four rectangular background areas NE, SE, SW, NW of the CS were chosen 
to avoid the prominent cardinal scattering pattern of the point spread 
function, and the background subtracted stellar spectrum was 
formed by subtracting the mean background value from the
star image. Fig. \ref{Fig:SpecCS} shows the resulting spectrum,
dereddened by the mean value for the integrated nebula (c=0.10).
Both absorption (\ion{H}{I} and \ion{He}{II}) and emission lines are
detected and Tab. \ref{Tab:SpecCS} lists the reliably 
identified lines by matching with the Atomic Line List of
\citet{vanHoof2018}.
Note that \ion{C}{IV} is present in the stellar spectrum and
as extended emission with a morphology similar to the \ion{He}{II} 
emission.
% IMAGE NOT SHOWN (see \ref{Fig:FluxImages3}).

\begin{table}
\caption{NGC~4361 stellar spectrum - detected lines}
\centering
\begin{tabular}{llrl}
\hline\hline
Obs $\lambda$  & Species & Rest $\lambda$ & Abs/Emis \\
~~~~(\AA)      &         & (\AA)~~~       & (A/E)~~   \\
 \hline
4861  &  H$\beta$ & 4861.3          & A \\
4931  &  \ion{O}{V} & 4930.3        & E \\
4945  & \ion{C}{V} & 4944.5, 4944.8 & E \\
5291  & \ion{O}{VI} & 5291.0        & E \\
5412  &  \ion{He}{II} & 5411.5      & A \\
5801  &  \ion{C}{IV}  & 5801.4      & E \\
5812  &  \ion{C}{IV}  & 5812.0      & E \\
6067  &  [\ion{Mn}{V}] ? & 6066.2   & E \\ 
6201  &   \ion{C}{VI} & 6200.6      & E \\
6275  &  \ion{N}{IV}  & 6274.8      & E \\
6563  & H$\alpha$ & 6562.8          & A \\
7063  & \ion{C}{IV} & 7063.0, 7063.2 & E \\
% Line not very convincing and no feasible high IP ID. DROP!
% 7276  &  ?          &               & E \\
7316  &  \ion{N}{V} & 7316.0        & E \\
7709  &  \ion{C}{III} & 7706.8      & E \\
7717  &  \ion{C}{VI} & 7716.8       & E \\
7726  &  \ion{C}{IV} & 7725.8       & E \\
8229  &  \ion{C}{IV} & 8229.5       & E \\
8282  &  \ion{C}{VI} & 8282.1       & E \\
% Dropped 8989  &  ? &            & E \\
% Dropped 9002  & \ion{C}{III} & E \\
\hline
\end{tabular}
% \tablefoot{}
\label{Tab:SpecCS}
\end{table}

A black body was fitted to the dereddened stellar spectrum in 
Fig. \ref{Fig:SpecCS} and a value of 89000\,K is indicated. However
beyond 8200\AA\ the black body fit deviates from the dereddened 
continuum flux:
this could arise from flux calibration issues (the flux standard 
observed with the observations was GD 108 \citep{Bohlinetal2014, 
Oke1990}) and/or aperture losses associated with the broader MUSE PSF 
to longer wavelengths. A match with the hot model atmospheres of 
\citet{Rauch2003} was also sought. \citet{Mendez1992} lists a
model atmosphere fit with T$_{eff}$ 82000\,K and log $g$ 5.5 from
non-LTE model atmosphere analysis.
The \citet{Rauch2003} T$_{eff}$ 80000\,K H+He models (log $g$ 5.5) fit
the continuum well but for a 10\% He atmosphere both the H$\beta$ and
\ion{He}{II} 5412\AA\ absorption model line profiles are too narrow 
and shallow (Fig. \ref{Fig:SpecCS}). 
A better match in terms of line width and absorption depth
to these two lines is found for a 20\% He atmosphere, which however
considerably exceeds the He/H abundance measured for the
nebula and expected for a typical PN central star. For the 90000\,K,
log $g$ 5.5, 10\% He model, the fit to the H$\beta$ profile shape
is good but the line too deep, while the \ion{He}{II} line fit is to
weak and narrow. The fit for the \ion{He}{II} line again improves 
for a 20\% He atmosphere but is still slightly too weak and narrow. We 
conclude that a value of T$_{eff}$ of around 85000\,K 
and log $g$ of 5.5 with a 10-15\% He contribution appears to be an 
adequate description of the CS matching this spectrum. However the
much higher T$_{eff}$ of 126kK listed by \citet{Ziegleretal2012} remains 
puzzling.

% And little table of atomic data used in Python2 version of PyNeb.
% \input{Appendix_AtomicData.tex}

\section{Atomic data used in PyNeb diagnostic CEL line ratios and abundances}
\label{App:AtomicData}

Table \ref{Tab:AtomicData} lists the sources for the collision strengths and 
transition probabilities used for the PyNeb diagnostic CEL ratios and the abundance 
calculations.

\begin{table}
\caption{References for Atomic Data}
\centering
\begin{tabular}{lll}
\hline\hline
Ion  & Collision  & Transition    \\
     & strengths  & probabilities \\
\hline

\ion{N}{+}   & \citet{Tayal2011} & \citet{FroeseFischer2004} \\
\ion{O}{+}   & \citet{Kisieliusetal2009} & \citet{FroeseFischer2004} \\
\ion{O}{++}  & \citet{StoreyZeippen2000} & \citet{FroeseFischer2004} \\
\ion{S}{+}   & \citet{Tayal2010} & \citet{Podobedovaetal2009} \\
\ion{S}{++}  & \citet{Tayal1999} & \citet{Podobedovaetal2009} \\
\ion{Cl}{++} & \citet{ButlerZeippen1989} & \citet{MendozaZeippen1982} \\
             &                           & \citet{KauffmanSugar1986} \\ 
\ion{Ar}{++} & \citet{MunosBurgos2009} & \citet{MunosBurgos2009} \\
\hline
\end{tabular}
% \tablefoot{}
\label{Tab:AtomicData}
\end{table}

\end{appendix}

\end{document}